\documentclass[iop]{emulateapj}

\usepackage{appendix,natbib}
\usepackage{amsmath}
\usepackage{subfigure}
\usepackage[backref,colorlinks=true,citecolor=blue,linkcolor=blue,urlcolor=blue]{hyperref}
\usepackage[all]{hypcap}

\usepackage{amsmath}
\usepackage{appendix}
\usepackage[utf8]{inputenc}
\usepackage{listingsutf8}
\usepackage{subfigure}
\usepackage{color}
\usepackage{xcolor}
\usepackage[normalem]{ulem} 
\renewcommand{\tabcolsep}{0.15cm}

\newcommand{\halpha}{H\ensuremath{\alpha}}
\newcommand{\hbeta}{H\ensuremath{\beta}}
\newcommand{\um}{$\mu$m}
\def\msun{\,${M_\odot}$}

\newcommand{\oh}{$12+\log({\rm O/H})$}

\def\R09{\hyperlink{R09}{R09}}
\def\S18{\hyperlink{S18}{S18}}
\def\L16{\hyperlink{L16}{L16}}
\def\K15{\hyperlink{K15}{K15}}
\def\B18{\hyperlink{B18}{B18}}
\def\PP04{\hyperlink{PP04}{PP04}}
\slugcomment{Draft: {\today}, accepted for publication in ApJ}

\begin{document}

\title{IR SED and Dust Masses of Sub-solar Metallicity Galaxies at $\MakeLowercase{z}\sim 2.3$}

\author{\sc Irene Shivaei\altaffilmark{1,2}, Gerg\"{o} Popping\altaffilmark{3}, George Rieke\altaffilmark{1}, Naveen Reddy\altaffilmark{4}, Alexandra Pope\altaffilmark{5}, Robert Kennicutt\altaffilmark{1,6}, Bahram Mobasher\altaffilmark{4},
Alison Coil\altaffilmark{7}, Yoshinobu Fudamoto\altaffilmark{8,9}, Mariska Kriek\altaffilmark{10}, Jianwei Lyu\altaffilmark{1}, Pascal Oesch\altaffilmark{11,12}, Ryan Sanders\altaffilmark{13,2}, Alice Shapley\altaffilmark{14}, Brian Siana\altaffilmark{4}
}
\altaffiltext{1}{Steward Observatory, University of Arizona, Tucson, AZ 85721, USA}
\altaffiltext{2}{NASA Hubble fellow}
\altaffiltext{3}{European Southern Observatory, Karl-Schwarzschild-Str. 2, D-85748, Garching, Germany}
\altaffiltext{4}{Department of Physics and Astronomy, University of California, Riverside, 900 University Avenue, Riverside, CA 92521, USA}
\altaffiltext{5}{Department of Astronomy, University of Massachusetts, Amherst, MA 01003, USA}
\altaffiltext{6}{George P. and Cynthia W. Mitchell Institute for Fundamental Physics \& Astronomy, Texas A\&M University, College Station, TX 77843, USA}
\altaffiltext{7}{Center for Astrophysics and Space Sciences, University of California, San Diego, 9500 Gilman Drive, La Jolla, CA 92093-0424, USA}
\altaffiltext{8}{Research Institute for Science and Engineering, Waseda University, 3-4-1 Okubo, Shinjuku, Tokyo 169-8555, Japan}
\altaffiltext{9}{National Astronomical Observatory of Japan, 2-21-1, Osawa, Mitaka, Tokyo, Japan}
\altaffiltext{10}{Astronomy Department, University of California, Berkeley, Berkeley, CA 94720, USA}
\altaffiltext{11}{Department of Astronomy, University of Geneva, Ch. Pegasi 51, 1290 Versoix, Switzerland}
\altaffiltext{12}{Cosmic Dawn Center (DAWN), Niels Bohr Institute, University of Copenhagen, Jagtvej 128, K\o benhavn N, DK-2200, Denmark}
\altaffiltext{13}{Department of Physics, University of California, Davis, One Shields Ave, Davis, CA 95616, USA}
\altaffiltext{14}{Department of Physics \& Astronomy, University of California, Los Angeles, 430 Portola Plaza, Los Angeles, CA 90095, USA}

\begin{abstract}
We present results from ALMA 1.2\,mm continuum observations of a sample of 27 star-forming galaxies at $z=2.1-2.5$ from the MOSFIRE Deep Evolution Field (MOSDEF) survey. These galaxies have gas-phase metallicity and star-formation rate measurements from {\hbeta}, [O{\sc iii}], {\halpha}, and [N{\sc ii}]. Using stacks of Spitzer, Herschel, and ALMA photometry  (rest-frame $\sim 8-400$\,{\um}), we examine the IR SED of high-redshift subsolar metallicity ($\sim 0.5\,Z_{\odot}$) LIRGs. 
We find that the data agree well with an average SED template of higher luminosity local low-metallicity dwarf galaxies (reduced $\chi^2$ of 1.8). When compared with the commonly used templates for solar-metallicity local galaxies or high-redshift LIRGs and ULIRGs, even in the most favorable case (with reduced $\chi^2$ of 2.8), the templates are rejected at $>98\%$ confidence level.
The broader and hotter IR SED of both the local dwarfs and high-redshift subsolar metallicity galaxies may result from different grain properties, a clumpy dust geometry, or a harder/more intense ionizing radiation field that heats the dust to higher temperatures.
The obscured SFR indicated by the FIR emission of the subsolar metallicity galaxies is only $\sim 60\%$ of the total SFR, which is considerably lower than that of the local LIRGs with $\sim 96-97\%$ obscured fractions.
Due to the evolving IR SED shape, the local LIRG templates fit to mid-IR data can overestimate the Rayleigh-Jeans tail measurements at $z\sim 2$ by a factor of $2-20$, and these templates underestimate IR luminosities if fit to the observed ALMA fluxes by $>0.4$\,dex.
At a given stellar mass or metallicity, dust masses at $z\sim 2.3$ are an order of magnitude higher than those at $z\sim 0$.
Given the predicted molecular gas mass fractions, the observed $z\sim 2.3$ dust-to-stellar mass ratios suggest lower dust-to-molecular gas masses than in local galaxies at the same metallicity. CO observations are necessary to better constrain the molecular gas content of sub-solar metallicity galaxies at $z>1$.

\end{abstract}

\maketitle

\section{Introduction}

The infrared (IR) emission of dust in galaxies accounts for a significant fraction of their bolometric luminosity and encodes critical clues to how it is produced. 
By mass, dust only represents $\sim 1\%$ of the ISM in typical galaxies. However, it reshapes galaxy spectral energy distributions (SEDs) by attenuating and absorbing UV-optical photons and reradiating that energy in the IR. The resulting IR emission accounts for approximately half of the cosmic extragalactic background \citep{dole06,finke10}, and the bulk of the cosmic star formation at $z\sim 0-3$ is detected in the IR \citep{madau14,planck14,casey18}. 

IR SEDs consist of a roughly Planckian and featureless far-IR (FIR) component plus emission features of aromatic molecules in the $\sim 6-20$\,{\um} range.  While the mid-IR spectra ($\lambda=5-30$\,{\um}) are dominated by the emission from small grains that are stochastically heated by single photons, the longer wavelength FIR and submm emission comes from larger grains that are in thermal equilibrium. The shape of the FIR/submm SED depends on the dust composition (which determines the submm spectral slope), the distribution of radiation field intensities on the dust, and the dust grain size distribution, which together determine the peak and width of the IR SED. In a comprehensive study of local galaxies, \citet{remyruyer15} showed that while there are many commonalities, distinct differences exist in the IR SEDs of low-metallicity dwarfs and metal-rich star-forming local galaxies. They found that, on average, the low-metallicity galaxies have broader IR SEDs that peak at shorter wavelengths compared  to  those  of  the  metal-rich  galaxies. 

\citet{remyruyer15} attributed the differences in the IR SEDs to a wider range of interstellar radiation field intensities ($\sigma U$), with a higher average radiation field intensity ($\langle U\rangle$\footnote{The intensity $U$ of the interstellar radiation field \citep{dale01} determines both the shape and normalization of the IR SED.}) in low-metallicity dwarfs due to their high specific star formation rates (sSFRs; sSFR$={\rm SFR}/M_*$). In a sample of local galaxies with oxygen abundances of \oh$\sim 8.2-8.8$, \citet{cortese14} also found a strong correlation between the wavelength-dependent emissivity index of the dust ($\beta$) and metallicity, with only a weaker anti-correlation with sSFR. These results illustrate that the effects of metallicity and of the radiation field on the integrated IR SEDs of galaxies cannot be easily separated, as the two parameters themselves are interconnected. Low-metallicity systems are often dominated by young stellar populations that have relatively hard and intense radiation fields. They also have lower dust-to-gas mass ratios \citep[e.g.,][]{draine07b, remyruyer14} that make their ISM more transparent to the stellar radiation, allowing massive star formation to impact the ISM over a larger volume. These effects make the ISM metallicity a good tracer of dust evolution processes in galaxies, as it traces both the elemental abundances in the ISM and the interstellar radiation field intensity that the dust grains are exposed to.

However, there are very few studies of the effect of metallicity on dust emission properties outside the local Universe, due both to the sensitivity limitations of the available IR-submm data and to the lack of robust metallicity measurements. Recent studies have shown that at a given stellar continuum reddening (UV slope), the IR to UV luminosity ratio (IRX) of $z\sim 2$ star-forming galaxies correlates with metallicity in the range of {\oh}$\sim 8.3-8.6$ \citep{shivaei20b} or with stellar mass \citep{reddy18a,fudamoto19}. This may indicate a change in grain properties or dust-star geometry with metallicity. Moreover, in \citet{shivaei17} we showed that the mid-IR aromatic (PAH) emission relative to the total IR luminosity of $z\sim 2$ galaxies decreases at \oh$\lesssim 8.3-8.4$, similar to the behavior seen in the local Universe but at \oh$\le 8.2$ \citep[][among many more]{engelbracht05,draine07b,hunt10,marble10,li20}. These studies suggest that the emission properties of dust at $z\sim 2$ vary significantly between solar and subsolar metallicity galaxies. 
The main goal of this work is to investigate the FIR emission of $z\sim 2.3$ subsolar metallicity galaxies and assess whether the broadening of the FIR SED seen locally in low metallicity galaxies occurs at sub-solar metallicities at high redshifts. Such behavior, if confirmed, would provide valuable insight into the dust and gas properties of high redshift galaxies. Additionally, it will have implications for the calibration of the JWST, ALMA, and other IR/submm measurements interpreted as SFR, IR luminosity, and dust mass indicators.
To reach this goal deep observations across the wavelengths from mid-IR to submm are required to detect the IR emission of the lower luminosity, and hence lower metallicity, galaxies at high redshifts.

In this work, we present the first results of a targeted ALMA band-6 (1.2\,mm) continuum survey, tracing rest-frame submm emission of a sample of 27 star-forming galaxies at $z=2.1-2.5$ drawn from the MOSFIRE Deep Evolution Field (MOSDEF) survey \citep{kriek15}. The galaxies have robust metallicity and star-formation rate (SFR) measurements from optical emission lines ({\halpha}, {\hbeta}, [O{\sc iii}], [N{\sc ii}]), and span a wide range in oxygen abundance from \oh$=8.1$ to 8.8. The availability of this detailed prior information enables this targeted survey of faint subsolar metallicity galaxies with ALMA. Compared to the local galaxies with similar metallicities, the $z\sim 2.3$ galaxies have higher SFRs and SFR surface densities, which make this analysis unique.

ALMA band-6 traces the Rayleigh Jeans (RJ) tail of the dust emission at rest-frame $\sim 360$\,{\um} for our sample. 
Combining the ALMA data with the shorter wavelength Spitzer and Herschel data at rest-frame 7 to 100\,{\um}, we can test whether the relatively broad FIR SEDs seen in local low metallicity galaxies are also typical at $z > 1$. 
By constraining the change in IR/submm SEDs with metallicity, we provide insight into the integrated dust properties of high-redshift galaxies (such as IR luminosity and obscured SFR), determine the masses of dust, study the dust mass fraction (dust-to-stellar mass) versus metal fraction (aka metallicity) at $z\sim 2.3$, and explore its redshift evolution by incorporating various $z\sim 0$ surveys from the literature. Our results extend over 0.7\,dex in metallicity and down to stellar masses of $\sim 10^{9.7}$\,{\msun}, exploring a new parameter space for galaxy evolution studies at these redshifts.

The outline of the paper is as follows. In Section~\ref{sec:sample-data}, we present the sample and data, and explain the data reduction. Section~\ref{sec:irfits} presents the analysis and construction of IR SEDs. Section~\ref{sec:shape_irsed} discusses outcome of the IR SED analysis, how it changes with metallicity and star-formation properties, and its physical interpretation. In Section~\ref{sec:implications}, we discuss the implications for measuring IR luminosity and SFR, as well as predicting submm fluxes based on shorter wavelength data. The evolution of dust mass fraction versus redshift and metallicity, and the comparison of dust-to-gas mass ratios with local galaxies are covered in Section~\ref{sec:dustmass}. Our results are summarized in Section~\ref{sec:summary}. We discuss the different models and IR templates that are used to fit the data in Appendix~\ref{app}. The systematic uncertainties in calculating dust masses from submm fluxes are discussed in Appendix~\ref{app:dmass}.
A cosmology with $H_0 = 70$\,km\,s$^{-1}$\,Mpc$^{-1}$, $\Omega_{\Lambda}=0.7$, and $\Omega_m=0.3$, and a \citet{chabrier03} IMF are adopted.

\section{Sample and Observations} \label{sec:sample-data}
Our sample is selected from the MOSDEF near-IR spectroscopic survey \citep{kriek15}. MOSDEF is a survey of $\sim 1500$ galaxies at $z=1.3-3.8$ with near-IR spectra from the MOSFIRE spectrometer on Keck, primarily in the COSMOS, GOODS-N and AEGIS fields (a small fraction of the MOSDEF galaxies is located in the GOODS-S and UDS fields). Using the catalogs of the 3D-HST survey \citep{skelton14}, the MOSDEF parent sample was selected based on the available photometric and spectroscopic redshifts in three redshift windows of $z=1.4-1.7$, $z=2.1-2.6$, and $z=3.0-3.8$ down to $H$-band (AB) magnitudes of 24.0, 24.5, and 25.0, respectively. In these redshift ranges, most of the prominent optical emission lines are observed in the $YJHK$ bands of MOSFIRE. 

\subsection{Sample Characteristics} \label{sec:sample-char}

\begin{figure}[t]
	\centering
		\includegraphics[width=.5\textwidth,trim={.2cm 0 0 0},clip]{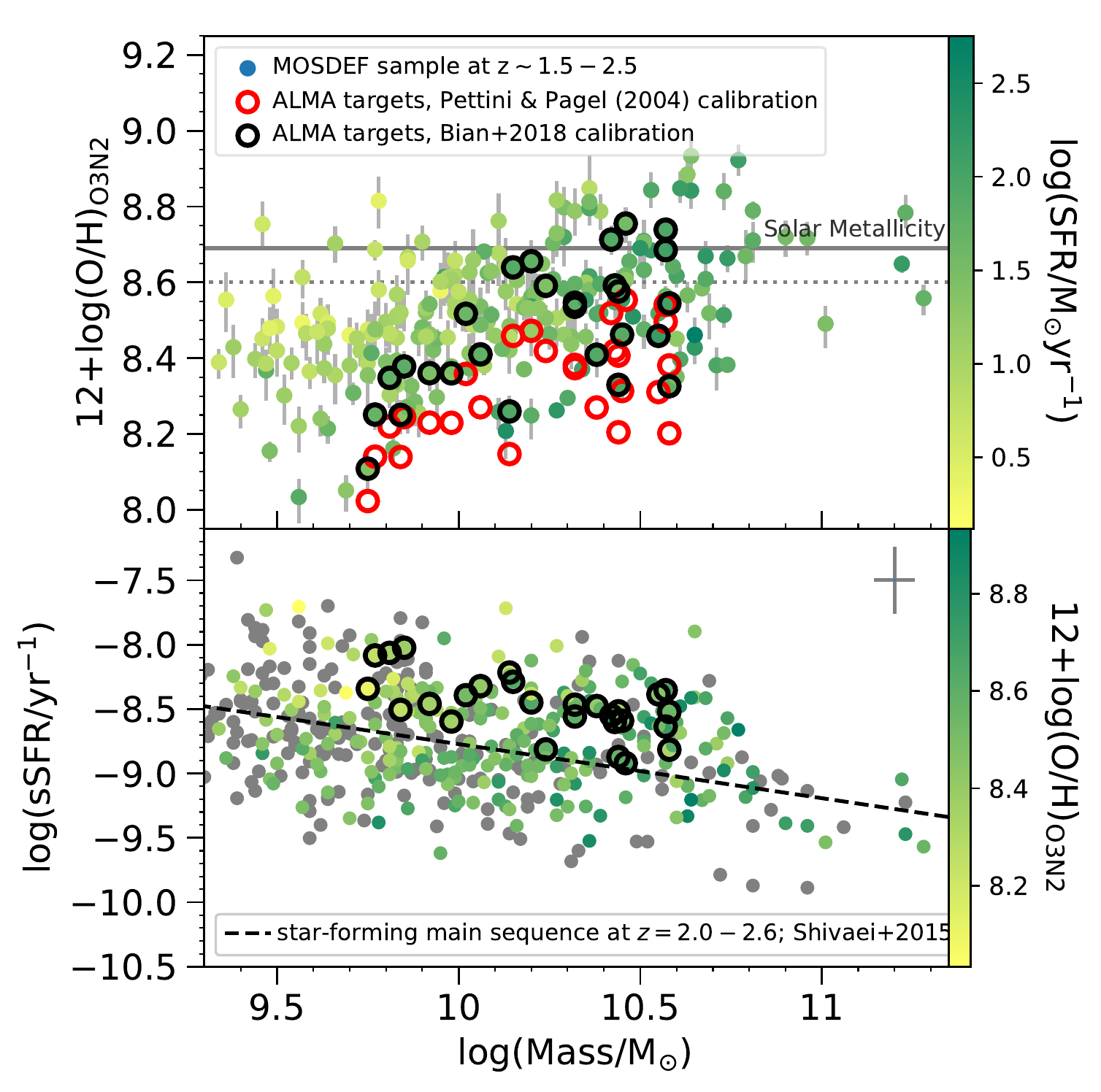} 

		\caption{The metallicity, mass, and specific SFR distributions of the ALMA targets (open black circles) with respect to those of the MOSDEF parent sample (small filled circles).
		Metallicities are derived from O3N2 line ratios adopting the \citet{bian18} calibration (the MOSDEF sample with filled circles and the ALMA sample with black open circles). The \citet{pp04} calibration estimates $\sim 0.15$\,dex lower metallicities from the same O3N2 line ratios. The \citet{pp04} metallicities for the ALMA sample are shown by red open circles (see Section~\ref{sec:ONIR_data_metal} for the discussion on metallicity).
		The solid horizontal line in the top panel indicates solar metallicity (\oh$=8.69$ from \citealt{asplund09}).
		We divide the sample into two bins for stacking at \oh$=8.6$ (dashed line), using the \citet{bian18} calibration (\oh$=8.43$ using the \citealt{pp04} calibration).
		The dashed line in the bottom panel is the star-forming main sequence relation from \citet{shivaei15b} for {\halpha}-inferred SFRs at $z=2.09-2.61$.
		SFRs are estimated from {\halpha}, assuming the conversion factors in Equations \ref{eq:sfr1} and \ref{eq:sfr2}, corrected for attenuation using the Balmer decrement (Section~\ref{sec:ONIR_data_SFR}). The grey circles in the bottom panel do not have metallicity estimates due to the lack of one or more of the lines used to compute O3N2.
		}
		\label{fig:mz-mssfr}
\end{figure}

The ALMA survey presented here (program 2019.1.01142.S; PI: I.~Shivaei) targeted MOSDEF galaxies in the COSMOS field for accessibility with ALMA. There are 629 MOSDEF-targeted galaxies in this field, of which 209 are within the correct redshift range ($2.0 < z < 2.5$), such that the ALMA 1.2\,mm and Spitzer MIPS 24\,{\um} spectral bands trace the cold dust and 7.7\,{\um} PAH emission, respectively. We further required (1) $> 3\sigma$ sigma detections in detection in the  {\halpha}, {\hbeta}, [O{\sc iii}], and [N{\sc ii}] lines to ensure robust estimates of SFR and metallicity \citep{reddy15,sanders15,shapley15}, which narrowed down the sample to 51; (2) unconfused MIPS 24\,{\um} photometry \citep{shivaei17}; (3) no evidence for AGN, based on X-ray emission, IRAC colors, and [N{\sc ii}]/{\halpha} line ratios \citep{coil15,azadi17,azadi18,leung19}; and (4) rest-frame $U-V$ and $V-J$ colors consistent those of star forming galaxies \citep{zick18}. Of the 40 galaxies remaining, we select the final sample of 27 to represent a wide range in mass and metallicity. The mass, metallicity, and sSFR distributions of galaxies in our sample are shown in Figure~\ref{fig:mz-mssfr}. 

The MOSDEF parent sample is largely representative of typical main-sequence star-forming galaxies \citep[e.g.,][]{shivaei15b}. However, as the parent sample is selected based on rest-frame optical photometry, there is a bias towards galaxies with redder UV and optical colors at a given UV magnitude compared to UV-selected samples \citep{reddy18b}. On the other hand, the subsample that is selected for ALMA followup in this work is selected based on strong detection in multiple optical emission lines, and may exclude highly dust-obscured galaxies with simple dust-star geometries\footnote{Depending on the dust-star geometry, it is possible to have highly dust-obscured systems that also exhibit optical lines \citep[e.g.,][]{shivaei16}}. 
The well-measured emission lines allow use of the Balmer decrement ({\halpha} to {\hbeta} flux ratio) to derive corrections for reddening of the nebular lines to recover the total (intrinsic) nebular emission. The success of this approach for a {\halpha}-{\hbeta} detected sample is illustrated in \citet{shivaei16} who compare {\halpha} SFR estimates corrected for reddening via the Balmer decrement with IR-based SFR estimates.

\begin{figure*}[t]
	\centering
		\includegraphics[width=1\textwidth,trim={.2cm 0 0 0},clip]{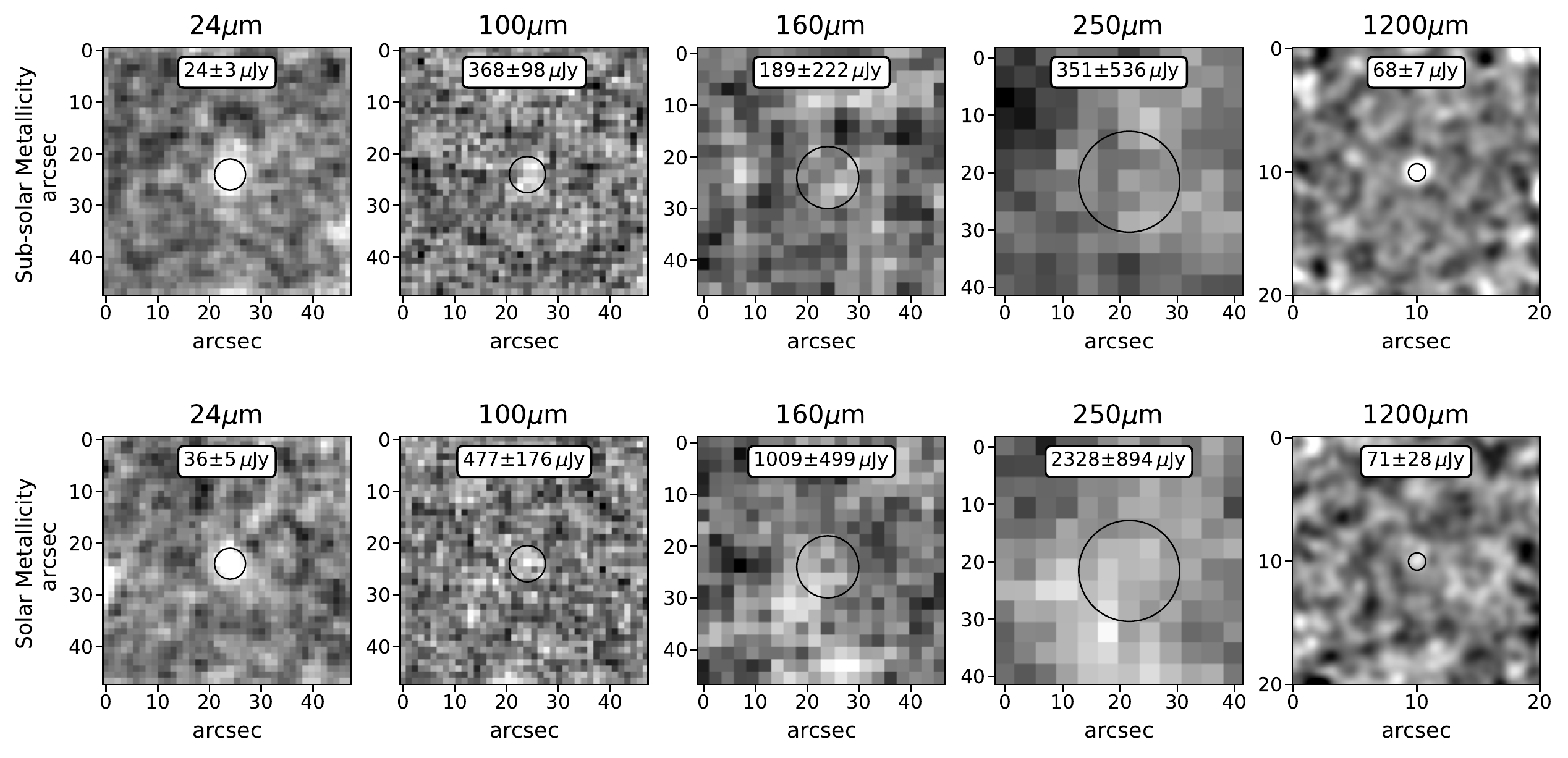} 
		\caption{Stacks of Spitzer/MIPS 24\,{\um}, Herschel/PACS 100 and 160\,{\um}, SPIRE 250\,{\um}, and ALMA band-6 (1200\,{\um}) images for the subsolar (top row) and solar (bottom row) metallicity bins described in Table~\ref{tab:sample}. The axes show the scale of the images in arcseconds. 
		The integrated flux densities ($f_{\nu}$) and background noise estimated errors are listed for each stacked image. Refer to Section~\ref{sec:irfits_bins} for details.
		Black circle in each image shows the FWHM of the respective PSF centered at the the subimage.
		}
		\label{fig:stacks}
\end{figure*}

\subsection{ALMA Data}
The ALMA continuum observations in band 6 (with a representative frequency of 242\,GHz) were executed in three different blocks with the 12m array in C43-2 and C43-3 configurations. 
Different depths were requested for the three blocks that included subsamples with different masses and metallicities, based on the expectation of what would have been detectable. The high mass/high metallicity bin, $M_*>10^{10}\,M_{\odot}$ and $12+\log({\rm O/H)_{PP04}}>8.35$ (where PP04 refers to the strong line metallicity calibrations of \citealt{pp04}; see Section~\ref{sec:ONIR_data}), has 13 targets with average ${\rm RMS}=26.0\,\mu$Jy/beam in the centers of the images (where the targets are located). The high mass/low metallicity bin, $M_*>10^{10}\,M_{\odot}$ and $12+\log({\rm O/H)_{PP04}}<8.35$, has 7 targets with average ${\rm RMS}=15.4\,\mu$Jy/beam. The low mass/low metallicity bin, $M_*<10^{10}\,M_{\odot}$ and $12+\log({\rm O/H)_{PP04}}<8.35$, has 7 targets with average ${\rm RMS}=9.8\,\mu$Jy/beam\footnote{We initially adopted three mass-metallicity bins for sensitivity calculations, however, later in Sections~\ref{sec:irfits} and \ref{sec:shape_irsed} we simply divide the full sample into two subsolar- and solar-metallicity bins to increase the SNR in the Spitzer and Herschel stacks. We go back to three bins in Section~\ref{sec:dustmass} again, as the analysis is based on the ALMA data alone.}. 
Due to the COVID-19 pandemic, the observations' progress on the low mass/low metallicity bin has been delayed and 94\% of the allocated integration time distributed over all sources in this bin was taken. However, as the available data reaches close to the desired RMS, we proceed with the analysis using the existing data.

Data are calibrated and imaged with the Common Astronomy Software Applications (CASA) package \citep{mcmullin07}. We create a cleaned image using the {\sc tclean} package with natural weighting and the multi-frequency synthesis (mfs) mode, and we apply tapering, which by suppressing the weights of the outer visibilities increases the beam size (lowers the resolution). The reason for using tapering is the effect of the non-Gaussian nature of the dirty beam on flux estimation (see \citealt{czekala21} and the appendix in \citealt{JvM95} for a complete discussion). In brief, in the {\sc tclean} process, the clean model is convolved with the clean beam and has units of Jy/clean beam, while the dirty image and the residual map that are created after running the {\sc tclean} task are in units of Jy/dirty beam.
If the dirty beam is a Gaussian, the two beams would have roughly the same area. However, if the dirty beam has shelves, as in the case of our data, the integrated dirty beam response would be larger than the integrated clean beam response. Therefore, in the latter case, once the residual map is added back to the convolved clean model, the fluxes will be overestimated. This effect is more pronounced for low signal-to-noise ratio (SNR) data, like ours, as the fraction of the flux in the residual map is higher. 
The solution can be either to convolve the clean model with a larger beam that matches the area of the dirty beam, or to scale the residuals following the prescription of \citet{JvM95}. The former sacrifices the resolution for correct flux and noise estimates. The latter retains the resolution, however it scales down the noise outside the regions containing emission, resulting in an incorrect noise determination  \citep{walter08}. Given that for this analysis we are not concerned about the resolution, we adopt the first solution by using tapered images, yielding a synthesized beam with FWHM of $1.4''\times1.4''$. In our sample, the fluxes estimated from the untapered natural-weighted images are higher by $\sim 15\%$ than the fluxes derived from the tapered (unaffected) images. This value is in agreement with the flux overestimate percentage calculated using the prescription of \citet{JvM95}.

\subsubsection{Flux Measurements} \label{sec:alma_phot}
Fluxes are extracted through aperture photometry, performed on the primary-beam-corrected images. We convert the image units from Jy/beam to Jy/pix using the number of pixels per beam (calculated from the image header keywords; {\sc bmajor}, {\sc bminor}, {\sc cdelt1}, {\sc cdelt2}). Then, the total integrated aperture flux is calculated as the sum of the pixel values in a given aperture. An aperture radius of 1.45$''$ is used to include $>99\%$ of the point-source Gaussian area (1.45$''$ is 2.1 times the half width at half maximum of the Gaussian beam). The aperture fluxes are cross-checked for consistency with those derived from 2D fitting on tapered images through CASA. We do not attempt to derive fluxes from the peak signal, as peak fluxes are highly uncertain for low SNR objects such as the ones in our sample.

Since the targets are all at the center of the images, the non-primary beam-corrected maps are used for error measurements \citep[e.g., as in][]{betti19}. The noise for the integrated flux measurements is estimated by taking the standard deviation of the integrated flux measurements in 100 apertures of the same size and offset randomly from the source position. 

Given the prior knowledge on the location of the galaxies in HST/F160W images\footnote{\citet{liu19} reported an offset between the HST/ACS $i$ band coordinates and ALMA coordinates of $<0.08$'' in the COSMOS field (see Appendix A of that paper). The offset is negligible compared to the resolution of our images, and hence we do not correct for it.}, a detection is defined as ${\rm SNR}>2$ in the integrated aperture flux measurements. According to this criterion, 10 out of 27 objects in the sample are detected. Five detections have $M_*>10^{10}\,M_{\odot}$ and $12+\log({\rm O/H)_{B18}}>8.5$\footnote{Metallicities are from O3N2 line ratios using the \citet{bian18} calibration, see Section~\ref{sec:ONIR_data_metal}.}. Three have $M_*>10^{10}\,M_{\odot}$ and $12+\log({\rm O/H)_{B18}}<8.5$, and the other two have $M_*<10^{10}\,M_{\odot}$ and $12+\log({\rm O/H)_{B18}}<8.5$. The fluxes and their errors are listed in Table~\ref{tab:almafluxes}.

\def\arraystretch{1.2}
\capstartfalse   
\begin{deluxetable*}{cllcccc}
\setlength{\tabcolsep} 
\tabletypesize{\footnotesize} 
\tablewidth{0pc}
\tablecaption{ALMA fluxes}
\tablehead{
\colhead{ID} &
\colhead{R.A.} &
\colhead{Decl.} &
\colhead{$f_{1.2{\rm mm}}$ [$\mu$Jy]} &
\colhead{$\sigma_{1.2{\rm mm}}$ [$\mu$Jy]} &
\colhead{$f_{24{\rm \mu m}}$ [$\mu$Jy]} &
\colhead{$\sigma_{24{\rm \mu m}}$ [$\mu$Jy]} 
}
\startdata
    24020 & 150.1151733 & 2.42553139 & 27 & 23 & -5 & 9\\
  6283 & 150.1152802 & 2.24191332 & 12 & 20 & 3 & 7\\
  25229 & 150.1084137 & 2.43971491 & 48 & 25 & 0 & 5\\
  8515 & 150.1844788 & 2.26626229 & 66 & 26 & 5 & 12\\
  3626 & 150.1047363 & 2.21573091 & 33 & 25 & -2 & 10\\
  3773 & 150.1981506 & 2.21658921 & 53 & 18 & -7 & 8\\
  4156 & 150.1793213 & 2.21977043 & 0 & 23 & 2 & 9\\
  9971 & 150.1435394 & 2.28179717 & 144 & 36 & 6 & 8\\
  9393 & 150.1635437 & 2.27508616 & 25 & 29 & 34 & 11\\
  19985 & 150.0603485 & 2.38277268 & 108 & 35 & 111 & 8\\
  3666 & 150.0775299 & 2.21603322 & 67 & 39 & 16 & 14\\
  19753 & 150.0757599 & 2.38064408 & 228 & 43 & 53 & 10\\
  6750 & 150.1630249 & 2.24749112 & 35 & 32 & 45 & 7\\
  19439 & 150.1015015 & 2.37672329 & 73 & 37 & 20 & 7\\
  5814 & 150.1691284 & 2.23839569 & 156 & 67 & 92 & 8\\
  5901 & 150.1894074 & 2.23814201 & 9 & 68 & 0 & 7\\
  16594 & 150.1249542 & 2.35021901 & 43 & 51 & 42 & 9\\
  22193 & 150.0854187 & 2.4059577 & 41 & 73 & 13 & 11\\
  24763 & 150.0567017 & 2.43466282 & -38 & 68 & 36 & 11\\
  4497 & 150.0714722 & 2.22387266 & 161 & 64 & 18 & 9\\
  5094 & 150.1403656 & 2.23018551 & 110 & 68 & 51 & 11\\
\hline
  3324 & 150.148407 & 2.21313357 & 163 & 60 & 47 & 8\\
  19013 & 150.1119995 & 2.37263298 & 4 & 74 & 32 & 8\\
  21955 & 150.0953522 & 2.40281034 & 81 & 62 & 20 & 9\\
  13701 & 150.1127167 & 2.31943941 & 53 & 72 & 58 & 9\\
  13296 & 150.1150971 & 2.31529069 & 216 & 66 & 24 & 12\\
  8280 & 150.0865021 & 2.2643168 & 149 & 64 & 293 & 9\\
\enddata
\tablenotetext{}{}
\tablenotetext{}{Columns left to right: 3D-HST v4 catalog ID, RA (deg, J2000.0), Dec (deg, J2000.0), ALMA 1.2\,mm flux density, its error, Spitzer 24\,{\um} flux density, and its error.  Galaxies below and above the horizontal line are below and above {\oh}=8.60 (in the {\B18} calibration), respectively. The Herschel data for the majority of sources are not individually detected, therefore we refrain from listing the individual Herschel measurements.}
\label{tab:almafluxes}
\end{deluxetable*}
\capstarttrue  

\subsubsection{ALMA 1.2\,mm Stacking} \label{sec:stack}
Despite the detections for 40\% of the sample, for this work we rely on image stacking for two main reasons: 1) the galaxies are not individually detected in Herschel (and in some cases in Spitzer) images, and hence stacking is required to extract the Herschel and Spitzer average fluxes, and 2) the IR SEDs of the detected objects may be biased relative to the underlying population, while stacking better represents the {\em average} behavior of the population. We start with the primary beam-corrected tapered images in Jy/pixel units with the sources at the center of subimages according to their optical coordinates. The optical coordinates are used for stacking, given that the targets are relatively faint and are unresolved in ALMA.
The stacks are constructed by taking the average flux values in each pixel. We measure stack fluxes using aperture photometry with aperture radius of 1.45$''$, as described in the previous section. The background and noise are measured in non-primary beam-corrected image stacks, in a similar manner as for individual galaxies. 

The emission of individual objects is corrected for different redshifts prior to stacking (K-correction). Following \citet[][equation 6]{scoville16}, we use the spectral slope of the Planck function with a temperature of 35\,K, and calculate the correction factor at the redshift of each galaxy relative to the average redshift of the sample. Assuming a different temperature does not change the results significantly, as the correction factors are $\lesssim 5$\% owing to the narrow redshift range of this sample.

\subsection{Spitzer and Herschel Data} \label{sec:spitzer_herschel}
We use Spitzer MIPS 24\,{\um} \citep{sanders07} and Herschel PACS 100 and 160\,{\um} and SPIRE 250 and 350\,{\um} images \citep{lutz11,oliver12} in this work to construct IR stacks, as the majority of our galaxies are not individually detected in any of these bands. Stacks are constructed following standard methods \citep[e.g.,][]{zheng06, reddy10, shivaei17}, as explained below. The signal to noise in these datasets is background limited and the dominant source of noise is source confusion.  Our stacking approach is designed to extract accurate values using proven approaches to minimize this noise source, as described below.

All images are converted to Jy/pixel. For each target, a subimage with the target in the center is constructed. The avoidance of regions near bright sources can significantly reduce confusion noise \citep{leiton15}. Therefore, the images at 24 and 100\,{\um} are inspected to ensure there are no bright sources near the target galaxy that could add noise even after nominal removal (no cases were found). The subimage is then cleaned of neighboring sources as follows, to provide an optimum smooth background for measuring the stacked flux. Prior lists for MIPS 24\,{\um} sources in the  field are generated based on galaxies that are detected at SNR $>3$ in the IRAC channels 1 and 2. For Herschel PACS images, we use a list of priors with SNR $>3$ in the MIPS 24\,{\um} data. \citet{magnelli13} show that at this depth ($\sim 20-25\,\mu$Jy), virtually full identification of the PACS sources will be achieved. For the SPIRE data, we tested priors at SNR of $>3$ and $>5$, and settled on the latter value because the density of 3-$\sigma$ priors made the removal ambiguous, given the large beams in these bands and the probability of more than a single prior within a beam area. The neighboring sources in each subimage are removed by simultaneously fitting scaled PSFs to all the prior sources and to the target, and removing all but the target (which is usually not detected). These cleaned subimages are aligned and then used to construct 3-$\sigma$  trimmed average stacks\footnote{The fluxes are conserved if average stacks are used}. Stack fluxes are derived by summing the pixel values of the stack image within an aperture. Appropriate aperture corrections are calculated for the MIPS photometry based on the 24\,{\um} PSF, and the aperture corrections for the SPIRE and PACS photometry are adopted from the Herschel Legacy Archive\footnote{\url{http://archives.esac.esa.int/hsa/legacy/ADP/PSF/SPIRE/SPIRE-P/SPIRE-P-EEF.csv}} and \citet{balog14}, respectively. To  determine the uncertainty in the stacked fluxes, we measure the flux in 1000 apertures that are randomly positioned in the cleaned subimages away from the center of the image (where the source is) by more than 1 FWHM of the image PSF. The apertures have the same radii as the source aperture radius. The standard deviation of these 1000 fluxes is taken as the stacked flux uncertainty. As this uncertainty is evaluated on the region around the source, it should be a measure of the net uncertainty including confusion noise. 

We now compare these results with standard estimates of confusion noise. A stacking approach virtually identical to ours has been simulated at 24 $\mu$m by \citet{zheng06}. That is, they first removed all individually detected sources in the surrounding area for each target, and then stacked the cleaned images. They measured the rms background fluctuations by placing apertures at random places on the stacked images. The rms fluctuations decreased accurately in proportion to the square root of the number of stacked images. They found that 12 images gave a 5-$\sigma$ detection at 24 $\mu$Jy; scaling to our 21 images for the low metallicity sample predicts a 7-$\sigma$ result, and we found a signal to noise of 8 at this flux level, in excellent agreement.  At the longer, Herschel, wavelengths our method uses the galaxy stack positions as priors; this is important because use of priors can reduce confusion effects by factors of 2 $-$ 3 \citep[e.g.,][]{rodighiero06, magnelli13, safarzadeh15}. For the PACS data, we take confusion noise measurements from the deep survey described by \citet{berta11}, who quote 1-$\sigma$ values of 270 and 920 $\mu$Jy respectively at 100 and 160 $\mu$m using 24 $\mu$m sources as priors. \citet{magnelli13} quote somewhat lower confusion noise limits, but we attribute this difference to the significantly longer integration than used by Berta or us and the resulting modest improvement in PSF fitting in crowded fields. Taking scaling as found by \citet{zheng06} with the square root of the number of stacked images, the noise we estimate is consistent for both the high and low metallicity samples (although slightly higher than the predicted confusion limit). For the SPIRE data, we start with the confusion noise values from \citet{smith17}; their values for the nebulized beam should be relevant for our situation. We assume that these values scale inversely with the square root of the number of images stacked, as in \citet{zheng06}. As these values were determined without priors, there should be some further reduction in our case as discussed above. To determine this gain in our case, we compare the noise predicted from confusion with that from our multi-position measurements. We find that the case of using the 5-$\sigma$ 24\,$\mu$m priors gives a consistent comparison if we assume an improvement in depth from the use of priors of a factor of 1.4. An exception is for the high metallicity sample at 350\,$\mu$m, where the measured noise is about half of the prediction. In any case, the predicted confusion levels are large enough for this band that it has little utility in constraining fits. We therefore have discarded both the high and low metallicity stacked results at 350\,$\mu$m.

To verify the accuracy of the aperture flux measurements on the stacked image of the target galaxies, we introduce fake sources following a similar methodology to that described in \citet{reddy12a}. We simulate 100 PACS sources with fluxes randomly drawn from a Gaussian distribution centered at the flux values of the real stacks (the width of the distribution is irrelevant for this test). Then, simulated sources in the same number as those that went into the real stack bins are randomly selected for stacking and aperture flux measurement. Based on this input and recovered fluxes, we find that a 3-$\sigma$ trimmed average stack is a more accurate estimator compared to a median stack. The recovered stack fluxes are  $\sim$ 10\% lower than the average input. This correction factor is applied to our stack fluxes. However, the correction is negligible compared to the measurement error.

A final source of uncertainties is the absolute calibration of the various datasets, but these errors are negligible and do not affect our results. For example, the 24\,{\um} absolute calibration of MIPS is accurate to $\sim 2$\% or better \citep{rieke08}. The Herschel PACS and SPIRE absolute calibrations are accurate to $4 - 5$\% \citep{bendo13, balog14, muller14, bertincourt16}. The nominal uncertainty in the absolute calibration of the ALMA observations is 5\% \citep{remjian19}, but it is likely to be twice as accurate \citep{farren21} particularly given the good conditions of our observations.

\subsection{Optical and Near-IR Data} \label{sec:ONIR_data}
The MOSFIRE spectral reduction and line flux extraction are fully described in \citet{kriek15}, \citet{reddy15}, and \citet{freeman19}.
In summary, emission line fluxes are measured from the MOSFIRE 1D spectra by fitting Gaussian functions on top of a linear continuum. The uncertainties are derived by perturbing the spectrum of each object according to its error spectrum and measuring the standard deviation of the new realizations. Slit-loss (also called path-loss) corrections are applied by normalizing the spectrum of a ``slit star'' placed on each MOSFIRE observing mask to match the 3D-HST total photometric flux. Additionally, the HST F160W images of the resolved targets were used to estimate and correct for the additional flux lost outside of the slit aperture, relative to the slit star. For details refer to \citet{kriek15}.

\subsubsection{SFRs} \label{sec:ONIR_data_SFR}
SFRs are estimated from {\halpha} luminosities, corrected for dust attenuation using Balmer decrements and the \citet{cardelli89} MW extinction curve.
The success of this approach to recover total SFRs for an {\halpha}-{\hbeta} detected sample is demonstrated in \citet{shivaei16}, where {\halpha} SFR estimates corrected for reddening via the Balmer decrement are compared with IR-based SFR estimates.
The assumption of a MW extinction curve to correct the observed nebular emission is supported by previous studies that showed the nebular attenuation curve is similar to the MW extinction curve \citep{reddy20,rezaee21}.

Following recent studies that have discussed the metallicity dependence of the {\halpha} luminosity to SFR conversion \citep[e.g.,][]{reddy18b,theios19}, we adopt two different conversion factors for galaxies with oxygen abundances at roughly the solar value ({\oh}$>8.6$\footnote{Using the \citet{bian18} calibration, see Section~\ref{sec:ONIR_data_metal}.}) and those with lower {\oh}. We assume the stellar and ISM gas metallicity are linearly correlated with each other. For galaxies with {\oh}$>8.6$ we adopt the conversion factor of \citet{hao11} converted to a \citep{chabrier03} IMF\footnote{\label{fn1}To adopt the conversion factors from the literature from Salpeter to Chabrier IMFs, we multiply by a constant factor of 0.63 \citep{madau14}.}:
\begin{eqnarray} \label{eq:sfr1}
\log({\rm SFR / M_{\odot}\,yr^{-1}}) = \log({\rm L(H\alpha)/erg\,s^{-1}})-41.26,
\end{eqnarray}
and for lower metallicity galaxies, we adopt the conversions derived from the \citet{bc03} stellar population models with constant star formation over 100\,Myr and $Z=0.004$ from \citet{reddy18b} and \citet{theios19}, once adjusted to the same IMF that is assumed in Equation~(\ref{eq:sfr1})\footnote{See footnote \ref{fn1}}:
\begin{eqnarray} \label{eq:sfr2}
\log({\rm SFR / M_{\odot}\,yr^{-1}}) = \log({\rm L(H\alpha)/erg\,s^{-1}})-41.41.
\end{eqnarray}
For reference, the \citet{kennicutt98} and \citet{kennicutt12} constants to convert $\log({\rm L(H\alpha)})$ to $\log({\rm SFR})$, for the same IMF assumed here, are $-41.30$ and $-41.26$, respectively.

\capstartfalse   
\begin{deluxetable*}{cc|ccc|ccc}[ht]
	\setlength{\tabcolsep} 
	\tabletypesize{\footnotesize} 
	\tablewidth{0pc}
	\tablecaption{Properties of the two metallicity samples}
	\tablehead{
	\colhead{Parameter}&\colhead{} & \multicolumn{3}{c}{Subsolar Metallicity ($N^*=21$)}  & \multicolumn{3}{c}{Solar Metallicity ($N^*=5$)} \\
	 \multicolumn{1}{c}{}&\colhead{} & \multicolumn{1}{c}{min} & \multicolumn{1}{c}{mean$\pm \sigma^{**}$} & \multicolumn{1}{c}{max} & \multicolumn{1}{c}{min} & \multicolumn{1}{c}{mean$\pm \sigma^{**}$} & \multicolumn{1}{c}{max}
	}	
	\startdata
	{redshift ($z$)} & {} & {2.09} & {$2.30\pm 0.14$} & {2.47} & {2.17} & {$2.31\pm 0.13$} & {2.47} \\
	{$ 12+\log({\rm O/H})_{\rm B18}$\footnote{O3N2 metallicity using the \citet[][B18]{bian18} calibration}} & {} & {8.11} & {$8.41\pm 0.13$} & {8.59} & {8.64} & {$8.71\pm 0.04$} & {8.76} \\
	{$ 12+\log({\rm O/H})_{\rm PP04}$\footnote{O3N2 metallicity using the \citet[][PP04]{pp04} calibration}} & {} & {8.02} & {$8.27\pm 0.11$} & {8.42} & {8.46} & {$8.51\pm 0.03$} & {8.55} \\
	{$ \log(M_*/M_{\odot})$\footnote{Stellar mass}} & {} & {9.75} & {$10.18 \pm 0.28$} & {10.58} & {10.15} & {$10.43 \pm 0.15$} & {10.57} \\
	{$ \log({\rm SFR}/M_{\odot}\,{\rm yr^{-1}})$\footnote{Dust-corrected SFR from {\halpha} and {\hbeta}}} & {} & {1.33} & {$1.75\pm 0.24$} & {2.16} & {1.54} & {$1.88\pm 0.22$} & {2.22} \\
	{$ \log({\rm sSFR}/{\rm yr^{-1}})$\footnote{specific SFR (SFR/$M_*$) from {\halpha} and {\hbeta}}} & {} & {$-8.82$} & {$-8.43 \pm 0.21$} & {$-8.02$} & {$-8.92$} & {$-8.55 \pm 0.22$} & {$-8.29$} \\
	{$\log (\Sigma_{\rm SFR}/M_{\odot}\,{\rm yr^{-1} kpc^{-2}})$\footnote{SFR surface density, $\Sigma_{\rm SFR}=$(SFR/2)/($\pi r^2$), from {\halpha} and {\hbeta}, where $r$ is the half-light radius in F160W filter from \citet{vanderwel14}}} & {} & {$-0.34$} & {0.22 $\pm 0.43$} & {1.11} & {$-0.32$} & {$0.12\pm 0.40$} & {0.81}\\
	{$\log(L_{\rm IR}/L_{\odot})$\footnote{Total IR luminosity of the subsolar and solar metallicity bins are based on the best-fit local low-metallicity template (Figure~\ref{fig:lowZ}) and best-fit {\R09} LIRG template (Figure~\ref{fig:highZ}), respectively.}}  & {} & {--} & {11.35} & {--}  & {--} & {11.46} & {--} \\
	{$\log({\rm SFR_{IR}}/M_{\odot}\,{\rm yr^{-1}})$\footnote{Obscured SFR estimated from total IR luminosity and calibrations of \citet{kennicutt12}.}}  & {} & {--} & {1.52} & {--}  & {--} & {1.63} & {--} \\
	{ALMA (1.2\,mm) stack $f_{\nu}$ [$\mu$Jy]} & {} & {--} & {$68\pm 7$} & {--}  & {--} & {$71\pm 28$} & {--} \\
	{Spitzer MIPS 24\,{\um} stack $f_{\nu}$ [$\mu$Jy]} & {} & {--} & {$24\pm 3$} & {--}  & {--} & {$36\pm 5$} & {--} \\
	{Herschel PACS 100\,{\um} stack $f_{\nu}$ [$\mu$Jy]} & {} & {--} & {$368\pm 98$} & {--}  & {--} & {$477\pm 176$} & {--} \\
	{Herschel PACS 160\,{\um} stack $f_{\nu}$ [$\mu$Jy]} & {} & {--} & {$189\pm 222$} & {--}  & {--} & {$1009\pm 499$} & {--} \\
	{Herschel SPIRE 250\,{\um} stack $f_{\nu}$ [$\mu$Jy]} & {} & {--} & {$351\pm 536$} & {--}  & {--} & {$2328\pm 894$} & {--} \\
	\enddata
	\tablenotetext{}{$^*N$ is the number of objects in each bin.  Object 8280 is removed from the stacking analysis (see Section~\ref{sec:spitzer_herschel}).\\
	$^{**} \sigma$ is the standard deviation (dispersion) of the properties in rows 1 to 7. For stack fluxes, $\sigma$ is the measurement error. }

	\label{tab:sample}
\end{deluxetable*}
\capstarttrue  

\subsubsection{Metallicities} \label{sec:ONIR_data_metal}
The metallicity of the gas is defined as {\oh}. It is common to estimate O/H using ratios of strong optical emission lines and adopting calibrations that have been constructed based on the observations of electron-temperature-sensitive lines (the so called ``direct'' method) in local galaxies and H{\sc ii} regions \citep[such as those of][]{pp04}.
At high redshifts, due to the evolving conditions of the ionized gas, the locally-calibrated relations may yield biases in the absolute values of the oxygen abundances \citep[e.g.,][]{kewley13,steidel14,shapley15,sanders16a,strom18,kashino19}. 
The degree of this bias will not be well known until large and representative samples of high-redshift galaxies with temperature-sensitive auroral lines are constructed to calibrate the strong emission lines for oxygen abundances.
\citet{bian18} used a sample of local analogs of $z\sim 2$ galaxies to derive empirical direct-method calibrations between {\oh} and strong optical line ratios. At the low-metallicity end, their calibrations for the oxygen and hydrogen line ratios agree with the median of 18 $z\sim 1.5-3.5$ galaxies with direct-method metallicities from \citet{sanders20a}. Unfortunately, the [N{\sc ii}] line was not available for the $z\sim 1.5-3.5$ direct-method sample in \citet{sanders20a} to compare with the N2 ($\frac{\rm [NII]\lambda 6585}{\rm H\alpha}$) and O3N2 ($\frac{\rm [OIII]\lambda 5008/H\beta}{\rm [NII]\lambda 6585/H\alpha}$) calibrations of \citet{bian18}.
However, the \citet{bian18} N2 and O3N2 calibrations are in good agreement with the local calibrations of \citet{maiolino08} and \citet{curti17} and the $z\sim 0$ relations of \citet{sanders21} at \oh$>8.0$. 

In this paper, we use O3N2 line ratios mainly for the practical reason that the four lines are available with high SNR for all of our targets. O3N2 strongly correlates with O/H inferred from strong line ratios involving only oxygen (such as R$_{23}$, which is the ratio of ([O{\sc iii}]$\lambda \lambda4959,5007+$[O{\sc ii}]$\lambda \lambda3726,3729)$ to {\hbeta}), and with stellar mass at $z\sim 2$ \citep[e.g.,][]{sanders18,strom18}. 
O/H inferred from O3N2 is less biased by N/O variations than when using N2, because the O3N2 spans a wider dynamic range in the line ratios. 
Additionally, all current calibrations for O3N2 show a nearly linear anti-correlation between the strong line ratios and O/H without any turnovers or plateaus at \oh$>8.0$ \citep{pp04,maiolino08,curti17,bian18,sanders21}. Following the methodology of \citet{sanders21}, in the small subset (6) of our galaxies with robust [O{\sc ii}], [O{\sc iii}], and [Ne{\sc iii}]$\lambda 3869$ lines, we compute the best-fit metallicity from the [O{\sc iii}]/[O{\sc ii}], [O{\sc iii}]/{\hbeta}, and [Ne{\sc iii}]/[O{\sc ii}] line ratios, adopting the \citet{bian18} calibrations. These metallicities agree well with the O3N2-derived metallicities. For reference, the O3N2 metallicities in our sample are also tightly correlated with the N2 metallicities, but are on average $\sim 0.05-0.10$\,dex lower.
Given the aforementioned uncertainties in estimating O/H at high redshifts, one should take caution in comparing metallicities derived using different line ratios and calibrations from different studies.

In Figure~\ref{fig:mz-mssfr}, we show the metallicities derived from both the \citet[][hereafter B18]{bian18} and \citet[][hereafter PP04]{pp04} O3N2 calibrations. These calibrations provide a linear relationship between O3N2 and oxygen abundance, and therefore the ordering of galaxies in oxygen abundance is preserved regardless of which calibration is used. However, the choice of calibration can affect comparisons from one sample to another. As a reference point, the metallicity of \oh$=8.35$ from the \hyperlink{PP04}{PP04} O3N2 calibration corresponds to \oh$=8.51$ in the \hyperlink{B18}{B18} O3N2 calibration. For the ease of comparing with other studies, we refer to the metallicities ({\oh}) from both calibrations in the rest of this paper. 

\subsubsection{Stellar Masses}
Stellar masses are derived from SED fitting to the photometry of the 3D-HST survey \citep{skelton14}. The photometry is corrected for the nebular emission lines from the MOSFIRE spectra \citep[details in][]{sanders21}. 
We use the FAST SED fitting code \citep{kriek09a} with the stellar population model library of \citet{conroy09} for a solar stellar metallicity, a Chabrier IMF, delayed exponentially declining star formation history, and the \citet{calzetti00} attenuation curve. The stellar masses are insensitive to the choice of the attenuation curve, as masses are dominated by the older stellar populations that emit in longer near-IR wavelengths where different dust attenuation curves are very similar and the total amount of attenuation ($A_{\lambda}$) is relatively small\footnote{The choice of attenuation curve in SED fitting can significantly affect the inferred SFRs, as the main difference between various attenuation curves is the curve slope in the UV, where the emission from recent star formation peaks. Studies have shown that low mass/low metallicity galaxies at $z\sim2$ have a steep SMC like curve while massive/metal rich galaxies have a shallower Calzetti-type curve \citep{reddy18a,shivaei20b,shivaei20a}. Therefore, in this work we do not use the UV/SED-inferred SFRs. The SFRs are derived from {\halpha} luminosity (Section~\ref{sec:ONIR_data_SFR}); hence, the choice of the stellar attenuation curve is irrelevant.}.

\section{Analysis} \label{sec:irfits}

In this section, we discuss our construction of the metallicity subsamples (Section~\ref{sec:irfits_bins}), and the adopted IR templates and SED fitting procedure (Sections~\ref{sec:irfits_template}).

\subsection{Metallicity Bins and Flux Determination} \label{sec:irfits_bins} 

By design, this study is based on lower-luminosity, and hence fainter, galaxies than have been reached by most
FIR/submm-selected samples at similar redshifts. To construct subsamples that contain enough galaxies to reliably yield a detection in the stacked images of Spitzer and Herschel data, we consider two metallicity bins (instead of the three mass-metallicity bins that were initially used in the design of the ALMA survey). The bins are selected to represent metallicities of $\sim$ solar and subsolar with average metallicities of {\oh}$=8.71$ and 8.41 on the {\B18} scale or 8.51 and 8.27 on the {\PP04} scale, respectively. In these two bins, our stacked images yield detections at 24\,$\mu$m and 1.2\,mm at high SNRs and $>2\sigma$ detections in one or more additional IR bands. We provide details of the metallicity subsamples and SED fitting in this section.

We first perform a jackknife resampling test to evaluate potential biases caused by outliers. 
In each metallicity subsample, we estimate stacked 24--1200\,{\um} fluxes by systematically leaving out one object at a time and compare the stacks with those of the full sample (stacking technique described in Sections~\ref{sec:stack} and \ref{sec:spitzer_herschel}). We find that the resampled stacks in the subsolar-metallicity bin are all consistent within 1$\sigma$ ($\sigma$ being the uncertainty of the stack flux) with the stacks of the full subsolar-metallicity bin. The solar-metallicity bin has a smaller sample size, but except for one object, there is no systematic bias at $>1\sigma$ level for more than one photometric band in the resampled stacks. The exception is Object 8280 (Table~\ref{tab:almafluxes}), whose IR emission behavior is significantly different from the rest of the solar-metallicity sample. This galaxy shows an elevated 24-160\,{\um} emission compared to its 250-1200\,{\um}, indicating the presence of warm dust, which can be due to a recent starburst or buried AGN.  Object 8280 has a metallicity of {\oh}$=8.66^{+0.05}_{-0.08}$ and its estimated age from SED fitting is $\sim 50-100\,$Myr depending on the star-formation history assumptions. Additionally, its SED inferred SFR (determined through the rest-frame UV continuum emission) is more than a factor of 2 larger than its {\halpha}-estimated SFR. The young age and the elevated UV to {\halpha} SFR both strengthen the argument that this galaxy has recently undergone a strong starburst.  As this peculiar galaxy can bias the results of the stacks, we remove it from the rest of the stacking analysis. The properties of the final subsolar and solar metallicity subsamples are listed in Table~\ref{tab:sample}.

\begin{figure}[ht]
	\centering
		\includegraphics[width=.5\textwidth,trim={.2cm 0 0 0},clip]{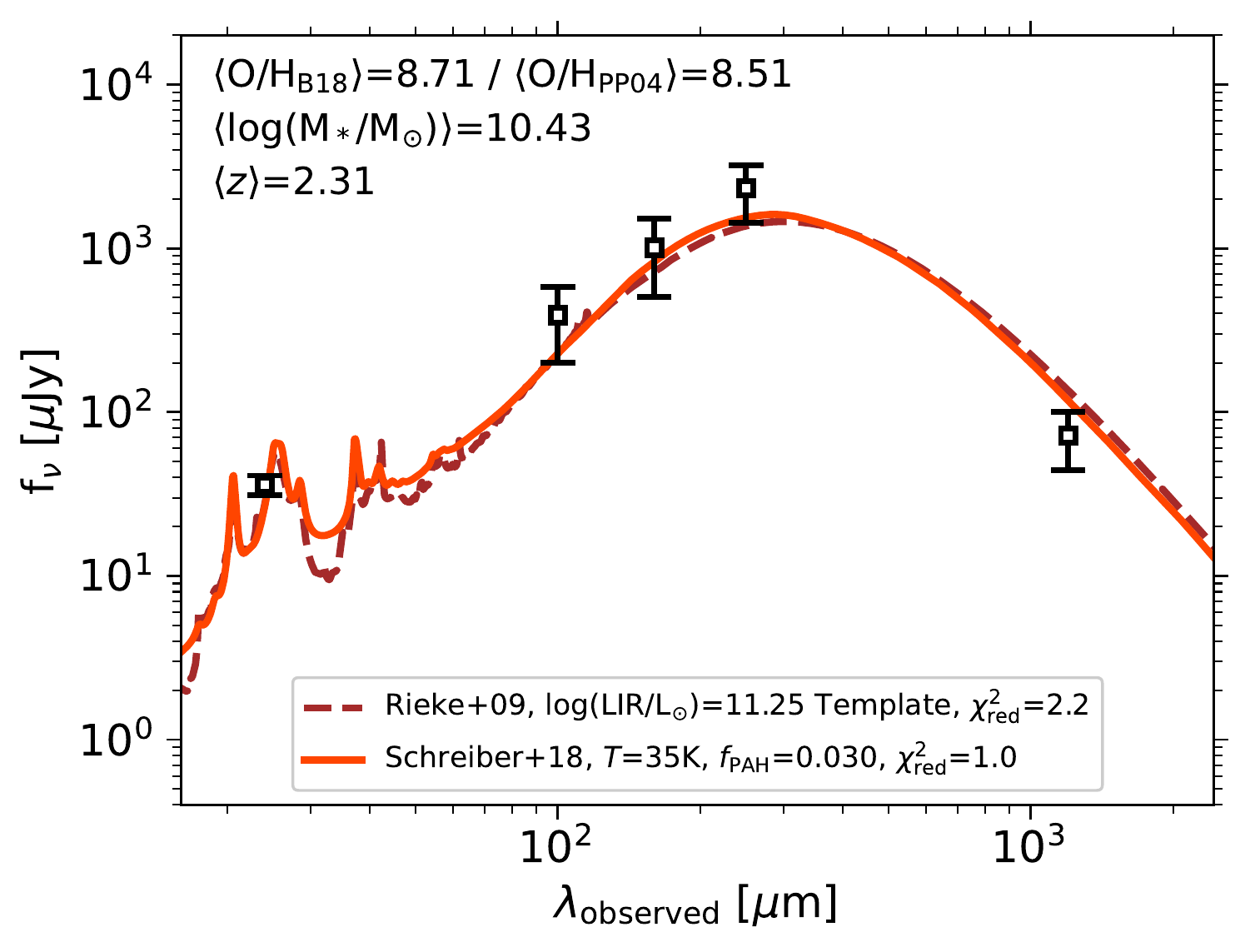} 
		\caption{Template fits to the solar-metallicity stacks of 24, 100, 160, 250, and 1200\,{\um} data (black squares).  Shown in the upper-left corner of the plot are the average metallicity (\oh, or O/H) estimated using the \citet[][B18]{bian18} and \citet[][PP04]{pp04} calibrations, stellar mass, and redshift of the sample. The conventional \citet{rieke09} $\log({\rm L(IR)/}L_{\odot}) =11.25$ template and the \citet{schreiber18} model with dust temperature of 35\,K and 3\% PAH fraction provide good fits to the photometry.}
		\label{fig:highZ}
\end{figure}

\begin{figure}[ht]
	\centering
		\includegraphics[width=.5\textwidth,trim={.2cm 0 0 0},clip]{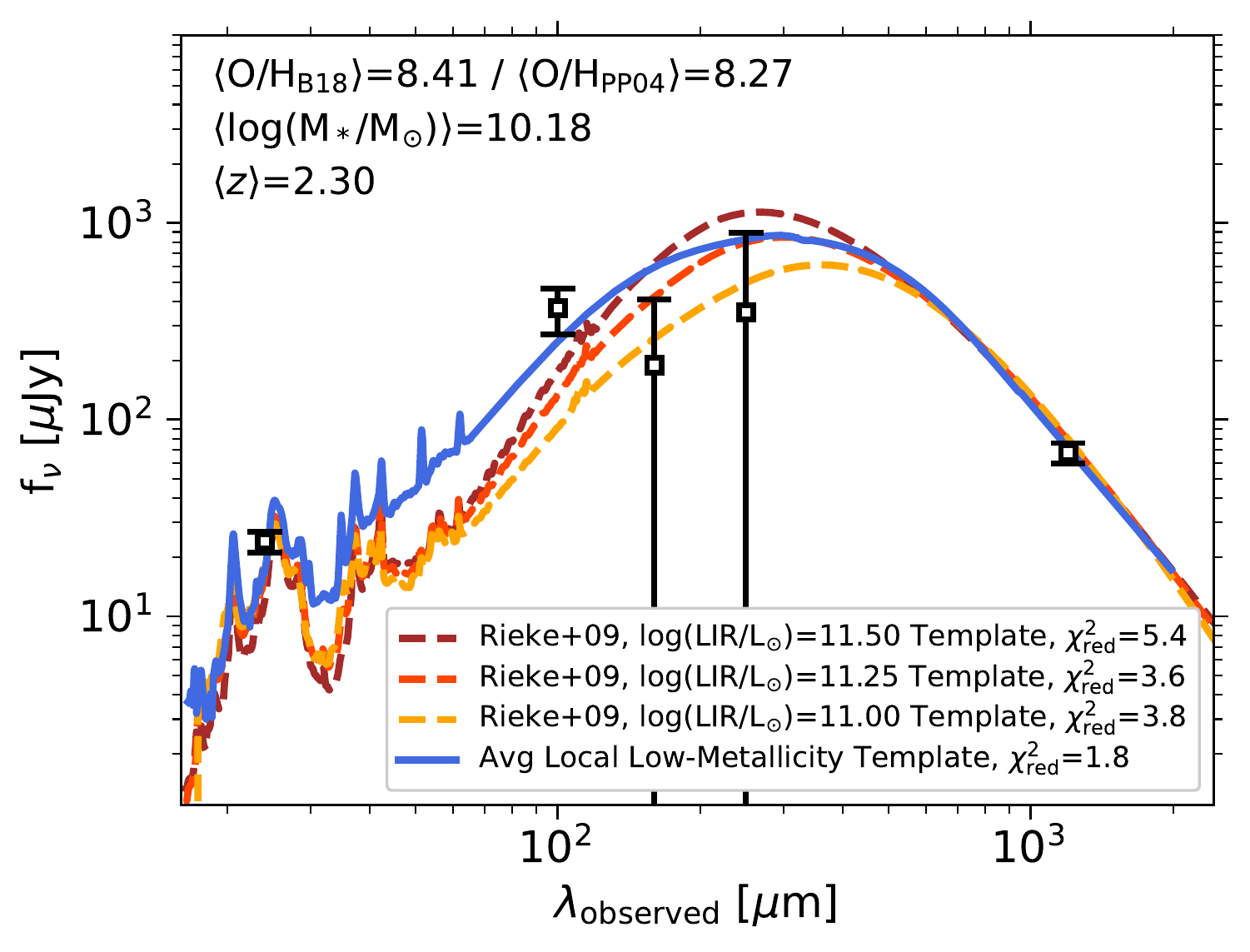} 
		\caption{Template fits to the subsolar-metallicity stacks of 24, 100, 160, 250, and 1200\,{\um} data (black squares). Shown in the upper-left corner of the plot are the average metallicity (\oh, or O/H) estimated using the B18 and PP04 calibrations, stellar mass, and redshift of the sample. 
		The fits using three templates of \citet{rieke09} with IR luminosities of $\log({\rm L(IR)/}L_{\odot}) =11.0$, 11.25, and 11.5, and the average local low-metallicity dwarf template (Section~\ref{sec:irfits_template}) are shown.
		The reduced $\chi^2$ values (with four degrees of freedom) are also indicated. The observations follow the behavior of the average local low-metallicity template most closely, indicating a broader and warmer IR SED than predicted by the local LIRG templates.}
		\label{fig:lowZ}
\end{figure}

\subsection{Template Fits} \label{sec:irfits_template}
We consider IR templates from the literature that are commonly used. We fit the models by weighted least squares, where the weights are the inverse of the variances of the measurements. 
As two examples, the \citet[][R09]{rieke09} and the \citet[][S18]{schreiber18} template fits to the solar-metallicity sample are shown in Figure~\ref{fig:highZ}\footnote{The {\R09} templates are adopted in the luminosity range recommended by \citet{rujopakarn13}, which selects the set of {\R09} models that most closely matches the shape of the IR SED of galaxies at $z>1$. These models should be applicable up to a total IR luminosity of $\sim 4 \times 10^{12}$ L$_\odot$ \citep{shipley16}, but the lower luminosity limit of applicability is not determined. We use the {\S18} IR template with temperature of 35\,K and PAH fraction of 3\%, as suggested by \citet{schreiber18} for $z\sim 2$ galaxies with stellar masses of $\sim 10^{10.0-11.5}$\,{\msun}.}. The only free parameter for each template is the normalization. Both of the {\R09} and {\S18} templates provide reasonably good fits to the measured IR SEDs for $2<z<4$ galaxies that are sufficiently luminous to have good detections in multiple FIR/submm bands \citep[e.g.,][]{schreiber18,derossi18}. Similarly good fits are also provided by the \citet{magdis12} $z \sim 2$ template. In agreement with previous studies on massive (and high metallicity) galaxies at $z\sim 2$ \citep[e.g.,][]{elbaz11,reddy12b,derossi18,shivaei18}, the photometry of the solar-metallicity galaxies agrees well with the local LIRG templates. Owing to the small sample size of the solar metallicity bin ($N=5$), we do not aim to draw conclusions based on the insignificant $\chi^2$ value differences among different fits.
That is, we can take the {\R09} sets of templates as proxies for other fits to the IR SEDs of solar-metallicity galaxies at $z\sim 2$. We now will use the stack measurements from Section~\ref{sec:sample-data} to test the FIR behavior of sub-solar metallicity galaxies at $z\sim 2.3$, and whether it exhibits the relatively broad FIR SED seen for local low-metallicity galaxies \citep{remyruyer15,lyu16}. 

As a starting point in the interpretation of our low metallicity measurements, we construct a {\em publicly available}\footnote{\href{http://www.ireneshivaei.com/shivaei22.html}{http://www.ireneshivaei.com/shivaei22.html}} IR SED template to represent the average behavior of {\em local} low-metallicity galaxies in a quantitative way, as follows. 
We take the photometry for the template from the comprehensive and homogeneous results in \citet{remyruyer15}. The FIR characteristics of low-metallicity galaxies depend on luminosity, such that above $\log{\rm(L(IR)/L_{\odot})} \sim 9.5$, the average SED is significantly warmer than below this value \citep{remyruyer15}. For this reason, we use only the galaxies at and above this threshold for the average template. This results in 11 galaxies. Before averaging their photometry for the FIR SED, we exclude three for the following reasons. 
Examination of the PACS 70\,$\mu$m image for the local galaxy UM 311 shows that, within the 115$''$ aperture used to extract photometry, the signal would have been dominated by the disk of a nearby galaxy, NGC 450 (it has even been suggested that UM 311 is an H{\sc ii} region in this galaxy), which artificially raises its stated luminosity. Moreover, the local galaxy HS0052+2536 is much fainter than the rest of the sample, and its SPIRE measurements have too low SNRs to be useful. Additionally, the IRS spectrum of the local galaxy Mrk 930 is too noisy and the slope of the spectrum conflicts with the photometry.
We take mid-IR spectra for all 9 (excluding UM 311 and Mrk 930) from \citet{lebouteiller11} through the NASA/IPAC Infrared Science Archive (IRSA). We average the measurements (photometry and spectra) in logarithmic space and fit the FIR photometry with polynomial series. We impose a small power law slope (linear in log space) to make the synthetic photometry on the averaged spectrum match the photometry in IRAC Band 4, while achieving a smooth join to the fit to the FIR photometry at 20\,{\um}. 

Figure~\ref{fig:lowZ} shows the fits to the sub-solar metallicity stacks ($N=21$) to the average local low-metallicity template, as well as three of the {\R09} LIRG templates in the range of $\log({\rm L(IR)/L_{\odot}})=11.0-11.5$, as recommended for high redshift galaxies \citep{derossi18}. None of the LIRG templates provide a good fit to the stacked photometry from 24 to 1200\,{\um}, as they either underestimate the 100\,{\um} emission or overestimate the 150-250\,{\um} emission significantly, showing that the templates are not broad enough to represent the sub-solar metallicity stacks. We also fit other commonly used templates to the stacks to investigate their goodness of the fit parameter (reduced $\chi^2$): the template of {\S18} with $T=35$\,K and 3\% PAH fraction ($\chi^2_{\rm red}=2.8$), the \citet{ce01} templates ($\chi^2_{\rm red}=3.2$), the \citet{elbaz11} starburst ($\chi^2_{\rm red}=13.4$) and main-sequence ($\chi^2_{\rm red}=14.5$) templates, the \citet{kirkpatrick15} $z\sim 2$ star-forming template ($\chi^2_{\rm red}=7.4$), the \citet{magdis12} $z\sim 2$ template ($\chi^2_{\rm red}=2.8$), and the \citet{dh02} templates ($\chi^2_{\rm red}=11.2$). None of these templates provide reasonable fits to the data and they are rejected at $>98\%$ confidence levels.

In comparison, the average local low-metallicity template better represents the behavior of the $z\sim 2.3$ sub-solar metallicity stacks with a broader and warmer FIR emission (solid curve in Figure~\ref{fig:lowZ}). It also shows a lower reduced $\chi^2$ of 1.8, compared to the other templates mentioned above.
The metallicities of the 9 dwarf galaxies that are used to build the average local low-metallicity template are in the range of {\oh}$=8.1-8.4$ with an average of 8.3 and standard deviation of 0.09\,dex.
On average, the $z\sim 2.3$ sample has a 0.1\,dex higher metallicity (on the {\B18} calibration scale; Table~\ref{tab:sample}) with half of the sample having metallicities larger than 8.4, which is the highest metallicity in the dwarf sample. However, the difference between the two strong-line metallicity calibrations adopted here for the high-redshift galaxies, the {\PP04} and {\B18}, introduces a systematic uncertainty of $\sim 0.1-0.2$\,dex (Section~\ref{sec:ONIR_data_metal}). In any case, the emergence of a warm dust component makes the IR SED of $z\sim 2.3$ subsolar metallicity galaxies more similar to that of the local low-metallicity dwarfs than the local LIRGs. We discuss the physical interpretation of this result further in Section~\ref{sec:physical_interp}.

\section{The IR SED Shape} \label{sec:shape_irsed}

In the preceding section, we found that the $z\sim 2.3$ sample with subsolar metallicity is fitted better by a template based on local low-metallcity galaxies than by templates based on local galaxies of $\sim$ solar metallicity. 
We now discuss the underlying behavior of this change in the IR SED shape.

\begin{figure*}[ht]
	\centering
		\includegraphics[width=.85\textwidth,trim={.2cm 0 0 0},clip]{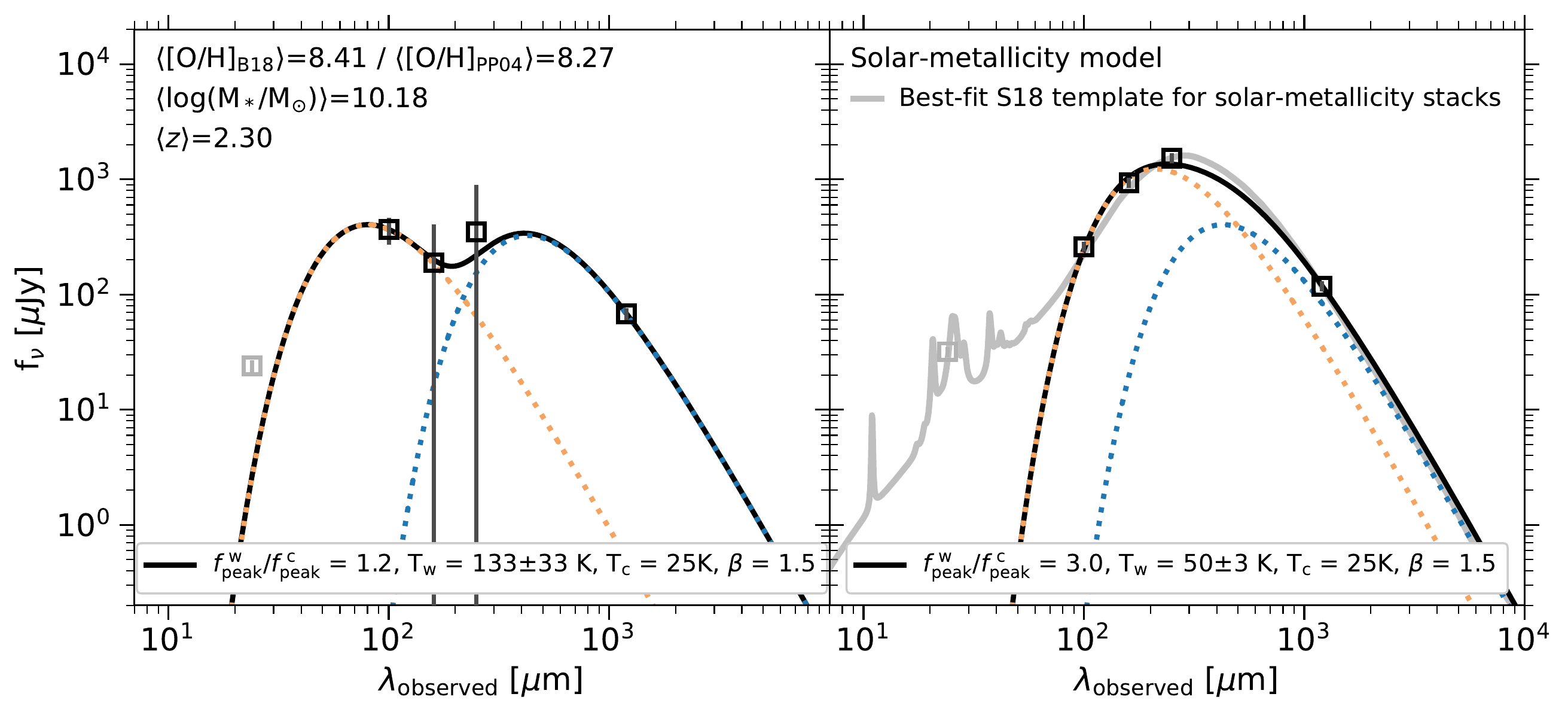}\\
		\includegraphics[width=.85\textwidth,trim={.2cm 0 0 0},clip]{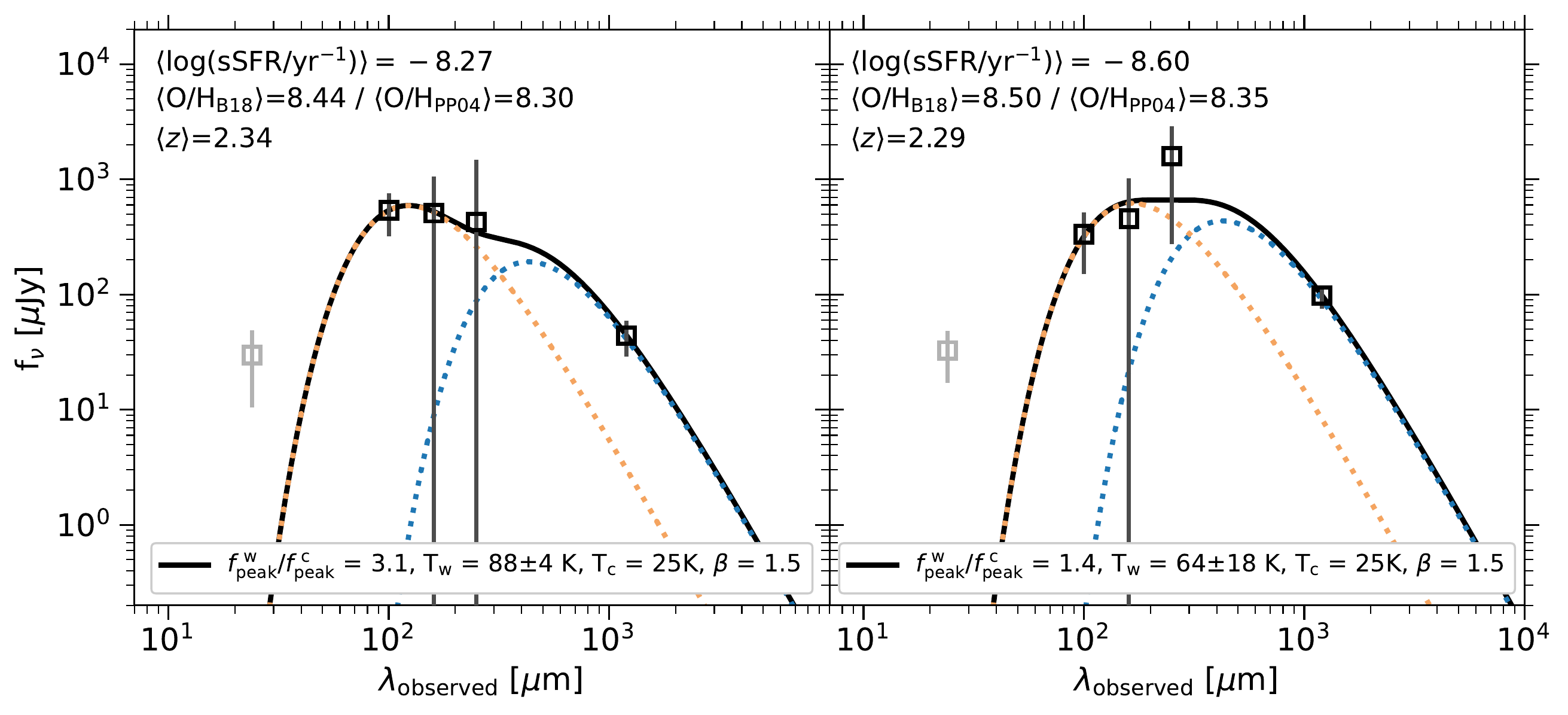} \\
		\includegraphics[width=.85\textwidth,trim={.2cm 0 0 0},clip]{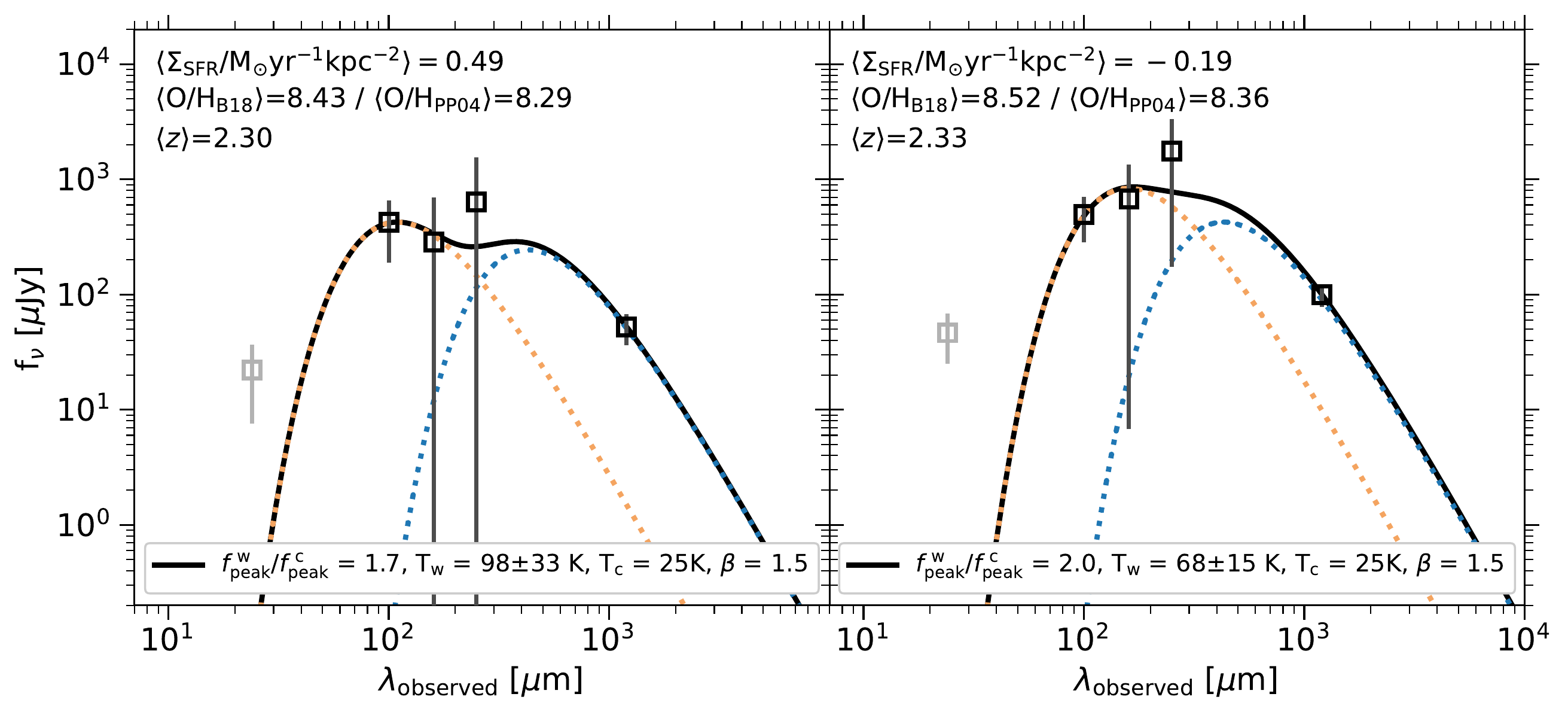}
		\caption{The two-temperature modified blackbody (2T-MBB) IR fits, as described in Section~\ref{sec:k15}, to the stacks of 100, 160, 250, and 1200\,{\um} data. First row shows the stacks of the subsolar-metallicity sample in the left panel, and the modeled solar-metallicity photometry based on the best-fit S18 template in Figure~\ref{fig:highZ} in the right panel. Middle and bottom rows show the stacks in two bins of sSFR and SFR surface density ($\Sigma_{\rm SFR}$) respectively.
		The light grey 24\,{\um} stacked measurement is not included in the fits.
		The warm and cold components of the fits are also shown separately with the dotted orange and blue curves, respectively.
		Parameters of the warm and cold dust components are shown in the bottom of the plots: $f^{\rm w}_{\rm peak}/f^{\rm c}_{\rm peak}$ is the warm to cold component peak flux ratio, and $T_{\rm w}$ and $T_{\rm c}$ are the warm and cold dust component temperatures, respectively. The average metallicity, redshift, and other relevant properties of the samples are shown in the top-left corners. The most significant change in the width of the IR SED is between the two metallicity bins.
		}
		\label{fig:2comp}
\end{figure*}

\subsection{The Evolution of the Warm Dust Component} \label{sec:warmdust}
To quantify the difference in the warm dust between the subsolar and solar metallicity bins, we use the simple 2T-MBB model of \citet[][K15]{kirkpatrick15} to fit the Herschel and ALMA stacks only. In brief, the model consists of two modified blackbodies with two different temperatures, which we designate as warm and cold dust components (more details in Appendix~\ref{sec:k15}).
The goal here is to investigate how the warm dust temperature and intensity (luminosity) changes between the subsolar and solar metallicity samples. Therefore, due to the lack of sufficient observational constraints on the cold dust component, we fix the cold dust temperature and $\beta$ and evaluate the change in the warm dust component temperature ($T_{\rm w}$) and its peak flux compared to that of the cold component. 
The cold dust temperature is set to 25\,K and the warm component temperature is a free parameter between $30-150$\,K, motivated by the findings of \citet{remyruyer15}, who showed that the cold dust temperature is fairly constant at $T\sim 25$\,K among local galaxies with different metallicities and that the warm dust component can be as high as 150\,K in a number of low-metallicity galaxies (see the discussion in Appendix~\ref{app:dmass}). 
We adopt a $\beta$ of 1.5 for the sub-solar metallicity samples, which is the average submm emissivity index of the local dwarf galaxies \citep{lyu16}.  For ease of comparison, we also adopt $\beta=1.5$ for the solar-metallicity model. The assumption of $\beta=2$ does not change any of the main results of the solar-metallicity model.

The fits are shown in the top panel of Figure~\ref{fig:2comp}, in which the 24\,{\um} data are not included in the fitting procedure. 
For the solar-metallicity model, we fit the 2T-MBB function to the model photometry at 100, 160, 250, and 1200\,{\um} extracted from the best-fit {\S18} template (Figure~\ref{fig:highZ}). The {\S18} template is also shown in grey in the solar-metallicity panel of Figure~\ref{fig:2comp}. For the sub-solar metallicity model, we fit the 2T-MBB function directly to the subsolar-metallicity stack fluxes at 100, 160, 250, and 1200\,{\um}.
The difference between the two metallicity samples is very clear. While the warm and cold dust components have different temperatures in both samples, the two components are easily distinguishable in the subsolar-metallicity bin. This is reflected in the temperature difference between the warm and cold components, as well as their peak flux ratios.
The 2T-MBB fit to the synthesized solar-metallicity photometry matches the original IR template very well. The average of the warm (50\,K) and cold (25\,K) components is the same as the {\S18} template temperature of 35\,K and the warm component profile dominates the IR peak width with a warm-to-cold peak flux ratio of 3.
In comparison, the subsolar-metallicity 2T-MBB fit indicates a $108\pm 33$\,K temperature difference between the cold and warm components, and comparable peak fluxes.\footnote{We caution that the exact value of the warm dust temperature depends on the choice of $\beta$ and $T_{\rm c}$. Therefore, the best-fit $T_{\rm w}$ values in this section should not be taken literally and are mainly for relative comparisons. However, altering the values of the cold dust temperature and $\beta$ within reasonable ranges does not affect our main conclusions. For example, a $T_{\rm c}=30$\,K for the subsolar-metallicity fit returns a $T_{\rm w}=150\pm 61$\,K and $f^{\rm w}_{\rm peak}/f^{\rm c}_{\rm peak}=0.9$, and for the solar-metallicity model returns a $T_{\rm w}=53\pm 2$\,K and $f^{\rm w}_{\rm peak}/f^{\rm c}_{\rm peak}=1.3$. As another example, keeping $T_{\rm c}=25$\,K but $\beta=2$, the subsolar (solar) metallicity fit results in $T_{\rm w}=135\pm 10$\,K ($49\pm 2$) and $f^{\rm w}_{\rm peak}/f^{\rm c}_{\rm peak}=0.8$ (1.2). Therefore, the overall conclusion that the warm component is hotter in the subsolar-metallicity fit does not change by varying $T_{\rm c}$ or $\beta$.}
Based on these fits we conclude that a) the temperature of the warm dust component increases with decreasing metallicity, and b) while the solar-metallicity IR SED width can be mainly represented by a single MBB, the subsolar-metallicity IR SED is broader as the difference in the temperatures of the two components is larger with similar peak fluxes (i.e., neither of the components dominates in terms of brightness). The solar-metallicity fit has $\Delta T=25\pm 3$\,K, corresponding to $\Delta\lambda= 213_{- 12}^{+14}$\,{\um}, while the subsolar-metallicity fit shows $\Delta T=108\pm 33$\,K, corresponding to $\Delta \lambda=347^{+26}_{-16}$\,{\um}, a factor of 1.6 wider. 
This behavior is similar to that found for local galaxies of subsolar and solar metallicity, e.g., \citet{remyruyer15}, and is discussed further below.

A hotter warm component at low metallicities has also been observed both in the $z\sim 0.3$ rest-frame UV analogs of $z > 5$ galaxies \citep{faisst17}, and in the local dwarf galaxies of the Dwarf Galaxy Survey \citep[DGS,][]{remyruyer15}. 
The three low-redshift analogs of $z>5$ galaxies in \citet{faisst17} with high sSFR and low metallicities similar to our subsolar metallicity sample, show luminosity-weighted temperatures of $\sim 70-90$\,K. \citet{faisst17} explained the presence of a hot dust component by possibly an optically thin ISM to UV radiation due to the low metallicities, as well as a strong UV radiation field due to high star formation densities.
In \citet{remyruyer15}, 11 DGS galaxies show an IR excess at $\lambda_{\rm rest}\sim 20-30$\,{\um}. These 11 galaxies have on average 0.25\,dex lower {\oh} compared to the average oxygen abundance of the rest of the sample.
\citet{remyruyer15} derived dust temperatures in the range of $100-150$\,K (average of $\langle T_{\rm w}\rangle=117$\,K) for the warm component, and average of $\langle T_{\rm c}\rangle=31$\,K for the cold component for those 11 galaxies, similar to the values found for the $z\sim 2.3$ subsolar-metallicity sample in this work. 
\citet{remyruyer15} attributed the warm component to a higher contribution from the hot H{\sc ii} regions heating the dust grains in dwarf galaxies, owing to the smaller physical sizes and lower dust attenuation of the local dwarf galaxies compared to local $L_*$ galaxies. This effect can also be explained by the higher sSFR of the dwarf galaxies in their sample, producing a wider equilibrium temperature distribution of dust grains, skewed towards higher dust temperatures (hence a hotter and wider IR SED). However, the sSFR distribution of our subsolar and solar metallicity samples are statistically indistinguishable (referring to the sSFR average and dispersion of the two subsamples in Table~\ref{tab:sample}), therefore a change in sSFR alone cannot explain the change in the IR SED of our two metallicity samples. Below, we explore possible causes for the elevated warm dust emission in the subsolar metallicity sample in detail.

\subsection{Possible Physical Causes for the Elevated Warm Dust Emission} \label{sec:physical_interp}

Dust evolution is the combination of grain formation, processing (e.g., grain size modification, structural modification, coagulation), and destruction that can be affected by the incident non-ionizing UV and ionizing radiation, cosmic rays, stellar ejecta, SNe shocks, and ISM elemental enrichment.
As a result, the shape of the IR SED can be altered by both the grain properties (size and composition) and the grains' heating sources.
The cold dust component resides in the diffuse ISM, is dominated by the emission from large grains, and constitutes most of the dust mass. On the other hand, the warm dust resides closer to star-forming regions or AGN.
The hotter and broader IR emission seen in the subsolar-metallicity galaxies in this work may be the result of an overabundance of small dust grains \citep[e.g.,][]{galliano18,ysard19}, an intense interstellar radiation field \citep[e.g.,][]{dale01,draine07a,galliano11b,faisst17}, and/or an overabundance of silicate grains \citep{derossi18}.
These possibilities can be probed by examining the metallicity, sSFR, SFR surface density ($\Sigma_{\rm SFR}$), the age of galaxies, and the AGN activity.
As the galaxies in our sample show no evidence of obscured AGNs based on the examination of the IRAC colors \citep{coil15,azadi17,azadi18,leung19}, we can assume the AGN contribution is negligible in the mid- to far-IR \citep[for a discussion on AGN dust emission see e.g.,][]{kirkpatrick12,kirkpatrick15,lyu17,lyu18}. Below we review the other factors (metallicity, sSFR, $\Sigma_{\rm SFR}$, and age) that may be responsible for the broad IR SED shown in Section~\ref{sec:irfits}.

\paragraph{Metallicity} Metallicity is a tracer of dust processing, owing to grain growth by accretion of gas-phase metals \citep[e.g.,][]{hirashita15}. However, the effect of elemental enrichment of the ISM on dust evolution becomes evident over long timescales of $\sim 1$\,Gyr \citep{galliano18}. 
Metallicity also traces the dust-to-gas ratio of a galaxy \citep{remyruyer14,devis19}. At low metallicities, the lower dust-to-gas ratio means the ISM is less dusty and more transparent, enabling the stellar radiation to heat a larger volume and deeper into the molecular cloud. Moreover, if the stellar and nebular metallicities correlate with each other \citep{fernandes05,gallazzi05,bresolin09,cipriano17}, a galaxy with a lower gas-phase metallicity may also have, on average, lower-metallicity stars that emit a harder ionizing spectrum. As a result, the harder and more intense radiation that affects larger volumes of the ISM can contribute to the hotter dust emission in low metallicity galaxies. This effect is in addition to the lower abundances and hence lower dust attenuation that enables heating the dust deeper into molecular clouds, or the higher abundance of smaller (hotter) grains at low metallicities\footnote{In a turbulent ISM, grain shattering increases the abundance of small grains; however its relative importance reduces as the metallicity decreases \citep{hirashita08,hirashita15}. Therefore, at very low metallicities, it is expected that the initial grain size distribution is conserved.}.

As demonstrated in Section~\ref{sec:irfits_template}, the stack photometry of subsolar metallicity galaxies at $z\sim 2.3$ follows very closely the average local low-metallicity template that we construct based on the IR photometry and spectra of higher luminosity dwarf galaxies. This average template has a broader and warmer FIR emission compared to local solar-metallicity LIRGs.
Half of the $z\sim 2.3$ subsolar-metallicity sample has oxygen abundances higher than the upper limit of the dwarf sample used to build the local low-metallicity template ({\oh}$=8.4$), with an average of 0.1\,dex higher oxygen abundance in the $z\sim 2.3$ sample compared to the dwarf sample.
If this difference is real, as the high-redshift oxygen abundances are subject to the assumed strong-line metallicity calibration (see discussion in Section~\ref{sec:ONIR_data_metal}), it indicates that galaxies at $z\sim 2.3$ can have similar ionization field properties as local galaxies of lower metallicity.
In fact, the rest-frame UV-optical studies have shown that the O/Fe ratio of $z\sim 2$ galaxies is $\sim 0.5-0.6$\,dex enhanced relative to the solar abundance \citep{steidel16,topping20a,topping20b,cullen21,reddy21}. In other words, the O/H abundances of $z\sim 2$ galaxies are higher than the O/H of $z\sim 0$ galaxies at a given Fe/H. It is thus plausible that the Fe/H that controls the ionizing spectral production is similar between the $z\sim 0$ dwarf galaxies and the $z\sim 2.3$ galaxies in this study, despite their different O/H values. As a result, a similar ISM ionizing radiation field intensity and hardness causes a similar IR SED shape between the $z\sim 0$ dwarfs and the $z\sim 2.3$ subsolar-metallicity LIRGs. 

It is also possible that the warmer IR SED of lower metallicity galaxies originates from a geometry effect, such that the low-metallicity galaxies have a more clumpy ISM where dust is spatially concentrated and heated to higher temperatures. Tentative evidence for a clumpy dust geometry at low metallicities at $z\sim 2$ has been discussed in \citet{shivaei20a}. In that study, the authors compared the nebular and stellar dust reddening ($E(B-V)$) and concluded that on average at low metallicities the two reddenings are not the same, which suggests a clumpy two-component dust geometry. We also discuss this possibility and its implications in Section~\ref{sec:dust_mass_evolution}. 

\paragraph{sSFR}
sSFR indicates the ratio of recent SFR to the SFR averaged during the lifetime of the galaxy. In studies of nearby galaxies it has been shown that the temperature of the warm dust correlates with both sSFR and metallicity \citep{galliano05,boselli10,smith12,remyruyer13,remyruyer15}, and that at high SFRs it is driven by recently-born young massive stars \citep{boquien11}.
\citealt{remyruyer15} attributed the warmer and wider SED of local dwarfs compared to the SED of local starbursts to differences in their sSFR.
A high sSFR, indicates the galaxy is undergoing an active phase of star formation and is likely to have a clumpier ISM structure as it has a larger number of massive hot stars embedded in their dusty birthclouds \citep{dale07,dacunha08}. The clumpier ISM allows for a wider range of interstellar radiation field intensities, leading to a wider equilibrium temperature distribution of the dust grains (hence, broader IR SED), and the hotter regions shift the temperature to higher values \citep{dacunha08,remyruyer15}. sSFR also correlates with SFR surface density \citep{elbaz11}, which is a proxy for an intense radiation field, and hence a hotter dust temperature (see below).
Moreover, a relatively higher supernovae rate in actively star forming galaxies allows for an increase in the population of small dust grains through shattering of large grains by grain-grain collisions in supernovae shock waves \citep{jones96}. 
The sSFR distributions of the two metallicity samples are similar to each other (with average $\log({\rm sSFR}/M_{\odot} {\rm yr^{-1}})=-8.43$ and $-8.53$, and standard deviation of $\sigma = 0.21$ and 0.21 for the subsolar and solar metallicity samples, respectively). Therefore, to assess the degree of the variation of the IR SED with sSFR, we define two new bins of galaxies below and above $\log(\rm{sSFR/yr^{-1}})=-8.45$ with 15 and 12 galaxies, respectively, and fit their Herschel and ALMA data with the 2T-MBB models (middle panel of Figure~\ref{fig:2comp}). The difference in the warm dust temperature of the two sSFR SEDs is similar to that of the two metallicity SEDs (top panel of Figure~\ref{fig:2comp}). However, the SEDs of both the low and high sSFR stacks are still relatively broad, as the peak fluxes of the warm and cold components in both bins are similar to each other. As a comparison, the warm component in the solar-metallicity SED is 20 times more intense than the cold component, making the overall shape of the solar-metallicity IR SED narrower than that of the subsolar-metallicity one.

\paragraph{SFR Surface Density}
The equilibrium temperature of dust grains can also be increased by high ISM radiation field intensity. A proxy for radiation field intensity is the SFR surface density, $\Sigma_{\rm SFR}$, indicating the compactness of the star forming region. The effect of $\Sigma_{\rm SFR}$ on dust temperature is shown in previous observational studies \citep{lehnert96,chanial07,burnham21}, as well as the theoretical models of \citet{derossi18}. In the latter study, a ``bluer'' SED is produced by increasing star formation efficiency, which is accompanied by a decrease in the virial radius, and hence an increase in luminosity density. It has also been shown that the luminosity surface density explains the similar mid-IR emission behavior of centrally concentrated local ULIRGs and that of high-redshift LIRGs \citep{elbaz11,rujopakarn11,rujopakarn13}. 

We use optical sizes, measured from HST/F160W images \citep{vanderwel14}, to calculate $\Sigma_{\rm SFR}$. There are recent studies that show the dust emission tend to be much more compact than the stellar emission at these redshifts \citep{rujopakarn19,popping21}. However, as we use sizes to only separate the galaxies into two bins of $\Sigma_{\rm SFR}$, the discrepancy between the absolute optical (tracing mass) and IR (tracing star formation) sizes would likely not change any of the following main results.

The $\Sigma_{\rm SFR}$ of galaxies in our sample ($\langle \log(\Sigma_{\rm SFR}/M_{\odot} {\rm yr^{-1} kpc^{-2}}) \rangle = 0.19$ and $\sigma = 0.43$) are consistent with those of the main-sequence $z\sim 2$ galaxies calculated from the parent MOSDEF sample ($\langle \log(\Sigma_{\rm SFR}/M_{\odot} {\rm yr^{-1} kpc^{-2}}) \rangle = -0.22$ and $\sigma = 0.82$), but smaller than $\Sigma_{\rm SFR}$ of local ULIRGs ($\log(\Sigma_{\rm SFR}/M_{\odot} {\rm yr^{-1} kpc^{-2}}) \sim 1.5-2.5$, from \citealt{tacconi13}). Within our sample, the subsolar-metallicity galaxies have on average a larger $\Sigma_{\rm SFR}$ than the solar-metallicity galaxies. However, the correlation between $\Sigma_{\rm SFR}$ and metallicity in the sample is weak (Pearson correlation coefficient of $-0.44$ with p-value of 0.02) and the standard deviation of $\Sigma_{\rm SFR}$ within each metallicity subsample is large, making the $\Sigma_{\rm SFR}$ distributions of the two metallicity bins not significantly distinct (Table~\ref{tab:sample}). To better evaluate the effect of $\Sigma_{\rm SFR}$ on the IR SED in our sample, we fit the 2T-MBB model with $T_{\rm c}=25$\,K and $\beta=1.5$ to the stacks of MIPS, Herschel, and ALMA data in two bins of $\Sigma_{\rm SFR}$ divided at $\log(\Sigma_{\rm SFR}/M_{\odot} {\rm yr^{-1} kpc^{-2}})=0$, with 12 and 15 galaxies in the low and high $\Sigma_{\rm SFR}$ bins, respectively. The fits are shown in the bottom panel of Figure~\ref{fig:2comp}. As expected, the best-fit model of the high $\Sigma_{\rm SFR}$ bin shows a warm component that is hotter than the one in the best-fit model of the low $\Sigma_{\rm SFR}$ stacks. The warm dust temperature difference between the two bins is $\Delta T_{\rm w}=23 \pm 33$, which is less significant than the warm dust temperature difference between the two metallicity bins. Equally important is the difference between the warm-to-cold luminosity ratios. Both the low and high $\Sigma_{\rm SFR}$ best-fit models have warm and cold components with similar luminosities, indicating a broad IR SED that does not change with $\Sigma_{\rm SFR}$ significantly.

\paragraph{Stellar Population Age}
Another parameter that affects dust emission properties is age. \citet{derossi18} showed that a silicate-rich mixture with amorphous carbon dust can explain a hot and broad IR SED that originates from the relatively high emission efficiency of silicates between $\sim 8-60$\,{\um}. A source of silicate-rich dust is massive AGB stars \citep[with masses above 3.5\,{\msun};][]{ventura12a,ventura12b}. A 3.5\,{\msun} star has a main-sequence lifetime of $\sim 400$\,Myr, suggesting that the silicate-rich dust composition would dominate the IR emission of galaxies younger than this age. We have age estimates for galaxies from the best-fit UV-to-near-IR SED models. However, as ages derived from exponentially rising star formation history models in SED fitting are ambiguous \citep[e.g., see Section 6.2 of][]{reddy12b}, a robust analysis of the effect of age on the IR SED shape is beyond the scope of this paper. Additionally, the luminosity-weighted ages are correlated with metallicity and anti-correlated with sSFR, which makes it difficult to disentangle the age effect from the other two parameters in this analysis.

\paragraph{Summary}
In conclusion, we find that a warm component is present across our sample, as seen in the wavelength separation of the warm and cold components' peaks in Figure~\ref{fig:2comp} compared to that of the {\S18} model. Even in the low sSFR and low SFR surface density bins, the two peaks have wavelength separations of $\sim 265\pm 32$\,{\um}, which is wider than the wavelength separation between the cold and warm components modeled for the {\S18} template is $213_{- 12}^{+14}$\,{\um}. This could be due to the high sSFR and $\Sigma_{\rm SFR}$ of these galaxies compared to the average values for their stellar mass at $z\sim 2$ (Figure~\ref{fig:mz-mssfr}), which makes this analysis distinct from many other studies at $z\sim 2$. For example, at $z=1.8-2.5$, the {\S18} lowest mass sample has $\log(M_*/M_{\odot})=10.0-10.5$, which is consistent with the stellar mass of our solar-metallicity sample ($\log(M_*/M_{\odot})\sim 10.2-10.6$). The average SFR of the {\S18} sample based on their IR luminosity measurements and a \citet{kennicutt12} relation (adopted for a Chabrier IMF) is $\sim 20$\,{\msun}/yr. At these masses, it is expected to have $>70$\% of the SFR in the obscured phase \citep{whitaker17}. Therefore, the $\log({\rm sSFR/yr^{-1}})$ of their low-mass sample is $\sim -8.8$ to $-8.9$, which is lower than the sSFR of the majority of our sample with the average and scatter of $\langle \log({\rm sSFR/yr^{-1}})\rangle=-8.4$ and $\sigma = 0.2$\,dex. 

Across our sample, we find that the IR SED gets broader and the temperature of the warm component increases with decreasing metallicity, increasing sSFR, and increasing SFR surface density. 
The subsolar metallicity, high sSFR, and high SFR surface density samples show wavelength separations between the peaks of their warm and cold components of $\Delta \lambda=347^{+26}_{-16}$\,{\um}, $\Delta \lambda=304^{+6}_{-5}$\,{\um}, $\Delta \lambda=317^{+55}_{-27}$\,{\um}, respectively. These are $\sim 1.5-1.6$ times higher than the wavelength separation of 213\,{\um} between the cold and warm components modeled for the best-fit {\S18} template.
The broadening effect is the most significant with metallicity, as the two dust components show the largest peak separation and temperature difference ($\Delta \lambda=347^{+26}_{-16}$\,{\um} and $\Delta T=108\pm33$\,K) at subsolar metallicities.

The high SFR surface density and subsolar metallicity samples studied here are useful analogs to higher redshift galaxies, given the expected redshift evolution in size \citep{mosleh12,vanderwel14}, SFR \citep{speagle14,tasca15}, and metallicity \citep{troncoso14,sanders21} at a fixed stellar mass. As will be discussed in Section~\ref{sec:submmestimation}, the commonly-used narrower and colder IR templates that are calibrated based on local LIRGs and ULIRGs, fit to shorter wavelength data alone ($\lambda_{\rm rest}\lesssim 60$\,{\um}), overestimate the RJ emission of high-redshift galaxies.
In the case of limited available data, our results suggest that templates with hotter and broader IR SEDs should be considered for {\rm typical} (i.e., LIRG and lower luminosity) galaxies at $z\gtrsim 2$, such as the average local low-metallicity template presented in this work. Similar results have been shown previously by \citet{derossi18} and \citet{faisst20}, who recommended using the hotter and broader IR template of the local low-metallicity galaxies at $z>4$.

\section{Implications for Integrated Quantities: IR Luminosity, SFR, IR Colors, and Dust Mass} \label{sec:implications}

In the coming decade, the primary means to estimate IR luminosities, obscured SFRs, and dust masses will be JWST and mm/submm facilities such as ALMA and LMT/TolTEC, operating respectively at $\sim 21$\,{\um} and $\sim 0.5-3$\,mm, which correspond to the rest-frame PAH emission and FIR/submm dust continuum emission at $z\sim 2.3$. Main-sequence typical galaxies similar to our subsolar-metallicity sample will be within reach for either approach and synergies between the observatories is a natural consequence. Since the results of this work indicate a change in the IR SED with metallicity, it is of concern how well either of the observed PAH or submm dust continuum measurements estimate IR luminosities and how well the PAH emission can estimate the submm fluxes and vice versa. The need for an improved prescription to predict the submm flux of high-redshift galaxies is underscored by the lower-than-expected detection rates of high-redshift galaxies in blind ALMA surveys (e.g., compare the predictions in \citealt{hatsukade16} and \citealt{fujimoto16} with the detection rate in \citealt{aravena16}). In the following subsections, we first describe the comparison samples at $z\sim 0$ and 2 (Section~\ref{sec:comparison_samples}), and then use the results of the previous sections on the evolution of IR SED with metallicity to discuss submm flux predictions and inferred IR luminosity (Section~\ref{sec:submmestimation}), obscured SFR (Section~\ref{sec:obssfr}), and inferred dust masses (Section~\ref{sec:dustmass}) of $z\sim 2.3$ galaxies. 

As the main results of this section and the next section (\ref{sec:dustmass}) rely primarily on ALMA photometry, we take advantage of the depth of our ALMA observations and construct stacks in more than two bins of metallicity. In this section, the uncertainties of the stacked photometry are only measurement uncertainties, and do not include the jackknife resampling as was performed in the previous sections. From a visual inspection, we exclude one of the 27 targets (ID 19753), which is likely a merger or a complex system with multiple star-forming components, from this part of the analysis. The 1.2\,mm and UV peak emission of this object are not co-spatial, and while the MOSFIRE slit is centered on the component with the peak UV emission, the IR emission (traced by ALMA) is mainly from another component. Excluding this source does not change the results of the best-fit SED models in the previous sections.

\subsection{Comparison samples at $z\sim 0$ and 2} \label{sec:comparison_samples}

Here we describe the comparison samples that are adopted from the literature.
At $z\sim 0$, we adopt four datasets with Herschel and Spitzer photometric observations that cover a wide range of galaxy populations in the local Universe from dwarfs to ULIRGs, as follows. 
\begin{itemize}
    \item The Dwarf Galaxy Survey \citep[DGS;][]{madden13} consists of 48 star-forming dwarf galaxies with metallicities from {\oh}$=7.14$ to 8.43. The Herschel and Spitzer photometric data and measured galaxy properties (metallicity, SFR, stellar and dust mass) are collected from \citet{madden13} and \citet{remyruyer15}. 
    \item The Key Insights on Nearby Galaxies: a Far-Infrared Survey with Herschel \citep[KINGFISH;][]{kennicutt11} is a survey of 61 nearby galaxies drawn from the Spitzer Infrared Nearby Galaxies Survey (SINGS), selected to span wide ranges in luminosity, optical-to-IR ratio, and morphology, with metallicities from {\oh}$=7.54$ to 8.77. The metallicities and AGN classification are taken from \citet{kennicutt11}. The Spitzer and Herschel photometry of the sample are listed in \citet{dale07} and \citet{dale12}, respectively. To be consistent with the DGS measurements, we adopt the SFR, stellar mass, and dust mass of KINGFISH galaxies from \citet{remyruyer15}. 
    \item The Herschel Reference Survey \citep[HRS;][]{boselli10} is a larger sample of 323 nearby galaxies that is complementary to the KINGFISH and DGS samples in terms of the coverage in luminosity, mass, and morphology. The IRAC, MIPS, and Herschel photometry of this sample are taken from \citet{ciesla14}, \citet{bendo12}, and \citet{ciesla12}, respectively. Stellar masses and SFRs are listed in \citet{hughes13}, and O3N2 metallicities are calculated using the optical line catalog of \citet{boselli13}. Dust masses \citep{ciesla14} are derived from SED fitting.
    \item To complete the sample of nearby galaxies, we also add data from Great Observatories All-sky LIRG Survey \citep[GOALS;][]{armus09}, which has over 200 of the most luminous infrared-selected galaxies in the local Universe. The Herschel photometry is listed in \citet{chu17}, and the Spitzer photometry and stellar masses are provided in a private communication based on the results published in \citet{diazsantos10}, \citet{diazsantos13}, and \citet{howell10}. When calculating SFRs, we correct for the AGN contribution in the IR luminosities by multiplying the IR luminosities by ($1-f_{\rm{AGN}}$), where $f_{\rm{AGN}}$ is bolometric AGN fraction \citep{diazsantos17}.
\end{itemize}

AGN are excluded from the KINGFISH sample \citep{kennicutt11} and the HRS sample \citep[based on optical lines;][]{hughes13}. The SFR and stellar masses of all the samples are either based on or converted to a Chabrier IMF.

As the main-sequence comparison sample at $z\sim 2$, we adopt data from two surveys, as follows. Neither of the two samples have metallicity measurements.
\begin{itemize}
    \item We adopt 10 galaxies from the ALMA Spectroscopic Survey in the Hubble Ultra Deep Field \citep[ASPECS;][]{walter16} that are detected in ALMA band-6 continuum and Spitzer/MIPS 24\,{\um}, and have redshifts of $z=1.6-2.2$. The Band-6 flux, stellar mass, SFR, and redshifts are taken from the survey's online public release\footnote{\href{https://www.aspecs.info/data/}{https://www.aspecs.info/data/}} \citep{aravena20,gonzalez20,boogaard20}. The Spitzer and Herschel fluxes are taken from the 3D-HST survey catalog \citep{skelton14}. Out of the 10 galaxies, only 4 are detected with PACS at 100\,{\um}. There are no metallicity measurements for these galaxies. For consistency with our measurements, we calculate the dust masses from ALMA band-6 fluxes in the same way as done for our sample (Equation~\ref{eq:mdust})
    \item The average SFR, stellar mass, and dust masses of \citet{santini14} at $z=2.0-2.5$ are also adopted. Dust masses in this study are derived from SED fitting to the PACS and SPIRE 100-to-500\,{\um} photometric stacks in bins of stellar mass and SFR. There are no metallicity measurements.

\end{itemize}

\begin{figure*}[hbt]
	\centering
		\includegraphics[width=.45\textwidth,clip]{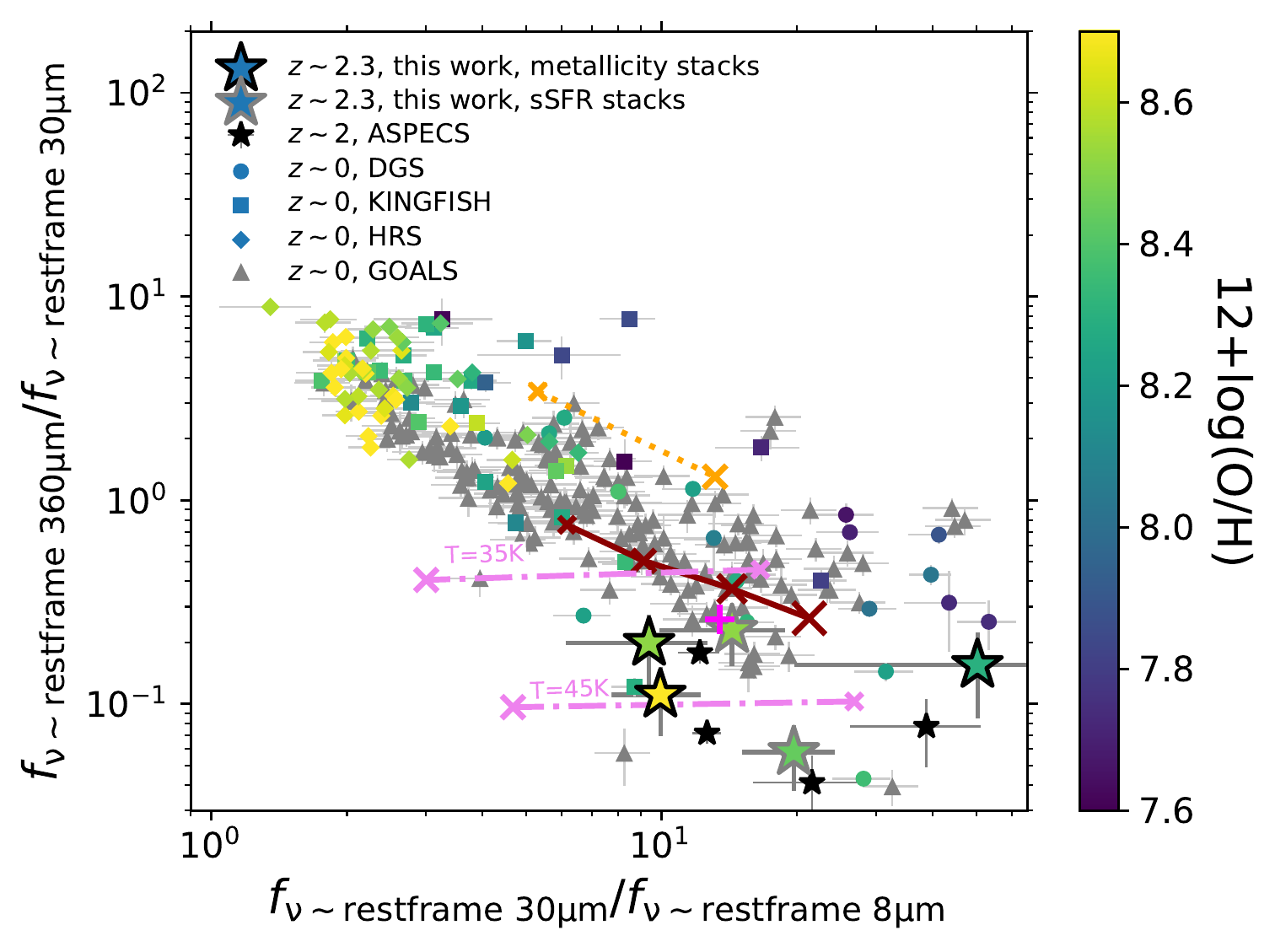}\\
		\includegraphics[width=.46\textwidth,clip]{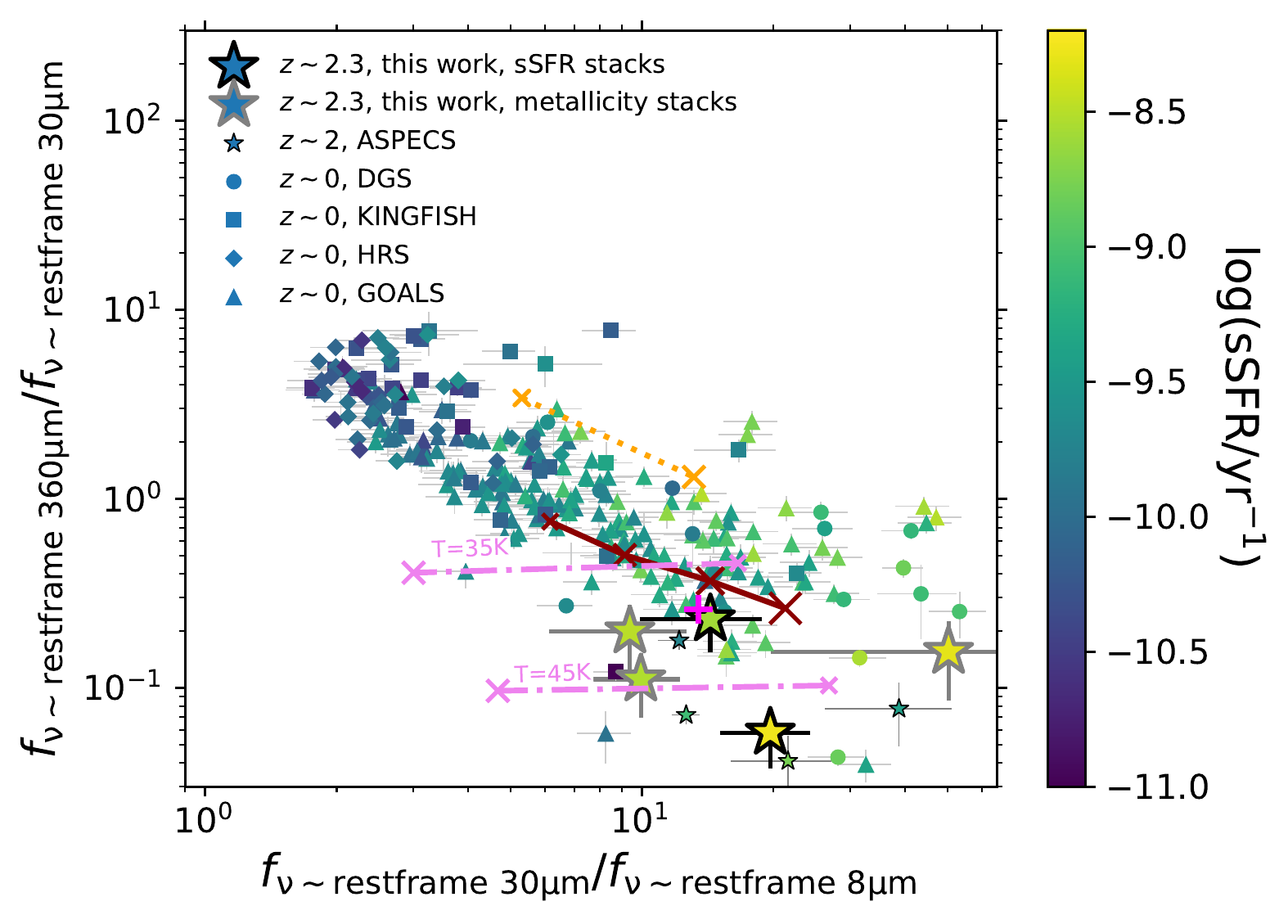}\\
		\includegraphics[width=.45\textwidth,clip]{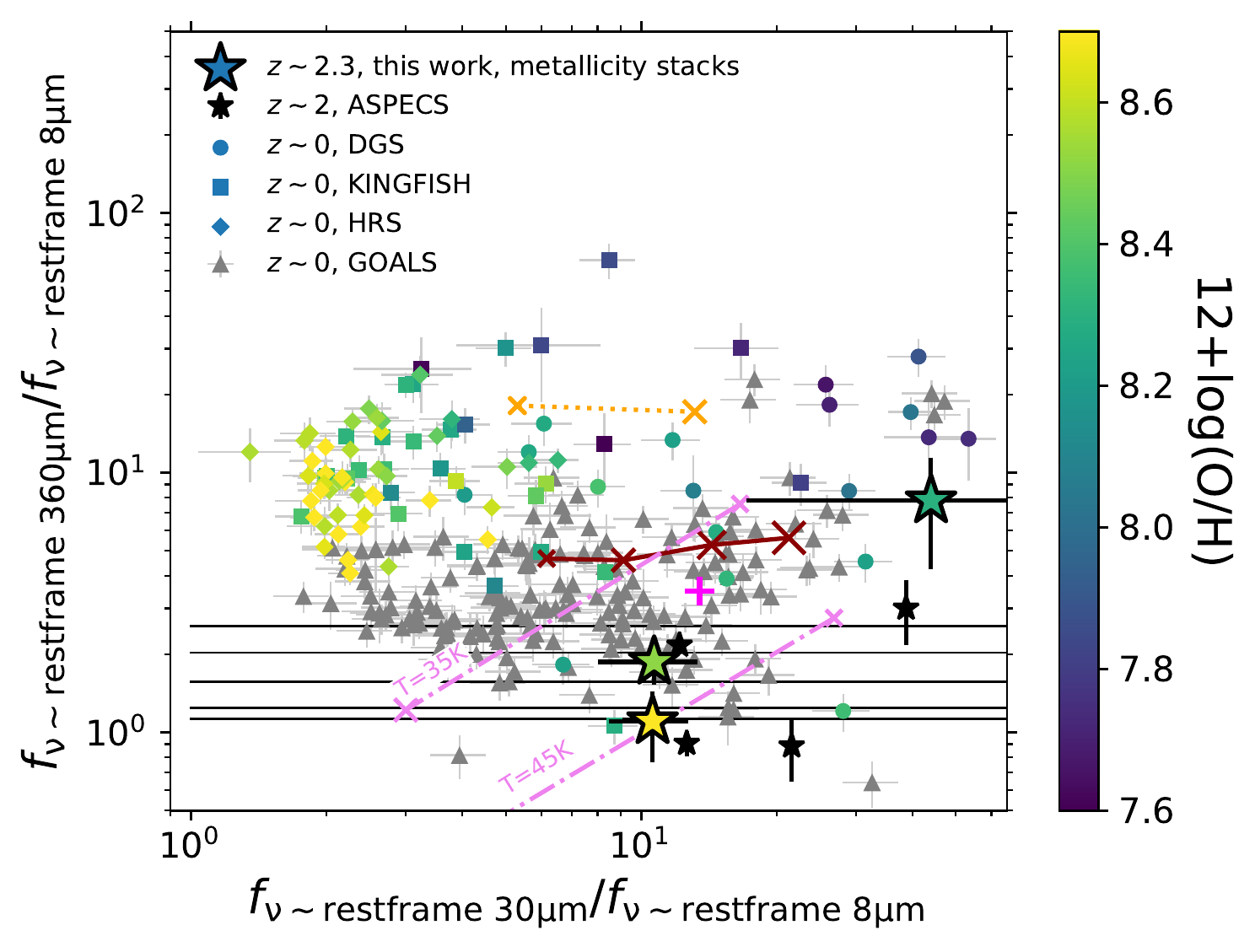}
	\caption{
	IR colors from rest-frame 360, 30, and 8\,{\um} emission for the $z\sim 2.3$ stacks in this work (large stars) compared to those of $z\sim 0$ surveys and the $z\sim 2$ ASPECS galaxies. Comparison samples are described in Section~\ref{sec:comparison_samples}.
	Crosses show model predictions: The magenta plus sign is the average local low-metallicity template, the small and large orange crosses (connected with a dotted line) show the main-sequence and starburst models of \citet{elbaz11}, respectively. 
	Dark red crosses (connected with a solid line) show the {R09} models with IR luminosities of $10^{11}$, $10^{11.25}$, $10^{11.5}$, and $10^{11.75}$\,$L_{\odot}$, in order of increasing size.
	Pink crosses (connected with dot-dashed lines) show the {S18} models for two temperatures (indicated on the plots) and two PAH fractions (0.01 and 0.1 for the small and large cross, respectively). 
	{\em Top:} 360-to-30\,{\um} versus 30-to-8\,{\um} flux ratios, color-coded by metallicity. The stars with black edges are stacks in three bins of metallicity, while the grey edge stars are two stacks in bins of sSFR, shown in the middle panel. 
	{\em Middle}: 360-to-30\,{\um} versus 30-to-8\,{\um} flux ratios, color-coded by sSFR. Stars with black edges are the stacks in two bins of sSFR. The stacks in three metallicity bins from the top panel are shown by stars with grey edges. 
	{\em Bottom:} 360-to-8\,{\um} versus 30-to-8\,{\um} flux ratios, color-coded by metallicity. Horizontal black lines show the 360-to-8\,{\um} color of the ASPECS $z\sim 2$ galaxies that do not have PACS 100\,{\um} (rest-frame 30\,{\um}) detections.
		}
		\label{fig:ircolor}
	\vspace{21pt}
\end{figure*}

\begin{figure}[ht]
	\centering
		\includegraphics[width=.45\textwidth,trim={.2cm 0 0 0},clip]{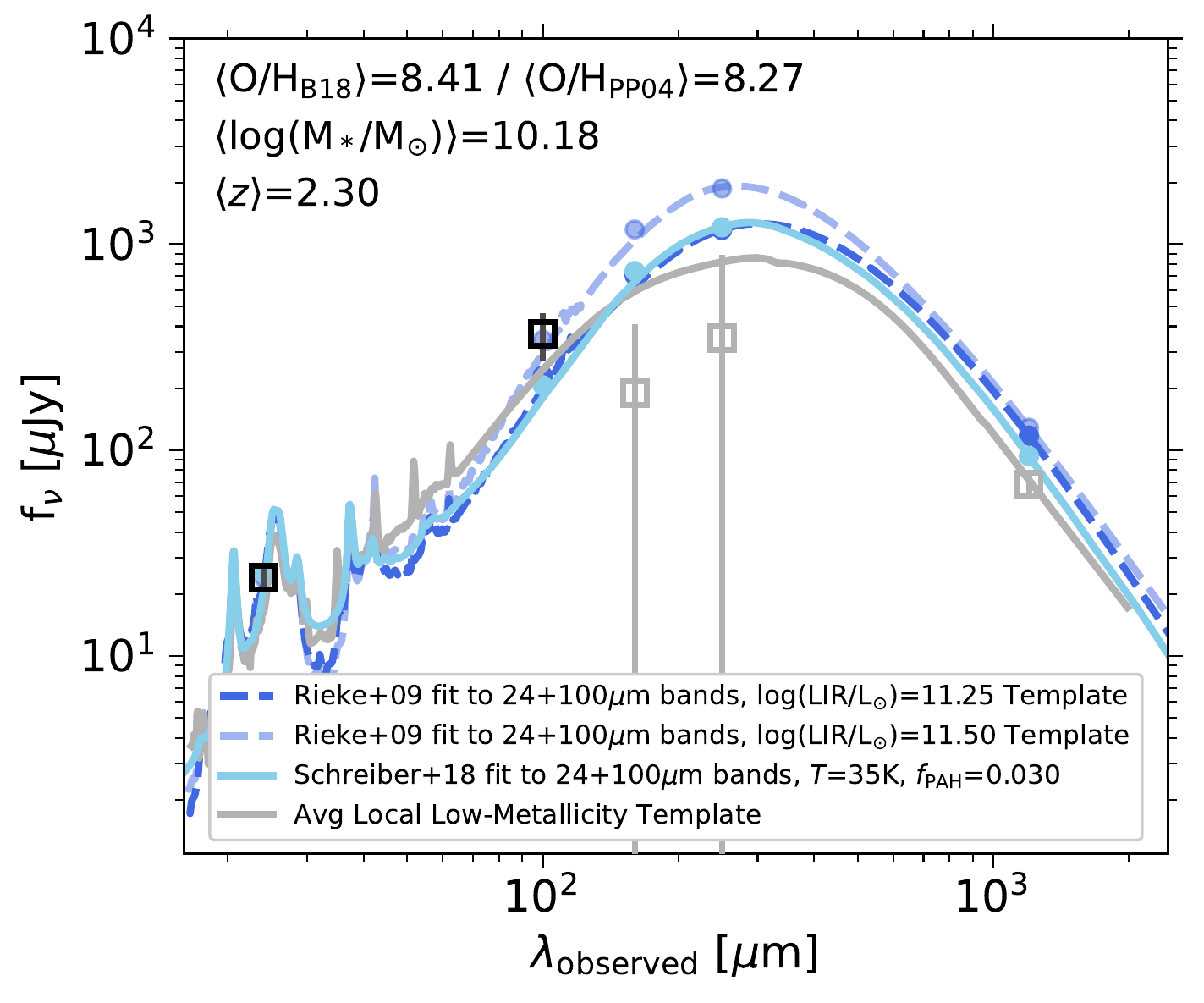} 
		\caption{The \citet{rieke09} $\log({\rm L(IR)}/L_{\odot})=11.50$ and 11.25 and the \citet{schreiber18} $T=35$\,K model fits to the 24 and 100\,{\um} stacks alone (blue curves) compared with the average local low-metallicity template fit to all data (grey curve).
		The R09 and S18 templates fit to the 24 and 100\,{\um} photometry alone overestimate the observed submm flux and the IR luminosity calculated from the local low-metallicity fit to 24-to-1200\,{\um} data.
		}
		\label{fig:only24a100}
\end{figure}

\subsection{Estimating submm fluxes and IR luminosities} \label{sec:submmestimation}

It is often the case at high redshifts that limited IR data are available to estimate IR luminosity or to predict flux densities of other parts of the IR SED. Therefore, for simplicity often a single locally-calibrated IR template is assumed and fit to the limited data. Here, we investigate how well the IR luminosities and IR colors of our $z\sim 2.3$ LIRG sample match with the commonly used IR templates of local LIRGs.

\paragraph{IR luminosity from ALMA}
The IR luminosity (8 to 1000\,{\um}) of our subsolar metallicity bin is $\log({\rm L(IR)/}L_{\odot})=11.35$, calculated from the best-fit average local low-metallicity template in Figure~\ref{fig:lowZ}. 
The {\S18} $T=35$ and the {\R09} $\log({\rm L(IR)/}L_{\odot})=11.25$ templates fit to the ALMA submm flux density alone underestimate the IR luminosity by $0.1-0.2$\,dex.
The colder templates (e.g., the {\S18} $T=30$\,K or the {\R09} $\log({\rm L(IR)/}L_{\odot})=11.00$ templates) underestimate the IR luminosity by $>0.4$\,dex. This is because these templates underestimate the elevated mid-IR emission in these galaxies. The IR luminosity of the low-metallicity $z\sim 2.3$ galaxies is generally less biased when the templates are fit to PACS data, or when the broader dwarf templates are used to fit the MIPS data \citep[e.g., see Appendix A in][]{shivaei20b}. The broader templates such as the one we constructed here from the local low-metallicity galaxies are necessary when only limited submm flux observations are available.

\paragraph{ALMA flux from IR luminosity}
From another point of view, we examine how well the submm fluxes can be predicted based on the IR luminosity of our galaxies. The observed ALMA flux to L(IR) ratio of our subsolar metallicity bin is $\log(f_{1200}/{\rm L(IR)}/\mu{\rm Jy}~L_{\odot}^{-1})= -9.52\pm 0.05$. This observed value is lower than predictions from locally-calibrated IR templates, such as those of {\R09} and {\S18}. As an example, the {\R09} $\log({\rm L(IR)/}L_{\odot})=11.25$ and 11.50 templates have submm flux-to-L(IR) ratios that are $\sim 0.2-0.3$\,dex larger than our observed value. The higher-than-observed submm flux to L(IR) ratio of the templates is due to the colder dust temperature of the templates at a given IR luminosity compared to the luminosity-weighted average temperatures of the galaxies in this work. 
The hotter average temperatures can be attributed to the higher sSFRs of the galaxies compared to the main-sequence. Similarly, the ASPECS galaxies that are located above the $z\sim 2$ main-sequence relation \citep{shivaei15b} show observed 1.2\,mm flux-to-L(IR) ratios\footnote{Here, the L(IR) of ASPECS sample is estimated from UV-to-IR SED fitting \citep{boogaard19}.} that are lower than those predicted by the aforementioned local LIRG templates by $>0.3$\,dex. 

\paragraph{ALMA flux from mid-IR and PAH emission}
Another common scenario for high-redshift surveys is that only photometry shortward of the IR emission peak is available (e.g., from Spitzer, Herschel, or future JWST/MIRI) to predict submm fluxes. In this case, we assess how well the rest-frame 30-to-8\,{\um}, 360-to-8\,{\um}, and 360-to-30\,{\um} IR colors of the $z\sim 2.3$ galaxies match with those of the $z\sim 0$ samples and local LIRG IR templates.

The $z\sim 2.3$ sample in this work has relatively good constraints on the observed 24, 100, and 1200\,{\um} photometry. These bands correspond to rest-frame 8, 30, and 360\,{\um}, respectively, which for a $z\sim 0$ sample are roughly traced by IRAC 8\,{\um}, MIPS 24\,{\um}, and SPIRE 350\,{\um}. In Figure~\ref{fig:ircolor}, we show the IR colors based on the aforementioned bands for the $z\sim 2.3$ sample and the $z\sim 0$ comparison samples (Section~\ref{sec:comparison_samples})\footnote{To correct for the slight offset in rest-frame wavelength of the $z\sim 0$ and 2 observations, we multiplied the $z\sim 0$ MIPS 24\,{\um}, and SPIRE 350\,{\um} fluxes by a factor of 2 and 0.9, respectively, to convert them to fluxes at rest-frame 30 and 360\,{\um}. These correction factors are estimated based on the local LIRG templates of {\R09}.}.
Overplotted on Figure~\ref{fig:ircolor} are also the IR color predictions from four IR templates: a) the {\R09} LIRG templates with $\log({\rm L(IR)}/L_{\odot})=11.00$, 11.25, 11.50, and 11.75 (dark red crosses with sizes increasing with increasing luminosity). These three models are in the range of recommended templates for galaxies at $z>1$ by \citet{rujopakarn13} and \citet{derossi18}, b) the {\S18} templates with $T=35$ and 45\,K and PAH fraction of 0.01 and 0.1 (pink crosses with sizes increasing with increasing PAH fraction). The $T=35$ template is the one that is recommended by \citet{schreiber18} to be used for high-redshift galaxies, c) the starburst and main-sequence templates of \citet[][orange crosses with the small and large ones for the main-sequence and starburst models, respectively]{elbaz11}, and d) the average local low-metallicity template (magenta plus sign, Section~\ref{sec:irfits}).

The rest-frame 30-to-8\,{\um} colors of the $z\sim 2.3$ galaxies vary by a factor of 5 between the lowest metallicity bin ($12+\log({\rm O/H})_{B18}=8.33$) and the rest, which shows suppressed PAH emission at low metallicities, as expected. The 24\,{\um} SNR of the lowest metallicity bin is only 2.3, which makes its 30-to-8\,{\um} color consistent with that predicted by the highest luminosity {\R09} template and the hotter {\S18} template within 1$\sigma$. However, we know from previous studies with larger samples and better constraints on the PAH emission of $z\sim 2$ galaxies, that the PAH emission at such metallicities at $z\sim 2$ is suppressed compared to that of the local LIRG templates \citep{shivaei17} and their IR SED shape resembles that of the local low-metallicity galaxies with low PAH fractions \citep{shivaei20b}. The IR colors of the lowest metallicity stack in this work also overlap with those of some of the local dwarfs (with {\oh}$\sim 7.8$), likely due to the low PAH emission and hot dust component of the local dwarfs that resemble the dust emission characteristics of the $z\sim 2$ galaxies with metallicity of $\sim 0.4\,Z_{\odot}$. 

In Figure~\ref{fig:ircolor}, the models that are matched to the 30-to-8\,{\um} color of the higher two metallicity stacks ($12+\log({\rm O/H})_{B18}\sim 8.5-8.7$) overpredict the rest-frame 360\,{\um} flux by at least a factor of 2.5 (4$\sigma$) using the {\R09} and the $T=35$\,K {\S18} models, and even more if either of the \citet{elbaz11} templates are adopted. This effect can also be seen in Figure~\ref{fig:only24a100} where the models are fit to the observed 24 and 100\,{\um} points alone, overpredicting the observed 1200\,{\um} emission (except for the local low-metallicity template). The middle panel of Figure~\ref{fig:ircolor} indicates that the discrepancy between the observed and model predicted 350-to-30\,{\um} colors increases with increasing sSFR, such that the higher sSFR bin has observed colors about an order of magnitude lower than those predicted by the {\R09} and the $T=35$\,K {\S18} templates and that match better with the higher temperature templates, while the lower sSFR bin is discrepant with the two mentioned models only at a $2-3\,\sigma$ level. Metallicity and sSFR are highly correlated with each other and separating their effects on the observed 360-to-30\,{\um} color is not possible with the current dataset.

The colors of the four ASPECS galaxies with detected observed 24, 100, 1200\,{\um} data are in agreement with our two higher metallicity stacks. The other 6 ASPECS galaxies are only detected in the observed 1200 and 24\,{\um} bands and are shown by horizontal lines (i.e., no constraints on the PACS photometry) -- again in agreement with the rest-frame 360-to-8\,{\um} colors of our two higher metallicity stacks.
These results indicate that, irrespective of the PACS observations, the locally calibrated LIRG templates anchored to the  rest-frame 8\,{\um} flux density may overestimate the submm flux of typical galaxies at $z\sim 2.3$ by $\sim 1.5$ up to an order of magnitude. 

In summary, using the local LIRG templates and the rest-frame 8\,{\um} emission alone or the combination of the 8\,{\um} and $\sim 30\,${\um} emission tends to overestimate the submm fluxes of the high-redshift LIRGs. This effect becomes more pronounced with increasing sSFR. Hotter templates (e.g., those with $T>40$\,K from the {\S18} library or the {\R09} templates with $\log({\rm L(IR)/L_{\odot}}>11.75$ {\R09}) predict more accurate submm fluxes based on shorter wavelength data. A similar conclusion holds for the local low-metallicity template derived in this work (Section~\ref{sec:irfits}), where the submm fluxes of $z\sim 2.3$ LIRGs are well predicted based on 8 or 30\,{\um} emission. However, all templates predict lower 30-to-8\,{\um} colors than the lowest metallicity $z\sim 2.3$ galaxies ({\oh}$\sim 8.3$). The IR colors of these low-metallicity $z\sim 2.3$ galaxies are similar to those of some of the lowest metallicity local dwarfs albeit their $0.2-0.6$\,dex higher oxygen abundances, suggesting very weak PAH emission at gas metallicities of {\oh}$\sim 8.3$ at $z\sim 2.3$ (similar to the findings of \citealt{shivaei17}).

\subsection{Obscured SFR fraction} \label{sec:obssfr}

We estimate the total SFR of the sample from dust-corrected {\halpha} luminosity using the Balmer decrement (Section~\ref{sec:ONIR_data_SFR}; Table~\ref{tab:sample}). As it has been shown elsewhere \citep{shivaei16}, this procedure is a good estimator of total SFR in galaxies that are not heavily dust-obscured, such as those in this study (see sample characteristics in Section~\ref{sec:sample-char}). Comparing the total SFR with the obscured SFR derived from L(IR) (using the calibrations of \citet{kennicutt12}) demonstrates that there is a significant fraction of unobscured star formation that does not contribute to the IR emission in both metallicity bins. The ratios of SFR(IR) to dust-corrected SFR({\halpha}) for the subsolar and solar-metallicity stacks in this work are 59\% and 57\%, respectively. 
Previous studies have shown that the obscured fraction of star formation (the ratio of SFR(IR) to bolometric SFR) decreases with decreasing mass \citep[e.g.,][]{reddy10,whitaker17}, however the $\sim 60\%$ obscured fraction of the our sample, with average stellar mass of $\log(\rm M_*/M_{\odot})=10.18$ and 10.43 in the subsolar and solar metallicity bins, is lower than that found previously. For example, \citet{whitaker17} predict an obscured fraction of $86\pm 2$\% at $\log(\rm M_*/M_{\odot})=10.18$ from their average fit to the data. However, the tail of their obscured fraction distribution at these masses extend to $\sim 50$\%. The discrepancy with the average prediction may be due to a bias towards less obscured star forming galaxies in our sample, as the sample is selected to have significant detection in optical nebular emission lines. However, the \citet{whitaker17} sample is also selected based on rest-frame optical continuum emission, and presumably biased against heavily obscured systems. The discrepancy may also originate from the way IR luminosity is estimated in \citet{whitaker17}. In that work, the authors convert 24\,{\um} fluxes to IR luminosity from a single log average of the \citet{dh02} templates. The use of a single template over all luminosities is an oversimplification.
For example, using a single IR template of {\R09} (e.g., the $\log({\rm L(IR)}/L_{\odot})=11.25$ or 11.50 template) to convert the 24\,{\um} flux to IR luminosity, overestimates the IR luminosity by a factor of 2 because the FIR/submm emission predicted by these templates is higher than the observations (Figure~\ref{fig:only24a100}).
If the unobscured SFR is about half of the obscured SFR (i.e., the obscured fraction is 0.67), then a factor of 2 overestimation in the obscured SFR results in an obscured fraction of $\sim 0.8$, which is similar to the discrepancy we see between the low-metallicity obscured fraction and that predicted by \citet{whitaker17}.
These results indicate that the unobscured SFR may be more significant in the subsolar metallicity and/or high sSFR galaxies at $z\sim 2.3$ than has been previously assumed.

\begin{figure*}[ht]
	\centering
		\includegraphics[width=.9\textwidth,clip]{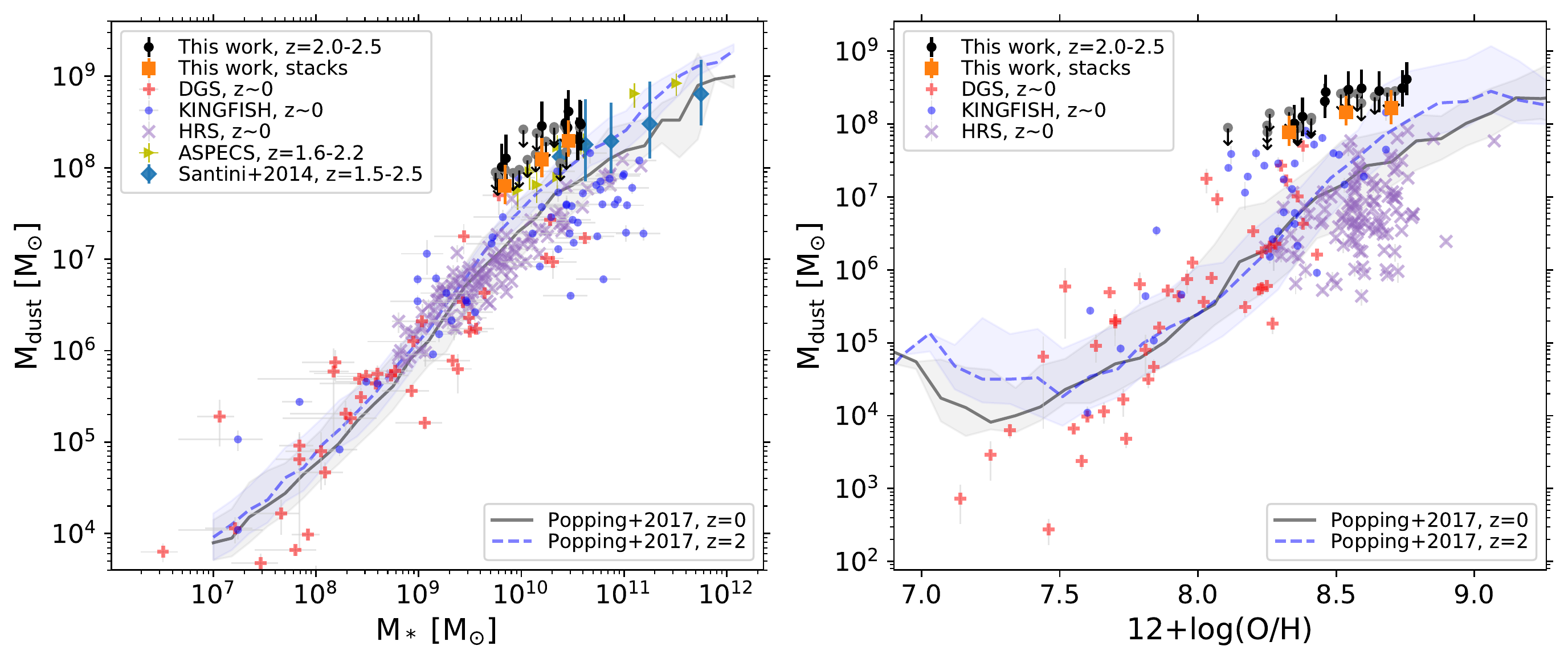}
		\caption{Dust mass versus stellar mass (left) and metallicity (right) for our sample in comparison with other surveys. 
		Dust masses for galaxies in this work are shown by black circles for ALMA detections (SNR$>2$) and gray circles with downward arrows as $2\sigma$ upper limits for the nondetections. Dust masses from the ALMA stacks in three bins of stellar masses (left) or metallicity (right) are shown with orange squares (Table~\ref{tab:gasmass}). 
		Samples from the left panel that do not have metallicity measurements are not shown in the right panel. High-redshift galaxies are on the metallicity scale of \citet{bian18}. Semi-analytic models of \citet{popping17} at $z=0$ and 2 are shown in both panels. At a given stellar mass or metallicity, the $z\sim 2.3$ data show about an order of magnitude higher dust mass compared to $z\sim 0$.
		} 
		\label{fig:Md_Ms}
	\vspace{21pt}
\end{figure*}

\begin{figure*}[ht]
	\centering
		\includegraphics[width=.7\textwidth,clip]{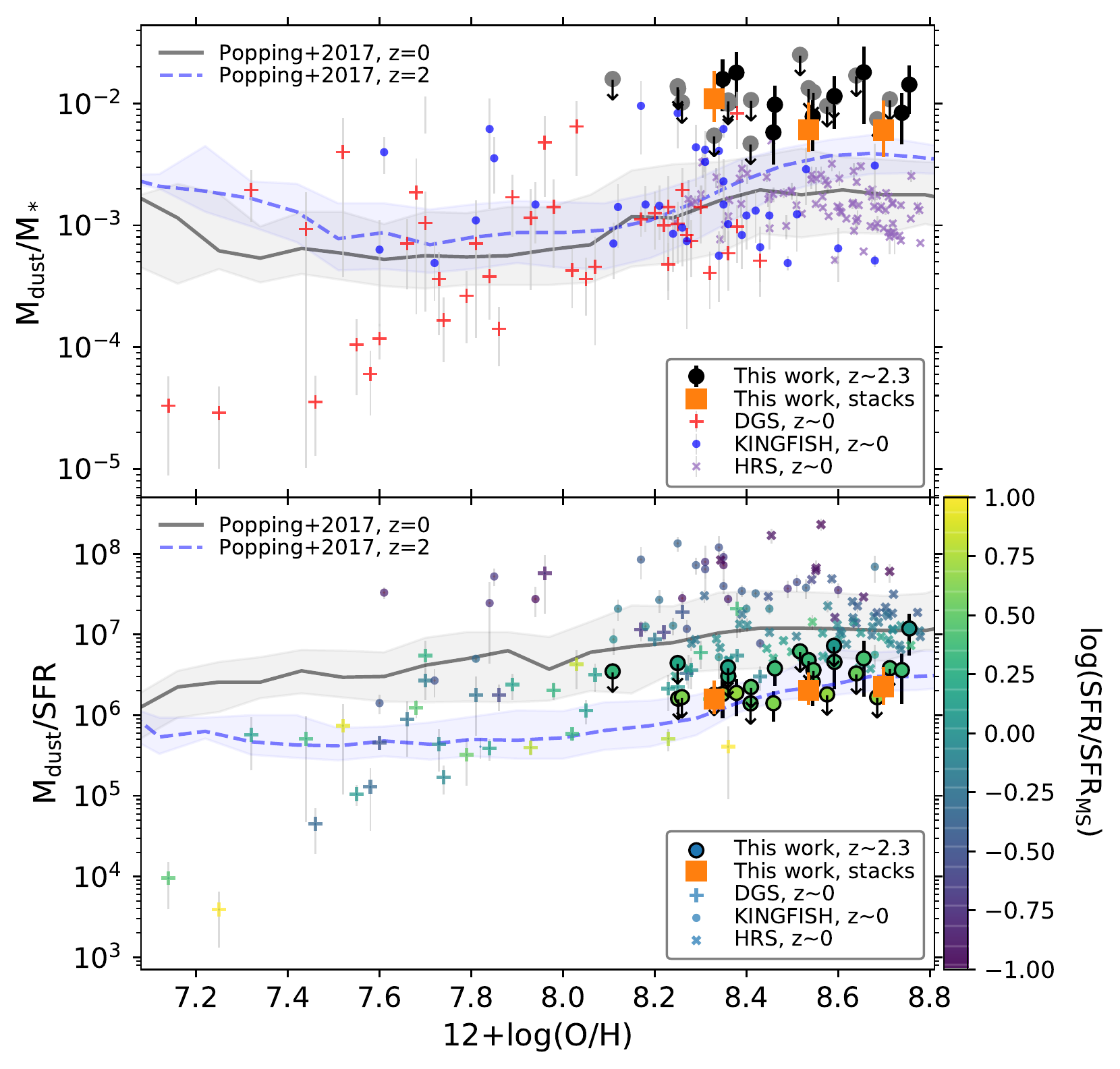}
	\caption{{\em Top:} Dust-to-stellar mass ratio as a function of metallicity for individual galaxies in this work and their stacked values in three bins of metallicity. Symbols are the same as in Figure~\ref{fig:Md_Ms}. AGN are excluded from the KINGFISH sample \citep{kennicutt11} and the HRS sample \citep{hughes13}.
	{\em Bottom:} Dust mass to SFR as a function of metallicity for the same objects as in the top panel. Symbols are color coded by the ratio of their SFR to the main-sequence SFR at the given mass. For the $z\sim 2.3$ sample, the main-sequence relation of \citet{shivaei15b} is adopted (modified for the same IMF as used in this work) and for the $z\sim 0$ objects, the main-sequence relation of \citet{salim07} is adopted. Semi-analytic models of \citet{popping17} at $z=0$ and 2 are shown in both panels.
		}
		\label{fig:Ms_Md_SFR_Z}
	\vspace{21pt}
\end{figure*}

\begin{figure*}[hbt]
	\centering
		\includegraphics[width=.9\textwidth,clip]{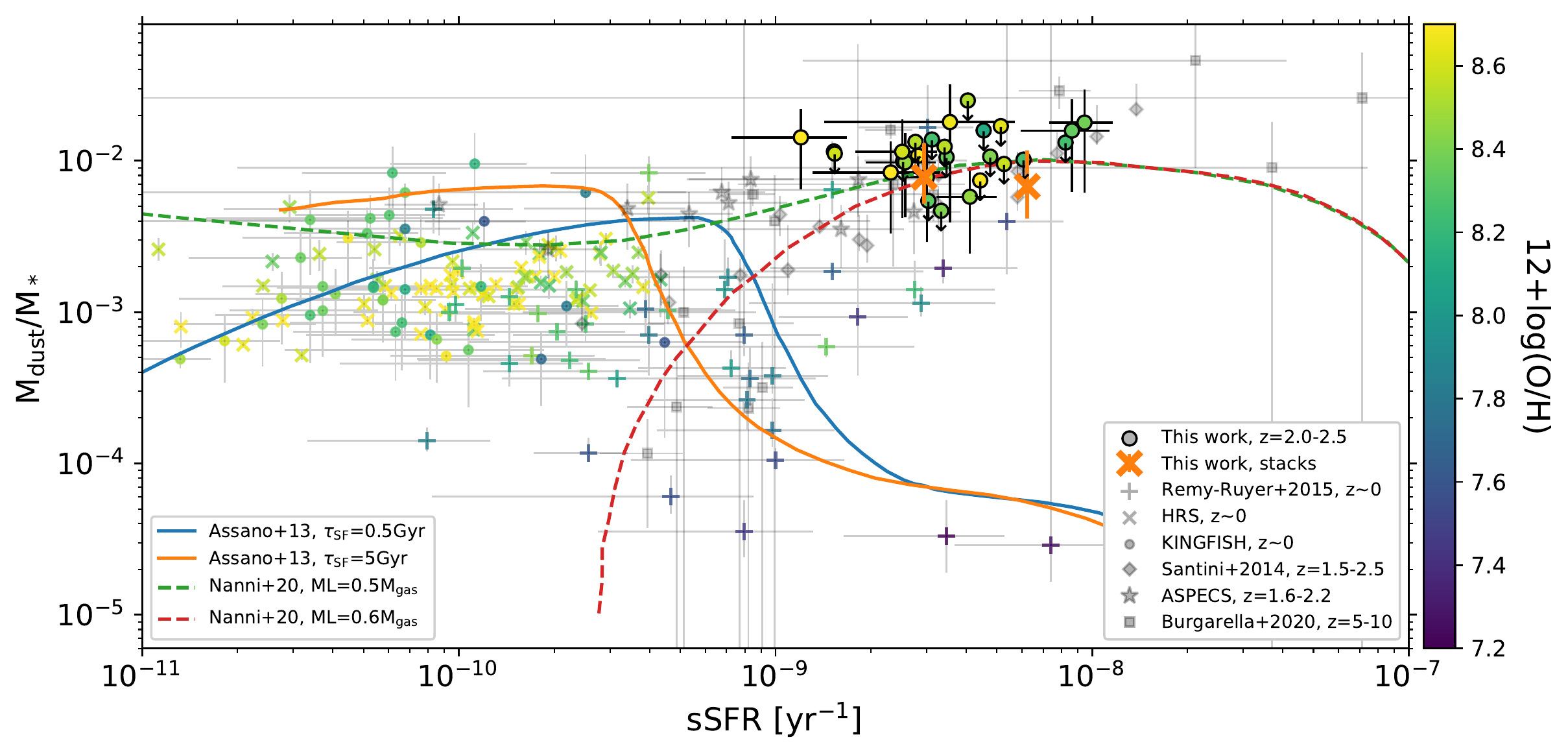}
	\caption{Dust-to-stellar mass ratios as a function of sSFR for our sample (circles with black edges for detections, circles with downward arrows for $2\sigma$ upper limits, and orange crosses for stacks in two bins of sSFR) and other samples in the literature at $z\sim 0-10$. Samples with metallicity measurements (including this work that adopts a B18 scale) are color-coded based on their {\oh}. Those without metallicity measurements are in grey.
	Modelled evolutionary tracks from \citet{asano13} (solid lines) and \citet{nanni20} (dashed and dot-dashed lines) are shown. The \citet{asano13} models are shown for two different star formation timescales ($\tau_{\rm SF}$, defined as ISM mass to SFR). The \citet{nanni20} models are shown for different outflow mass-loading factors (ML) and gas-to-stellar mass ratios, assuming a top-heavy IMF (with $\alpha=1.35$). These models are shown as a few examples from the literature. A detailed analysis of theoretical dust evolution frameworks that fit various observations across redshifts will be addressed in future work.
		}
		\label{fig:sMd_sSFR}
	\vspace{21pt}
\end{figure*}

\capstartfalse   
\begin{deluxetable*}{cccc}[ht]
	\setlength{\tabcolsep} 
	\tabletypesize{\footnotesize} 
	\tablewidth{35pc}
	\tablecaption{Estimates of dust-to-gas mass ratios based on the observed dust masses and metallicities in three bins of metallicity}
	\tablehead{
	\colhead{Parameter} &
	\colhead{Bin 1} &
	\colhead{Bin 2} &
	\colhead{Bin 3} 
	}
	\startdata
    {12+log(O/H)$_{\rm B18}$\footnote{The \hyperlink{B18}{B18} scaling relation (Section~\ref{sec:ONIR_data_metal}).}} & {8.33$\pm 0.03$} & {8.54$\pm 0.02$} & {8.70$\pm 0.02$}\\
    {12+log(O/H)$_{\rm PP04}$\footnote{The \hyperlink{PP04}{PP04} scaling relation (Section~\ref{sec:ONIR_data_metal}).}} & {8.20$\pm 0.02$} & {8.37$\pm 0.02$} & {8.51$\pm 0.01$}\\
    {redshift} & {2.22$\pm 0.05$} & {2.39$\pm 0.04$} & {2.38$\pm 0.06$}\\
    {$\log(M_*/M_{\odot})$} & {9.85$\pm 0.07$} & {10.38$\pm 0.05$} & {10.44$\pm 0.07$}\\
    {SFR [{\msun}\,yr$^{-1}$]\footnote{SFR is derived from {\halpha}, corrected for dust attenuation using Balmer decrement (Section~\ref{sec:ONIR_data_SFR}).}} & {49$\pm 8$} & {71$\pm 11$} & {73$\pm 17$}\\
    {$\log(M_{\rm dust}/M_{\odot})$]} & {7.89$_{-0.16}^{+0.31}$} & {8.16$_{-0.15}^{+0.30}$} & {8.22$_{-0.17}^{+0.32}$} \\
    {$\log(M_{\rm molgas,~T18}/M_{\odot})$\footnote{From the \citet{tacconi18} molecular gas fraction equation (Eq. 6 in that paper), which is a function of redshift, sSFR offset from main-sequence, stellar mass, and optical effective radius. The main-sequence and size evolution relations adopted in the \citet{tacconi18} equation are those from \citet{speagle14} and \citet{vanderwel14}, respectivelly. The errors are propagated measurements errors in SFR, stellar mass, and radius.}} & {10.39$\pm0.04$} & {10.64$\pm0.06$} & {10.67$\pm0.09$}\\
    {$\log(M_{\rm molgas,~L11}/M_{\odot})$\footnote{From the \citet{leroy11} relation: $\log_{10}(M_{\rm dust}$/$M_{\rm gas})=-9.4+0.85\times(12+\log(\rm O/H))$. Errors are propagated from dust mass measurement errors.}} & {10.15$_{-0.16}^{+0.31}$} & {10.25$_{-0.15}^{+0.30}$} & {10.17$_{-0.17}^{+0.32}$}\\
    {$\log(M_{\rm gas,~RR14}/M_{\odot})$\footnote{From the \citet{remyruyer14} relation: $\log_{10}(M_{\rm dust}$/$M_{\rm gas})=-2.21-( x_{\odot}-x)$, where $x$ is 12+log(O/H) and $x_{\odot}=8.69$. Errors are propagated from dust mass measurement errors.}} & {10.41$_{-0.16}^{+0.31}$} & {10.47$_{-0.15}^{+0.30}$} & {10.37$_{-0.17}^{+0.32}$}\\
    {$\log(M_{\rm gas,~dV19}/M_{\odot})$\footnote{From the \citet{devis19} relation: $\log_{10}(M_{\rm dust}$/$M_{\rm gas})=2.15\times(12+\log(\rm O/H))-21.19$. Errors are propagated from dust mass measurement errors.}} & {11.12$_{-0.16}^{+0.31}$} & {10.95$_{-0.15}^{+0.30}$} & {10.65$_{-0.17}^{+0.32}$}\\
    {$\log(M_{\rm gas,~KS}/M_{\odot})$\footnote{HI+H$_2$ gas mass estimate from the Kennicutt-Schmidt star formation law \citep{kennicutt21}: $\log(\Sigma_{\rm SFR})=1.5\log(\Sigma_{M_{\rm HI+H_2}})-3.87$. The errors are propagated measurement errors of $\Sigma_{\rm SFR}$ and radius and the uncertainty on the star formation law coefficient (1.5$\pm 0.05$) as reported in \citet{kennicutt21}.}} & {10.11$\pm0.15$} & {10.39$\pm0.19$} & {10.42$\pm0.28$}\\
	\enddata
	\tablenotetext{}{Reported errors in mass, SFR, redshift, and metallicities are errors of the mean. Systematic uncertainties of the adopted gas mass scaling relations are not included in the reported errors.}

	\label{tab:gasmass}
\end{deluxetable*}
\capstarttrue  

\subsection{Dust Mass} \label{sec:dustmass}

The bulk of dust mass ($M_{\rm dust}$) in galaxies is from the cold dust population that dominantly emit at FIR/submm wavelengths, making the RJ emission a good diagnostic for dust masses.
Following the discussion in \citet{scoville16}, we derive dust masses from the ALMA 1.2\,mm flux densities, assuming an optically-thin MBB, as:
\begin{equation}\label{eq:mdust}
    M_{\rm dust} = \frac{(S_{\nu}/f_{\rm CMB})~D_{\rm L}^2(z)}{B_{\nu}(T)~(1+z)~\kappa_{\nu}(\beta)},
\end{equation}
where $S_{\nu}$ is the 1.2\,mm flux density (in the observed frame), $f_{\rm CMB}$ is the correction factor for the cosmic microwave background (CMB) effect on the background at the redshift of the targets \citep[Equation 18 in][]{dacunha13}\footnote{At the dust temperatures and redshifts of the galaxies under study the additional dust heating by CMB is negligible, but due to the effect of the CMB background at the observed wavelength, the observed flux against CMB is $\sim 80\%$ of the intrinsic flux (see Figure 3 in \citealt{dacunha13}).}, $D_{\rm L}(z)$ is the luminosity distance to redshift $z$, $B_{\nu}(T)$ is the Planck function with dust temperature $T$, $\kappa_{\nu}(\beta)$ is the dust grain absorption cross section per unit mass at frequency $\nu$ with a functional form of $\kappa_0(\frac{\nu}{\nu_0})^{\beta}$, where $\kappa_0$ is the opacity at $\nu_0$ and $\beta$ is the submm emissivity index. The main assumptions that enter this calculation and contribute to the uncertainties on dust masses are the temperature of the cold component and the submm emissivity and opacity parameters. For the cold dust temperature, we perform a MC simulation by drawing dust temperatures from a Gaussian distribution with a mean of 25\,K and $\sigma=5$\,K. We assume an emissivity index of $\beta=1.5$, as our subsolar-metallicity data suggest a $\beta$ shallower than 2 (wider IR peak), in agreement with previous studies (see Appendix~\ref{app:dmass}). Based on the choice of $\beta$, an opacity of $\kappa_0=0.232$\,m$^2$\,kg$^{-1}$ at 250\,{\um} is adopted \citep{draine03,bianchi13}. The $\kappa_0$ value is the most uncertain factor in dust mass estimations. Different assumptions of $\kappa_0$ from the literature affect dust mass estimations by up to a factor of 3. In Appendix~\ref{app:dmass}, we discuss our assumptions on dust temperature, $\beta$, and $\kappa_0$ and their systematic uncertainties in detail.

\subsubsection{Dust mass evolution} \label{sec:dust_mass_evolution}
Figure~\ref{fig:Md_Ms} shows the individual dust mass measurements and the stacked values in bins of stellar mass and metallicity of the galaxies in this work, compared with other samples at $z\sim 0$ to 2 (for the description of the comparison samples refer to Section~\ref{sec:comparison_samples}). 
The $z\sim 2$ samples show consistent dust masses where they overlap in stellar mass.
At a given stellar mass, dust mass increases from $z\sim 0$ to 2 by about a factor of 10.
An increase in dust mass at a given stellar mass has been previously seen at higher stellar masses \citep{santini14,kirkpatrick17}, and is expected owing to the increase in gas-to-stellar mass ratio from $z\sim 0$ to 3 \citep{schinnerer16,tacconi18,decarli20}. We also see higher dust masses at a given metallicity at $z\sim 2.3$ compared to $z\sim 0$ in the right panel of Figure~\ref{fig:Md_Ms}. If the PP04 metallicity scale is used instead of the B18 scale for the $z\sim 2$ sample, the redshift evolution of dust mass at a fixed O/H would be even larger. Therefore, the systematic uncertainties associated with the metallicity estimates at high redshifts are unlikely to artificially induce the observed trend in dust mass evolution.

To better understand the redshift evolution of dust mass fractions, we plot dust-to-stellar mass ratio (D/$M_*$) and dust mass to SFR ratio (D/SFR) as a function of metallicity in Figure~\ref{fig:Ms_Md_SFR_Z}. As also seen in the $z\sim 0$ sample \citep{remyruyer15} and the simulations \citep{popping17}, there is no statistically strong correlation between the D/$M_*$ and metallicity at $z\sim 2$. However, there is about an order of magnitude increase in D/$M_*$ from $z\sim 0$ to 2 across all metallicities that are covered in this work. In Section~\ref{sec:dustmass_gas}, we estimate gas masses from star-formation law scaling relations to compare with our dust mass measurements and discuss the possible explanation and implications of the dust mass evolution.

The galaxy formation semi-analytic models of \citet{popping17} at $z=0$ and 2 are shown in Figure~\ref{fig:Md_Ms}. These models predict less dust mass at $z\sim 2$ compared to the observations, particularly at lower stellar mass and metallicities. This discrepancy originates from an underproduction of molecular gas masses of the $z\sim 2$ main-sequence galaxies in the models compared to observations \citep{somerville15,popping19}. Given that in the \citet{popping17} model the dust growth mechanism at these metallicities is dominated by the accretion of metals onto grains in the dense ISM, an underestimation of molecular gas mass compared to the observations also results in a lower than observed dust masses. The discrepancy is most pronounced at lower metallicities in the right panel of Figure~\ref{fig:Md_Ms}, which stems from a known outcome of the model that dust-to-metal ratios are lower than the observations at {\oh}$\lesssim 8.5$ (Popping \& P\'eroux, in prep). 

In Figure~\ref{fig:Ms_Md_SFR_Z}, the \citet{popping17} semi-analytic models are in good agreement with D/SFR-metallicity relation at $z\sim2$, even though they underestimate dust masses at lower metallicities in Figure~\ref{fig:Md_Ms}. In these models, the main mechanism of dust formation at these metallicities is through dust growth in the ISM, which is a strong function of molecular hydrogen surface density and metallicity. On the other hand, SFR is also regulated by $H_2$ mass through empirical relations. Therefore, although the $H_2$ gas mass predicted by these models is less than that observed at $z\sim 2$, the ratio of dust mass to SFR stays in agreement with observations. In other words, in the models, dust to molecular gas mass and SFR to molecular gas mass relations with metallicity are in agreement with observations, yet dust mass and SFR are lower than those observed at these metallicities.

An increase in dust masses from $z=0$ to 2 at a given stellar mass is in tension with the apparent lack of redshift evolution in the obscured SFR fraction and IR to UV luminosity ratio (IRX) or UV obscuration (A$_{1600}$) at a fixed stellar mass over the same redshift range \citep[][but \citealt{shivaei20b} found an evolution in IRX at a given UV slope with mass and metallicity, see below]{bouwens16c,whitaker17,reddy18a,shapley21}. 
One possible explanation is a change in the dust-star geometry of main-sequence galaxies at $z\sim 2$ compared to the local galaxies. The bulk of dust mass is determined by the cold dust population, while the IR emission is dominated by the emission from warmer dust grains in the actively star-forming regions. A two disjoint dust component geometry, where one component is the hot dust in the birth clouds of recently formed stars and the other is the colder dust that resides in the diffuse ISM \citep{charlotfall00}, would reconcile the high D/$M_*$ of $z\sim 2$ galaxies with the lack of evolution in their obscured SFR fraction (and IRX) vs stellar mass relation. There is observational evidence for the emergence of such a two-component dust geometry in subsolar metallicity galaxies at $z\sim 2$ from a comparison of the dust reddening of ionized nebular gas and that of stellar continuum \citep[Figure 8 in][]{shivaei20a}. A higher average nebular dust reddening compared to the stellar reddening for subsolar metallicity $z\sim 2$ galaxies suggests two different dust populations affect the emission from massive young stars and older stars. 
Spatially resolved observations (on the scales of a few tens of pc) are required to verify this physical picture.
Different dust characteristics (composition and size) with different radiation efficiencies may also play a role. For example,  \citet{shivaei20b} showed that at a given UV continuum slope $\beta$ (i.e., stellar continuum reddening, not to be confused with the $\beta$ that characterizes submm dust emissivity), the lower metallicity galaxies have lower IRX compared to the higher metallicity ones, which may indicate different dust characteristics in low-metallicity galaxies. Other studies have shown similar results that younger and lower mass galaxies at $z\sim 2$ have a lower IRX than high-mass galaxies at a fixed UV slope \citep[e.g.,][]{reddy12a,reddy18a,fudamoto19}.
In this picture, when comparing the $z\sim 0$ and 2 galaxies at a given stellar mass, the $z\sim 2$ galaxies have lower metallicities, and although they have higher dust masses, dominated by large grains emitting at FIR/submm wavelengths, their light-weighted IR luminosity is not proportionally higher as it is dominated by the re-emitted light from the smaller grains whose characteristics have changed.
Another important factor that might have contributed to the lack of a redshift evolution in the IRX (or obscured SFR)-stellar mass relations in the literature is the uncertainties in measuring L(IR) and obscured SFR of low-mass/low-metallicity galaxies at $z\sim 2$. For example, in the presence of a significant warm dust component, a cold-dust IR template tends to underestimate the IR luminosity (as discussed in Section~\ref{sec:implications}). On the other hand, obscured SFRs derived from PAH emission could be underestimated if the reduced intensity of PAH emission relative to L(IR) at low metallicities is not considered \citep[Figure 7 in][]{shivaei17}. 

\subsubsection{Gas Mass Estimates} \label{sec:dustmass_gas}

\paragraph{Gas masses from star-formation law scaling relations}
We first estimate gas masses from the scaling relations of \citet{tacconi18}, independent of our dust mass measurements (Table~\ref{tab:gasmass}).
\citet{tacconi18} used a combination of CO observations, dust SEDs, and MBB fits to the RJ tail to estimate molecular gas masses. The gas masses of galaxies with $M_*<10^{10.5}$\,{\msun} at $z>1$ in that work are dominantly from either of the latter two methods (no CO observations). For those galaxies, the molecular gas mass is derived from the dust mass, assuming a linear relation between dust to molecular gas ratio and metallicity\footnote{The \citet{tacconi18} molecular gas mass estimates at low stellar masses is similar to that in \citealt{leroy11}, and perhaps the reason for the better agreement between the two dust to molecular gas ratio estimates at lower metallicities in Table~\ref{tab:gasmass}.}, where metallicities are estimated from a mass-metallicity scaling relation. The \citet{tacconi18} scaling relations at higher metallicities (higher stellar masses) are based on direct CO observations and not dust mass estimates from IR/submm data. These estimates are subject to uncertainties in CO-to-H$_2$ conversions \citep{bolatto13}. A more detailed discussion on the CO conversion uncertainties is beyond the scope of this work and will be addressed in future work where direct CO observations are available.

D/$M_*$ is proportional to dust-to-gas mass ratio (D/G) multiplied by gas-to-stellar mass ratio (G/$M_*$). 
Using the scaling relations of \citet{tacconi18} between molecular gas mass, $M_*$, and SFR over a redshift range of $z=0-4$, we estimate the molecular gas masses of our galaxies based on their redshift, offset from the main sequence\footnote{Here we adopt the \citet{speagle14} main-sequence relation to be consistent with that assumed in \citet{tacconi18}.}, and stellar mass. 
For consistency, we also calculate molecular gas masses in the same manner for a subset of the DGS, KINGFISH, and HRS samples that have similar average metallicities as our three metallicity bins ({\oh}$\sim 8.3$, 8.5, and 8.7, for the DGS, KINGFISH, and HRS subsamples, respectively).

Assuming no redshift evolution in D/G at a given metallicity between $z\sim 2.3$ to 0 (shown at solar metallicities in \citealt{shapley20} and Popping et al., in prep) and using the estimated molecular G/$M_*$ ratios from the \citet{tacconi18} relations as described, D/$M_*$ is expected to be $\sim 20-50$ times higher in the $z\sim 2.3$ sample compared to the $z\sim 0$ comparison samples.
However, our observations show a factor of $\sim 2-10$ smaller change in D/$M_*$ at a given metallicity. This implies that dust to molecular gas mass is lower at $z\sim 2.3$ compared to that at $z\sim 0$ by a factor of $\sim 2-10$. 

A drastic change in D/G is anticipated for galaxies with different star formation efficiencies (SFE; SFR per unit molecular gas mass), when the metallicities are close to a ``critical'' metallicity defined by \citet{asano13}. Critical metallicity is the metallicity at which dust mass growth in the ISM becomes equal to dust production from stars (AGB stars and SNe).
According to the theoretical dust evolution models of \citet{asano13}, the critical metallicity itself depends on SFE, and becomes larger with shorter star formation timescales ($\tau_{\rm SF}$, defined as gas mass to SFR). Based on their simulations, at {\oh}$=8.4$, the DtG of a model with $\tau_{\rm SF}=0.5$\,Gyr is 13 times lower than that of a model with $\tau_{\rm SF}=5$\,Gyr. 
In the bottom panel of Figure~\ref{fig:Ms_Md_SFR_Z}, the clear trend of decreasing D/SFR at a given metallicity with increasing SFR offset from the main-sequence (SFR/SFR$_{\rm MS}$; color-coding in Figure~\ref{fig:Ms_Md_SFR_Z}, bottom panel) is a result of increasing SFE -- both with increasing offset from the main sequence and with increasing redshift.
In summary, the higher D/$M_*$ at $z\sim 2.3$ compared to $z\sim 0$ at a given metallicity in Figure~\ref{fig:Ms_Md_SFR_Z} can be explained by the higher molecular gas fractions along with the lower D/G (due to the higher SFEs) of the $z\sim 2.3$ galaxies.

As a comparison, Table~\ref{tab:gasmass} also shows gas masses derived from the Schmidt-Kennicutt star formation law adopted from \citet{kennicutt21}. The single power-law relation in \citet{kennicutt21} (with slope $n=1.5$) represents the overall relation for the nearby starbursts and nonstarbursting disk galaxies. However, the resolved Schmidt-Kennicutt relation has not been tested for main-sequence galaxies at higher redshifts.
There are resolved CO and dust observations of $z\sim 1-3$ galaxies that indicate a different spatial distribution of molecular gas and dust compared to the stellar emission \citep[e.g.,][]{calistro-rivera18,kaasinen20}. These observations are typically limited to more massive and/or more actively star-forming galaxies than those in this work. At last, whether the F160W (rest-frame optical) sizes that we adopt here are representative of where recent star formation activity occurs, is another source of uncertainty in the gas masses derived from the Schmidt-Kennicutt relation in Table~\ref{tab:gasmass}. Resolved CO and dust observations of larger samples of main-sequence galaxies at $z>1$ would help to shed light on these uncertainties.

\paragraph{Gas masses from D/G-metallicity relations}
Using the metallicity of our galaxies, we also calculate the expected D/G ratios based on the D/G-metallicity relations of \citet{leroy11}, \citet{remyruyer14}, and \citet{devis19} in Table~\ref{tab:gasmass}: the \citet{leroy11} relation between molecular gas and dust mass is derived based on the resolved CO, H{\sc i}, and IR maps of five nearby galaxies that span an order of magnitude in metallicity from {\oh}$=8.0-9.0$. 
\citet{remyruyer14} used integrated CO, H{\sc i}, and IR data for 126 local galaxies from the DGS, KINGFISH, and a subsample of the sample presented in \citet{galametz11}, covering two orders of magnitude in metallicity. The work of \citet{devis19} is based on 466 local galaxies from the DustPedia project\footnote{\href{http://dustpedia.astro.noa.gr/}{http://dustpedia.astro.noa.gr/}} with metallicities from optical lines, H{\sc i} observations (H$_2$ is calculated from a scaling relation between the H$_2$-to-H{\sc i} ratio and the H{\sc i}-to-stellar mass ratio), and IR data. 

In Table~\ref{tab:gasmass}, the \citet{tacconi18} and \citet{leroy11} estimates are for molecular gas fractions, while the \citet{remyruyer14} and \citet{devis19} relations include atomic gas fractions as well. While the molecular gas is expected to dominate the total gas fraction in the more massive and metal-rich galaxies at $z\sim 2$ \citep{tacconi18}, the atomic-to-molecular gas fraction increases with decreasing metallicity \citep{remyruyer14}. This may explain the large difference between the \citet{devis19} predictions and the \citet{tacconi18} and \citet{leroy11} values at low metallicities, although the latter two are consistent with the estimates of \citet{remyruyer14}. Such discrepancies underscores the importance of CO observations of subsolar metallicity galaxies at $z>1$ to robustly constrain their dust to molecular gas ratios, even though the estimates will be subjected to the CO-to-H$_2$ conversion uncertainties \citep{bolatto13}.

\subsubsection{D/$M_*$ versus sSFR (dust formation rate diagram)}

Several studies have attempted to model D/$M_*$ evolution in a galaxy  \citep[e.g.,][]{asano13,nanni20} and its evolution with redshift \citep[e.g.,][]{calura17,popping17}.
To track the time evolution of dust within a galaxy, we look at the D/$M_*$ versus sSFR diagram (or dust formation rate diagram) as shown in Figure~\ref{fig:sMd_sSFR}. We show dust evolutionary tracks from two different models, those of \citet{asano13} and \citet{nanni20}. The models assume various dust production (stellar origin and dust mass growth in the ISM) and destruction (SN shocks) channels. In the \citet{asano13} models, dust production at low metallicity and high sSFR is dominated by stellar sources. Once the galaxy reaches its critical metallicity in the ISM, dust mass growth by metal accretion on the dust grains in the ISM dominates the dust production. At that point, dust mass increases more rapidly than the rate of star formation (hence the steep rise in the DtS). After that, when the majority of metals is locked up in grains, dust mass growth saturates but the stellar mass build up continues, resulting in a decrease in sSFR and DtS. 

Figure~\ref{fig:Ms_Md_SFR_Z} shows that our data, with dust and metallicity measurements, occupies a distinct parameter space with higher sSFR and higher DtS compared to the majority of the $z\sim 0$ population. The $z\sim 2$ data of \citet{santini14} and the ASPECS program overlap with our measurements (but without metallicity constraints).
One can argue that an \citet{asano13} model with a lower depletion time than that assumed in their work ($\tau=$500\,Myr shown by the blue solid curve is the model with the shortest timescale in that work) may cover the $z\sim 2$ parameter space. However, typical depletion timescales for these galaxies are not expected to be much shorter than a couple of 100\,Myr \citep[e.g.,][]{tacconi18,boogaard19}.

The $z\sim 2.3$ data can also be explained by the \citet{nanni20} models that have outflows and a higher SN condensation efficiency compared to the \citet{asano13} models. \citet{nanni20} assumed a top-heavy IMF as their fiducial model. A typical IMF (e.g., a Chabrier IMF) would shift the curves downward in this diagram. Dissecting the exact theoretical dust evolutionary model that best describes our dataset is beyond the scope of this paper. However, it is clear that given the large number of uncertain parameters in the theoretical models, the combination of dust mass, metallicity, and sSFR at $z>0$ opens a new parameter space to constrain the theoretical models with observational evidence. 

\section{Summary} \label{sec:summary}
In this work, we present band-6 (1.2\,mm) ALMA continuum observations of a sample of 27 star-forming galaxies at $z=2.1-2.5$, with robust metallicity and SFR measurements from rest-frame optical emission lines ({\hbeta}, [O{\sc iii]}$\lambda5008$, {\halpha}, [N{\sc ii}]$\lambda6585$), carefully selected to represent a wide range in metallicity, {\oh}$_{\rm B18}$\,$\sim 8.1-8.8$, and stellar mass, $\log(M_*/M_{\odot})\sim 9.7-10.6$. The galaxies are selected from the rest-frame optical spectroscopic MOSDEF survey and are located in the COSMOS field with a wealth of photometric data covering rest-frame UV to FIR.

Using the Spitzer, Herschel, and ALMA data, we constrain the IR SED and dust masses of the $z\sim 2.3$ subsolar metallicity galaxies and compare them with those of local galaxies spanning a wide range in galaxy populations from low-metallicity dwarfs to starbursts. The low metallicity and high sSFR of the $z\sim 2.3$ galaxies in this work make them ideal analogs for typical star-forming galaxies in the early Universe. The main conclusions of the paper are as follows.

We find that the average IR SED of the $z\sim 2.3$ subsolar metallicity galaxies ($\langle 12+\log({\rm O/H)_{B18}}\rangle = 8.4$) has an additional warm dust component (with peak temperature of about $100$\,K) that broadens the IR SED, a behavior similar to that of the local dwarf galaxies but different from the commonly adopted LIRG and ULIRG templates (Section~\ref{sec:irfits} and Figure~\ref{fig:lowZ}). The width of the IR SED of the $z\sim 2.3$ solar-metallicity galaxies is similar to that of the local LIRGs, as expected (Section~\ref{sec:irfits} and Figure~\ref{fig:highZ}). The IR SED also becomes broader and warmer with increasing sSFR, and increasing SFR surface density (Section~\ref{sec:physical_interp} and Figure~\ref{fig:2comp}). However, the broadening effect is the most significant when the sample is divided by gas metallicity, such that the $\sim 30-80$\,{\um} IR SED changes from a dominant single warm MBB component with $T\sim 50$\,K at solar metallicity to two distinct components with $T\sim 25$ and $130$\,K at subsolar metallicity (Section~\ref{sec:warmdust} and Figure~\ref{fig:2comp}).

The warm and broad IR SED of the $z\sim 2.3$ subsolar metallicity galaxies can be attributed to a more transparent ISM as well as a harder ionizing radiation that increase the dust temperature to higher values and in larger volumes, similar to that of the local dwarf galaxies with $\sim 0.1-0.6$\,dex lower oxygen abundances (O/H). It is also partly due to the higher star formation surface density and sSFR of the $z\sim 2.3$ galaxies compared to the $z\sim 0$ ones at a given stellar mass, which result in a wider range of interstellar radiation field intensities that dust grains are exposed to, and a higher relative abundance of hot grains in the vicinity of H{\sc ii} regions, emitting at shorter wavelengths.A more clumpy dust geometry of the low metallicity galaxies may also contribute to the broadening effect (Section~\ref{sec:physical_interp}).

The presence of a significant warm dust component in the subsolar metallicity and high sSFR galaxies has important implications for deriving IR luminosity and obscured SFR of high-redshift galaxies, as well as for predicting submm fluxes with limited optical to mid-IR data. We show that the locally calibrated LIRG templates overestimate submm ALMA fluxes of $z\sim 2.3$ LIRGs by a factor of $\sim 1.5$ up to an order of magnitude, if anchored to the PAH emission and/or mid-IR continuum emission (rest-frame $\sim 30$\,{\um}; Figures~\ref{fig:ircolor} and \ref{fig:only24a100}).
The overestimation increases with increasing sSFR. 
Adopting a broader SED with a prominent warm dust component, similar to those of local dwarf galaxies, alleviate the overprediction of the submm fluxes. 
On the other hand, the IR luminosity can be underestimated by $>0.4$\,dex if the cold LIRG templates are used and scaled to a single submm ALMA continuum flux. This is often the case at $z>3$, where only a single FIR datapoint is available to estimate IR luminosities. As such galaxies have relatively lower metallicities and are often selected to have vigorous star formations, we recommend to adopt warmer and broader templates to mitigate the underestimation of IR luminosity (Section~\ref{sec:submmestimation}). The average local low-metallicity template constructed in this work (Section~\ref{sec:irfits_template}) is publicly available\footnote{\href{http://www.ireneshivaei.com/shivaei22.html}{http://www.ireneshivaei.com/shivaei22.html}}.

Using the best-fit IR models with the ALMA constraint, we estimate total L(IR) and obscured SFR for the sample. Comparing the obscured SFR to total SFR derived from optical emission lines, we find that the obscured SFR of the subsolar metallicity galaxies with average stellar mass of $\log({\rm M/M_{\odot}})=10.18$ is $\sim 60\%$ of the total SFR. This is 1.5 times lower than the average value indicated in previous studies for the general population of galaxies at this stellar mass (Section~\ref{sec:obssfr}). 

We find that dust masses (derived from ALMA 1.2\,mm data) are about an order of magnitude higher at $z\sim 2.3$ compared to $z\sim 0$ at a given stellar mass and metallicity. 
An order of magnitude higher dust-to-stellar mass implies an order of magnitude higher gas-to-stellar mass fractions at $z\sim 2.3$ compared to $z\sim 0$, if dust-to-gas mass ratio stays constant. However, CO-based studies \citep{tacconi18} predict a larger dust-to-stellar mass evolution for the star-forming galaxies in this work, which suggests that dust in these galaxies is depleted (i.e., lower dust to molecular gas mass ratios) by a factor of $> 2$. A lower dust-to-gas ratio can be a result of a high star-formation efficiency in the galaxies in this sample (Section~\ref{sec:dustmass}). 

The higher dust-to-stellar mass ratios of $z\sim 2.3$ galaxies compared to $z\sim 0$ is potentially in tension with the apparent lack of redshift evolution in the obscured luminosity (or SFR) fraction at a given stellar mass. Given that the bulk of dust mass is from the cold dust population, while the IR emission is dominated by the emission from warmer dust in the vicinity of actively star-forming regions, this finding hints at different dust-star geometry in $z\sim 2.3$ main-sequence galaxies compared to that of the local galaxies (Section~\ref{sec:dustmass}). 

Many of the existing empirical IR SED templates and gas mass calibrations at $z>1$ are based on high mass, high metallicity galaxies. This study pushes into the important territory of lower metallicities and finds key differences in the dust properties of subsolar metallicity galaxies at $z\sim 2.3$ compared to their more metal rich counterparts. These differences are crucial to be considered for future observations with IR facilities and motivate further detailed studies of gas and dust in the lower metallicity regime at high redshifts.
For instance, better sampling of the FIR/submm emission of galaxies with ALMA and LMT/TolTEC, over a larger wavelength range, would help to set observational constraints on the cold dust temperature and dust emissivity index of subsolar metallicity galaxies, which are not well constrained by the single ALMA band observations presented in this work. Moreover, the relative contribution of PAHs and very small grains to the mid-IR spectra remains unconstrained with only a single photometric data point from Spitzer/MIPS 24\,{\um}. Future studies with JWST/MIRI will be able to break this degeneracy owing to the multi-band imaging and spectroscopy capabilities of MIRI. Additionally, future and ongoing CO observations of subsolar metallicity galaxies at $z\sim 2$ are crucial to better constrain their molecular gas mass fractions and the redshift evolution of dust-to-gas mass ratios as a function of metallicity, providing key constraints for theoretical galaxy evolution models.

\acknowledgments
We thank the referee for a very constructive review.
We thank Luke Maud from the European ALMA Regional Center at the European Southern Observatory (ESO) for valuable discussions and his help with running the ALMA calibration and imaging pipeline. We thank Loreto Munoz and Ryan Loomis at the North American Alma Science Center for valuable input on image construction and flux measurements of the ALMA data. We also thank Tanio D{\'\i}az-Santos for providing the GOALS survey data catalogs.
Support for this work was provided by NASA through the NASA Hubble Fellowship grant \#HST-HF2-51420, awarded by the Space Telescope Science Institute, which is operated by the Association of Universities for Research in Astronomy, Inc., for NASA, under contract NAS5-26555,  to IS.
This paper makes use of the following ALMA data: ADS/JAO.ALMA\#2019.1.01142.S. ALMA is a partnership of ESO (representing its member states), NSF (USA) and NINS (Japan), together with NRC (Canada), MOST and ASIAA (Taiwan), and KASI (Republic of Korea), in cooperation with the Republic of Chile. The Joint ALMA Observatory is operated by ESO, AUI/NRAO and NAOJ.
The National Radio Astronomy Observatory is a facility of the National Science Foundation operated under cooperative agreement by Associated Universities, Inc.
PAO acknowledges support from the Swiss National Science Foundation through the SNSF Professorship grant 190079.
The Cosmic Dawn Center (DAWN) is funded by the Danish National Research Foundation under grant No.\ 140.
YF further acknowledges support from NAOJ ALMA Scientific Research Grant number 2020-16B
Funding for the MOSDEF survey was provided by NSF AAG grants AST-1312780, 1312547, 1312764, and 1313171 and archival grant AR-13907, provided by NASA through a grant from the Space Telescope Science Institute.
The data presented herein were obtained at the W. M. Keck Observatory, which is operated as a scientific partnership among the California Institute of Technology, the University of California and the National Aeronautics and Space Administration. The Observatory was made possible by the generous financial support of the W. M. Keck Foundation.
We are grateful to the MOSFIRE instrument team for building this powerful instrument, and to Marc Kassis at the Keck Observatory for his many valuable contributions to the execution of the MOSDEF survey. 
The authors wish to recognize and acknowledge the very significant cultural role and reverence that the summit of Maunakea has always had within the indigenous Hawaiian community. We are most fortunate to have the opportunity to conduct observations from this mountain.

\appendix
\begin{appendices}

\section{A. IR Fitting Methodologies}\label{app} \label{sec:fittingmethods}

In this appendix we describe the IR templates that are used to fit the 24-to-1200\,{\um} data by weighted least squares technique, where the weights are the inverse of the variances of the measurements. These templates are chosen to represent those commonly used in the literature.

\subsection{A1. The R09 IR templates} \label{sec:r09}
The \citet[][R09]{rieke09} SEDs are average templates for purely star-forming galaxies with L(IR) between $5\times 10^{9}$ to $10^{13}\,L_{\odot}$. These empirical templates are constructed based on the IR emission of local star-forming galaxies, LIRGs and ULIRGs.
On average, from lower to higher luminosity templates, the mid-IR aromatic emission becomes stronger and the far-IR peak wavelength (proportional to luminosity-weighted dust temperature) shifts to shorter values (higher temperatures). 
It has been previously shown that the rescaled \hyperlink{R09}{R09} L(IR)$=10^{11.00}-10^{11.75}\,L_{\odot}$ templates describe well the IR emission of ULIRGs at $z\sim 1-4$ \citep[e.g.,][]{rujopakarn13,derossi18,shivaei20b}. 
Following the results of these studies, we use the \hyperlink{R09}{R09} $L({\rm IR})=10^{11}-10^{12}\,L_{\odot}$ templates in this work. 

\subsection{A2. The K15 2-temperature MBB fitting} \label{sec:k15}
The two-temperature modified blackbody (2T-MBB) fitting of \citet[][K15]{kirkpatrick15} assumes two 
MBB components with different temperatures for the far-IR ($>20$\,{\um}) emission. We do not use these models to fit the mid-IR emission. This simplified model is useful for quantifying the relative temperature and strength of the warm and cold dust components. However, in case of limited data such as ours, a simplified 2T MBB approach would give us insight on the physical parameters governing the far-IR emission, i.e., the properties of the average cold and warm dust components. The 2T MBB has the form of
\begin{eqnarray}
S_{\nu}(\nu) = a_{\rm w} ~ \nu^{\beta} ~ B_{\nu}({\nu},T_{\rm w}) + a_{\rm c} ~ \nu^{\beta} ~ B_{\nu}({\nu},T_{\rm c}),
\end{eqnarray}
where $\beta$ is emissivity index and $B_{\nu}({\nu})$ is the Planck function, 
\begin{equation}
B_{\nu}({\nu}) = \frac{2h\nu^3}{c^2} ~ \frac{1}{e^{\frac{h\nu}{kT_{\rm w}}}-1}.
\end{equation}
$T$ is the temperature in K, $a$ is the normalization, and the w and c subscripts indicate the warm and cold components, respectively. 

We fix $\beta$ and $T_c$ to simplify the model, as there is only a single ALMA point which is insufficient to constrain these two parameters. 
We set $\beta$ to 1.5 as suggested by {\K15}.
There is evidence that the temperature of the cold dust component that determines the RJ emission tail is almost constant in different types of galaxies and redshifts. \citet{remyruyer15} calculated median dust temperatures of 25.9\,K for the local low-metallicity galaxies of the Dwarf Galaxy Survey \citep[DGS;][]{madden13}, while they found evidence for an additional hotter dust component with average temperature of 105\,K for a fraction of the DGS galaxies. They found a median dust temperature of 25.7\,K for the local higher metallicity sample of KINGFISH, without any additional warm component.
{\K15} also showed a cold dust $T=25.8-28.1$\,K component in their star-forming dominated templates at $z\sim 1$. 
Based on these findings, we set the cold dust temperature to $T_c=25$\,K in our fits. As our goal in using the {\K15} 2T MBB fitting procedure is to quantify differences in the shape of the IR SED between the low and high metallicity bins, and not to derive a template, adopting a different $T_c$ does not change our conclusions. For example, adopting a $T_{\rm c}=30$\,K increases the best-fit $T_{\rm w}$ by $\sim 5-7$\,K, keeping the qualitative conclusions intact.

\subsection{A3. The L16 low-metallicity templates} \label{sec:l16}
To complete the set of IR templates, we also use the IR SED library of 19 local low-metallicity galaxies from \citet[][L16]{lyu16}. The data are from the DGS survey \citep{madden13}. \hyperlink{L16}{L16} selected 19 DGS galaxies that had high enough SNR in Spitzer/IRS spectra and constructed the templates by combining the IRS spectra with Herschel and WISE photometry (Table~7 in \hyperlink{L16}{L16}). The main difference between their SEDs and those presented in \citet{remyruyer15} is that \hyperlink{L16}{L16} adopted the powerlaw$+$MBB fitting procedure of \hyperlink{C12}{C12}, while \citet{remyruyer15} used the semi-empirical dust SED models of \citet{galliano11b}. The advantage of the \hyperlink{C12}{C12} model in this case is that, owing to the well-sampled mid-IR emission of these 19 galaxies, \hyperlink{L16}{L16} were able to relax the bounding condition of the powerlaw turnover wavelength, $\lambda_c$ (see the discussion in Section~\ref{sec:irfits_template}). As a result, they were able to successfully capture the flux density excess at $\lambda\lesssim 50$\,{\um} with their fits, while \citet{remyruyer15} had to invoke a second MBB with a hotter temperature to fit the flux excess.

\section{B. Derivation of Dust Masses} \label{app:dmass}
To derive dust masses from RJ emission based on Equation~\ref{eq:mdust}, we need to assume a cold dust temperature, the opacity at a reference wavelength $\kappa_0$, and a submm emissivity index $\beta$. Here, we discuss each of these assumptions in our calculations and their associated systematic uncertainties.

\paragraph{Cold dust temperature} For the cold dust temperature, we perform a MC simulation by drawing dust temperatures from a Gaussian distribution with a mean of 25\,K and $\sigma=5$\,K. This assumption takes into account the scatter in mass-weighted dust temperatures based on the results of previous studies with better submm wavelength coverage both at $z\sim 0$ down to low metallicities and for massive galaxies at higher redshifts \citep{cortese14,kirkpatrick15,remyruyer15,scoville16,faisst17}.
These studies show that the mass-weighted temperature varies less than the luminosity-weighted dust temperature derived from shorter wavelength data. As the mass-weighted dust temperature is the relevant quantity to derive dust masses from RJ emission, we use $T=25\pm 5$\,K in Equation~\ref{eq:mdust} and do not consider the hotter dust component that appears in Figure~\ref{fig:2comp}. Dust mass inferred from the RJ emission is only linearly dependent on dust temperature ($M_{\rm{dust}}\propto 1/T$), and hence, the relative contribution of a possible warm component dominating the FIR luminosity would not be significant. For example, in the top-right panel of Figure~\ref{fig:2comp}, in which the warm component has the most significant flux contribution to the RJ emission compared to the other bins, if we assume half of the 1.2\,mm flux originates from the warm component with $T=54$\,K, the dust mass would be lower by only 35\%, which does not change any of the main conclusions in this paper.

\paragraph{Emissivity Index $\beta$} For $\beta$, we take the average of our best fits to the solar and subsolar metallicity stacks from the earlier sections in the paper, $\beta=1.5$. An emissivity index of $1.5-2.0$ has been previously observed in moderately lower metallicity local galaxies, for example in the outer disk of M33 \citep{tabatabaei14}, in the Magellanic Clouds \citep{planck11}, in local galaxies \citep{cortese14,faisst17}, and in a few higher redshift galaxies \citep{faisst20}.
Our fits to the subsolar-metallicity stacks strongly prefer an even lower $\beta=1$. The emissivity index derived from SED fitting is in fact an effective emissivity index, which can be different from the intrinsic emissivity of the cold grains \citep{dunne00,galliano11b}. 
If there is a distribution of temperatures (which is realistically always the case for integrated observations of high-redshift galaxies), the SED will be broadened and the effective $\beta$ will be lower. In our case of subsolar metallicity observations, because of the presence of a significant warm component, if we use the flexible models of \citet{casey12}, the best-fit model with the  least $\chi^2$ is the one with lowest allowed $\beta$, as it provides the widest possible SED to account for the excess emission at $\sim 30$\,{\um}. The best-fit $\beta$ value in the \citet{casey12} models also depends on the pivot wavelength assumption ($\lambda_{\rm c}$), which determines the transition point between the MBB and power-law components. $\lambda_{\rm c}$ is fixed in our fits based on the suggested value in \citet{casey12} that is derived from a sample of GOALS galaxies. Therefore, the validity of this assumption for our galaxies is in question, which in turn affects the best-fit $\beta$. These degeneracies can only be diminished by a better sampling of the IR SED. In short, the best-fit emissivity index of 1 in our subsolar-metallicity fits is a lower limit on the intrinsic emissivity, which is likely lower than that of the solar-metallicity galaxies. Although it is common practice to assume the Galactic emissivity index of $\beta=2$, it is shown in the resolved studies of LMC that the intrinsic $\beta$ can be lower than 2 \citep{galliano11b}.

\paragraph{Opacity Coefficient $\kappa_0$} The absorption cross-section per dust mass or dust mass opacity coefficient is the main source of uncertainty in dust mass estimates from FIR/submm wavelengths. The absorption cross-section is a function of wavelength in the form of $\kappa = \kappa_0 (\frac{\lambda}{\lambda_0})^{-\beta}$, where $\kappa_0$ is the normalization at a reference wavelength $\lambda_0$. The reference normalization value is notoriously uncertain. It depends on the grain composition, size distribution, and many other factors that make the results of models that are even based on the same dataset significantly different from each other. As a result, dust masses derived from different models can vary by up to a factor of $\sim 3$. In addition to the $\beta$ dependence in the functional form of $\kappa$, $\kappa_0$ itself depends on $\beta$ \citep{galliano11b, bianchi13}. Grains with smaller emissivity index have larger submm opacity. Therefore, the assumption of $\beta$ in the dust model (or the MBB functional shape) should be consistent with the assumption of $\beta$ that the $\kappa_0$ is based on. In fact, \citet{bianchi13} shows that if $\beta$ is kept consistent, dust masses are not significantly dependent on the assumption of $\beta$. Another source of uncertainty in $\kappa_0$ is the possible variations in the ISM and dust emission properties of the systems under study compared to those in the local MW, on which the $\kappa_0$ calibrations are often based \citep{dwek97,li01,draine03,zubko04}. 
In these studies, the absorption cross-section per H atom is derived from the emission properties of the MW cirrus, and then converted to an absorption cross-section per dust mass by adopting a MW gas-to-dust ratio. \citet{bianchi19} uses a different method \citep[also used by][]{clark16} to derive dust masses from integrated measurements of gas masses, metallicities, and FIR observations for a local sample from the DustPedia project. The study shows a large range of $\kappa_0$ values that vary galaxy-to-galaxy, and are mildly dependent on metallicity.
Their results are conditional on the assumptions made for the fraction of metals in dust, the dust heating conditions, and dust-to-gas mass relation with metallicity. 

We adopt the $\kappa_0=0.4$\,m$^2$\,kg$^{-1}$ value of \citet{draine03} at 250\,{\um} for $\beta=2.08$ \citep{bianchi13} and use the conversion factor of 0.58 from \citet{bianchi13} to calibrate it for $\beta=1.5$. That is, we assume a $\kappa_0=0.232$\,m$^2$\,kg$^{-1}$ as the opacity at 250\,{\um} for an emissivity index of 1.5. 
A $\beta=2$ assumption would decrease the estimated dust masses by $\sim 30\%$.
The HRS dust masses and the dust masses of \citet{santini14} are also based on the \citet{draine03} models, so the $\kappa_0$ assumption should be consistent. The DGS and KINGFISH dust masses are derived based on $\kappa_0(100\mu{\rm m})=4.5$\,m$^2$\,kg$^{-1}$ for $\beta=2$ \citep{remyruyer15}. 
A $\kappa_0=0.192$\,m$^2$\,kg$^{-1}$ at 350\,{\um} from the \citet{draine03} models would increase the dust masses by a factor of 1.9.

It is previously claimed that dust masses derived from a single MBB tend to be lower than those derived from IR SED fitting \citep[e.g., see][]{remyruyer15}. To assess the degree of such bias in this work, we estimate the dust masses of the $z\sim 0$ DGS, KINGFISH, and HRS galaxies from their observed SPIRE 350\,{\um} fluxes using Equation~\ref{eq:mdust}, and compare them with the estimated dust masses of the local samples derived from the IR SED fitting in \citet{remyruyer15} and \citet{cortese14}.
Following the previous discussion, we make the same assumptions as done for the $z\sim 2.3$ sample: a $T=25\pm 5$\,K and two different $\beta$ (and $\kappa0$) assumptions of $\beta=1.5$ ($\kappa0(250\,\mu{\rm m})=0.232$\,m$^2$\,kg$^{-1}$) and $\beta=2.08$ ($\kappa0(250\,\mu{\rm m})=0.4$\,m$^2$\,kg$^{-1}$). 
The assumption of $\beta=1.5$ ($\beta=2.08$) result in dust masses that are on average 0.15 (0), $-0.02$ ($-0.17$), and $-0.01$ ($-0.16$) dex offset from the reported dust masses for the DGS \citep{remyruyer15}, KINGFISH \citep{remyruyer15}, and HRS \citep{ciesla14} samples based on IR SED fitting, respectively.
These offsets are smaller than the typical uncertainties of our dust mass measurements ($0.15-0.30$\,dex). 

\end{appendices}

\bibliographystyle{apj}

\begin{thebibliography}{}
\expandafter\ifx\csname natexlab\endcsname\relax\def\natexlab#1{#1}\fi

\bibitem[{{Aravena} {et~al.}(2016){Aravena}, {Decarli}, {Walter}, {Bouwens},
  {Oesch}, {Carilli}, {Bauer}, {Da Cunha}, {Daddi}, {G{\'o}nzalez-L{\'o}pez},
  {Ivison}, {Riechers}, {Smail}, {Swinbank}, {Weiss}, {Anguita}, {Bacon},
  {Bell}, {Bertoldi}, {Cortes}, {Cox}, {Hodge}, {Ibar}, {Inami}, {Infante},
  {Karim}, {Magnelli}, {Ota}, {Popping}, {van der Werf}, {Wagg}, \&
  {Fudamoto}}]{aravena16}
{Aravena}, M., {Decarli}, R., {Walter}, F., {et~al.} 2016, ApJ, 833, 71

\bibitem[{{Aravena} {et~al.}(2020){Aravena}, {Boogaard},
  {G{\'o}nzalez-L{\'o}pez}, {Decarli}, {Walter}, {Carilli}, {Smail}, {Weiss},
  {Assef}, {Bauer}, {Bouwens}, {Cortes}, {Cox}, {da Cunha}, {Daddi},
  {D{\'\i}az-Santos}, {Inami}, {Ivison}, {Novak}, {Popping}, {Riechers}, {van
  der Werf}, \& {Wagg}}]{aravena20}
{Aravena}, M., {Boogaard}, L., {G{\'o}nzalez-L{\'o}pez}, J., {et~al.} 2020,
  ApJ, 901, 79

\bibitem[{{Armus} {et~al.}(2009){Armus}, {Mazzarella}, {Evans}, {Surace},
  {Sanders}, {Iwasawa}, {Frayer}, {Howell}, {Chan}, {Petric}, {Vavilkin},
  {Kim}, {Haan}, {Inami}, {Murphy}, {Appleton}, {Barnes}, {Bothun}, {Bridge},
  {Charmandaris}, {Jensen}, {Kewley}, {Lord}, {Madore}, {Marshall},
  {Melbourne}, {Rich}, {Satyapal}, {Schulz}, {Spoon}, {Sturm}, {U}, {Veilleux},
  \& {Xu}}]{armus09}
{Armus}, L., {Mazzarella}, J.~M., {Evans}, A.~S., {et~al.} 2009, PASP, 121, 559

\bibitem[{{Asano} {et~al.}(2013){Asano}, {Takeuchi}, {Hirashita}, \&
  {Inoue}}]{asano13}
{Asano}, R.~S., {Takeuchi}, T.~T., {Hirashita}, H., \& {Inoue}, A.~K. 2013,
  Earth, Planets, and Space, 65, 213

\bibitem[{{Asplund} {et~al.}(2009){Asplund}, {Grevesse}, {Sauval}, \&
  {Scott}}]{asplund09}
{Asplund}, M., {Grevesse}, N., {Sauval}, A.~J., \& {Scott}, P. 2009, Annual
  Review of Astronomy and Astrophysics, 47, 481

\bibitem[{{Azadi} {et~al.}(2017){Azadi}, {Coil}, {Aird}, {Reddy}, {Shapley},
  {Freeman}, {Kriek}, {Leung}, {Mobasher}, {Price}, {Sanders}, {Shivaei}, \&
  {Siana}}]{azadi17}
{Azadi}, M., {Coil}, A.~L., {Aird}, J., {et~al.} 2017, ApJ, 835, 27

\bibitem[{{Azadi} {et~al.}(2018){Azadi}, {Coil}, {Aird}, {Shivaei}, {Reddy},
  {Shapley}, {Kriek}, {Freeman}, {Leung}, {Mobasher}, {Price}, {Sanders},
  {Siana}, \& {Zick}}]{azadi18}
{Azadi}, M., {Coil}, A., {Aird}, J., {et~al.} 2018, ApJ, 866, 63

\bibitem[{{Balog} {et~al.}(2014){Balog}, {M{\"u}ller}, {Nielbock}, {Altieri},
  {Klaas}, {Blommaert}, {Linz}, {Lutz}, {Mo{\'o}r}, {Billot}, {Sauvage}, \&
  {Okumura}}]{balog14}
{Balog}, Z., {M{\"u}ller}, T., {Nielbock}, M., {et~al.} 2014, Experimental
  Astronomy, 37, 129

\bibitem[{{Bendo} {et~al.}(2012){Bendo}, {Galliano}, \& {Madden}}]{bendo12}
{Bendo}, G.~J., {Galliano}, F., \& {Madden}, S.~C. 2012, MNRAS, 423, 197

\bibitem[{{Bendo} {et~al.}(2013){Bendo}, {Griffin}, {Bock}, {Conversi},
  {Dowell}, {Lim}, {Lu}, {North}, {Papageorgiou}, {Pearson}, {Pohlen},
  {Polehampton}, {Schulz}, {Shupe}, {Sibthorpe}, {Spencer}, {Swinyard},
  {Valtchanov}, \& {Xu}}]{bendo13}
{Bendo}, G.~J., {Griffin}, M.~J., {Bock}, J.~J., {et~al.} 2013, MNRAS, 433,
  3062

\bibitem[{{Berta} {et~al.}(2011){Berta}, {Magnelli}, {Nordon}, {Lutz}, {Wuyts},
  {Altieri}, {Andreani}, {Aussel}, {Casta{\~n}eda}, {Cepa}, {Cimatti}, {Daddi},
  {Elbaz}, {F{\"o}rster Schreiber}, {Genzel}, {Le Floc'h}, {Maiolino},
  {P{\'e}rez-Fournon}, {Poglitsch}, {Popesso}, {Pozzi}, {Riguccini},
  {Rodighiero}, {Sanchez-Portal}, {Sturm}, {Tacconi}, \&
  {Valtchanov}}]{berta11}
{Berta}, S., {Magnelli}, B., {Nordon}, R., {et~al.} 2011, \aap, 532, A49

\bibitem[{{Bertincourt} {et~al.}(2016){Bertincourt}, {Lagache}, {Martin},
  {Schulz}, {Conversi}, {Dassas}, {Maurin}, {Abergel}, {Beelen}, {Bernard},
  {Crill}, {Dole}, {Eales}, {Gudmundsson}, {Lellouch}, {Moreno}, \&
  {Perdereau}}]{bertincourt16}
{Bertincourt}, B., {Lagache}, G., {Martin}, P.~G., {et~al.} 2016, A\&A, 588,
  A107

\bibitem[{{Betti} {et~al.}(2019){Betti}, {Pope}, {Scoville}, {Yun}, {Aussel},
  {Kartaltepe}, \& {Sheth}}]{betti19}
{Betti}, S.~K., {Pope}, A., {Scoville}, N., {et~al.} 2019, ApJ, 874, 53

\bibitem[{{Bian} {et~al.}(2018){Bian}, {Kewley}, \& {Dopita}}]{bian18}
{Bian}, F., {Kewley}, L.~J., \& {Dopita}, M.~A. 2018, ApJ, 859, 175

\bibitem[{{Bianchi}(2013)}]{bianchi13}
{Bianchi}, S. 2013, A\&A, 552, A89

\bibitem[{{Bianchi} {et~al.}(2019){Bianchi}, {Casasola}, {Baes}, {Clark},
  {Corbelli}, {Davies}, {De Looze}, {De Vis}, {Dobbels}, {Galametz},
  {Galliano}, {Jones}, {Madden}, {Magrini}, {Mosenkov}, {Nersesian}, {Viaene},
  {Xilouris}, \& {Ysard}}]{bianchi19}
{Bianchi}, S., {Casasola}, V., {Baes}, M., {et~al.} 2019, A\&A, 631, A102

\bibitem[{{Bolatto} {et~al.}(2013){Bolatto}, {Wolfire}, \& {Leroy}}]{bolatto13}
{Bolatto}, A.~D., {Wolfire}, M., \& {Leroy}, A.~K. 2013, \araa, 51, 207

\bibitem[{{Boogaard} {et~al.}(2019){Boogaard}, {Decarli},
  {Gonz{\'a}lez-L{\'o}pez}, {van der Werf}, {Walter}, {Bouwens}, {Aravena},
  {Carilli}, {Bauer}, {Brinchmann}, {Contini}, {Cox}, {da Cunha}, {Daddi},
  {D{\'\i}az-Santos}, {Hodge}, {Inami}, {Ivison}, {Maseda}, {Matthee}, {Oesch},
  {Popping}, {Riechers}, {Schaye}, {Schouws}, {Smail}, {Weiss}, {Wisotzki},
  {Bacon}, {Cortes}, {Rix}, {Somerville}, {Swinbank}, \& {Wagg}}]{boogaard19}
{Boogaard}, L.~A., {Decarli}, R., {Gonz{\'a}lez-L{\'o}pez}, J., {et~al.} 2019,
  ApJ, 882, 140

\bibitem[{{Boogaard} {et~al.}(2020){Boogaard}, {van der Werf}, {Weiss},
  {Popping}, {Decarli}, {Walter}, {Aravena}, {Bouwens}, {Riechers},
  {Gonz{\'a}lez-L{\'o}pez}, {Smail}, {Carilli}, {Kaasinen}, {Daddi}, {Cox},
  {D{\'\i}az-Santos}, {Inami}, {Cortes}, \& {Wagg}}]{boogaard20}
{Boogaard}, L.~A., {van der Werf}, P., {Weiss}, A., {et~al.} 2020, ApJ, 902,
  109

\bibitem[{{Boquien} {et~al.}(2011){Boquien}, {Calzetti}, {Combes}, {Henkel},
  {Israel}, {Kramer}, {Rela{\~n}o}, {Verley}, {van der Werf}, {Xilouris}, \&
  {HERM33ES Team}}]{boquien11}
{Boquien}, M., {Calzetti}, D., {Combes}, F., {et~al.} 2011, AJ, 142, 111

\bibitem[{{Boselli} {et~al.}(2013){Boselli}, {Hughes}, {Cortese}, {Gavazzi}, \&
  {Buat}}]{boselli13}
{Boselli}, A., {Hughes}, T.~M., {Cortese}, L., {Gavazzi}, G., \& {Buat}, V.
  2013, A\&A, 550, A114

\bibitem[{{Boselli} {et~al.}(2010){Boselli}, {Ciesla}, {Buat}, {Cortese},
  {Auld}, {Baes}, {Bendo}, {Bianchi}, {Bock}, {Bomans}, {Bradford},
  {Castro-Rodriguez}, {Chanial}, {Charlot}, {Clemens}, {Clements}, {Corbelli},
  {Cooray}, {Cormier}, {Dariush}, {Davies}, {de Looze}, {di Serego Alighieri},
  {Dwek}, {Eales}, {Elbaz}, {Fadda}, {Fritz}, {Galametz}, {Galliano},
  {Garcia-Appadoo}, {Gavazzi}, {Gear}, {Giovanardi}, {Glenn}, {Gomez},
  {Griffin}, {Grossi}, {Hony}, {Hughes}, {Hunt}, {Isaak}, {Jones}, {Levenson},
  {Lu}, {Madden}, {O'Halloran}, {Okumura}, {Oliver}, {Page}, {Panuzzo},
  {Papageorgiou}, {Parkin}, {Perez-Fournon}, {Pierini}, {Pohlen}, {Rangwala},
  {Rigby}, {Roussel}, {Rykala}, {Sabatini}, {Sacchi}, {Sauvage}, {Schulz},
  {Schirm}, {Smith}, {Spinoglio}, {Stevens}, {Sundar}, {Symeonidis}, {Trichas},
  {Vaccari}, {Verstappen}, {Vigroux}, {Vlahakis}, {Wilson}, {Wozniak},
  {Wright}, {Xilouris}, {Zeilinger}, \& {Zibetti}}]{boselli10}
{Boselli}, A., {Ciesla}, L., {Buat}, V., {et~al.} 2010, A\&A, 518, L61

\bibitem[{{Bouwens} {et~al.}(2016){Bouwens}, {Aravena}, {Decarli}, {Walter},
  {da Cunha}, {Labb{\'e}}, {Bauer}, {Bertoldi}, {Carilli}, {Chapman}, {Daddi},
  {Hodge}, {Ivison}, {Karim}, {Le Fevre}, {Magnelli}, {Ota}, {Riechers},
  {Smail}, {van der Werf}, {Weiss}, {Cox}, {Elbaz}, {Gonzalez-Lopez},
  {Infante}, {Oesch}, {Wagg}, \& {Wilkins}}]{bouwens16c}
{Bouwens}, R.~J., {Aravena}, M., {Decarli}, R., {et~al.} 2016, ApJ, 833, 72

\bibitem[{{Bresolin} {et~al.}(2009){Bresolin}, {Gieren}, {Kudritzki},
  {Pietrzy{\'n}ski}, {Urbaneja}, \& {Carraro}}]{bresolin09}
{Bresolin}, F., {Gieren}, W., {Kudritzki}, R.-P., {et~al.} 2009, ApJ, 700, 309

\bibitem[{Bruzual \& Charlot(2003)}]{bc03}
Bruzual, G., \& Charlot, S. 2003, MNRAS, 344, 1000

\bibitem[{{Burnham} {et~al.}(2021){Burnham}, {Casey}, {Zavala}, {Manning},
  {Spilker}, {Chapman}, {Chen}, {Cooray}, {Sanders}, \& {Scoville}}]{burnham21}
{Burnham}, A.~D., {Casey}, C.~M., {Zavala}, J.~A., {et~al.} 2021, \apj, 910, 89

\bibitem[{{Calistro Rivera} {et~al.}(2018){Calistro Rivera}, {Hodge}, {Smail},
  {Swinbank}, {Weiss}, {Wardlow}, {Walter}, {Rybak}, {Chen}, {Brandt},
  {Coppin}, {da Cunha}, {Dannerbauer}, {Greve}, {Karim}, {Knudsen},
  {Schinnerer}, {Simpson}, {Venemans}, \& {van der Werf}}]{calistro-rivera18}
{Calistro Rivera}, G., {Hodge}, J.~A., {Smail}, I., {et~al.} 2018, ApJ, 863, 56

\bibitem[{{Calura} {et~al.}(2017){Calura}, {Pozzi}, {Cresci}, {Santini},
  {Gruppioni}, {Pozzetti}, {Gilli}, {Matteucci}, \& {Maiolino}}]{calura17}
{Calura}, F., {Pozzi}, F., {Cresci}, G., {et~al.} 2017, MNRAS, 465, 54

\bibitem[{Calzetti {et~al.}(2000)Calzetti, Armus, Bohlin, Kinney, Koornneef, \&
  Storchi-Bergmann}]{calzetti00}
Calzetti, D., Armus, L., Bohlin, R.~C., {et~al.} 2000, ApJ, 533, 682

\bibitem[{Cardelli {et~al.}(1989)Cardelli, Clayton, \& Mathis}]{cardelli89}
Cardelli, J.~A., Clayton, G.~C., \& Mathis, J.~S. 1989, ApJ, 345, 245

\bibitem[{{Casey}(2012)}]{casey12}
{Casey}, C.~M. 2012, Monthly Notices of the Royal Astronomical Society, 425,
  3094

\bibitem[{{Casey} {et~al.}(2018){Casey}, {Zavala}, {Spilker}, {da Cunha},
  {Hodge}, {Hung}, {Staguhn}, {Finkelstein}, \& {Drew}}]{casey18}
{Casey}, C.~M., {Zavala}, J.~A., {Spilker}, J., {et~al.} 2018, ApJ, 862, 77

\bibitem[{Chabrier(2003)}]{chabrier03}
Chabrier, G. 2003, PASP, 115, 763

\bibitem[{{Chanial} {et~al.}(2007){Chanial}, {Flores}, {Guiderdoni}, {Elbaz},
  {Hammer}, \& {Vigroux}}]{chanial07}
{Chanial}, P., {Flores}, H., {Guiderdoni}, B., {et~al.} 2007, \aap, 462, 81

\bibitem[{{Charlot} \& {Fall}(2000)}]{charlotfall00}
{Charlot}, S., \& {Fall}, S.~M. 2000, ApJ, 539, 718

\bibitem[{Chary \& Elbaz(2001)}]{ce01}
Chary, R., \& Elbaz, D. 2001, ApJ, 556, 562

\bibitem[{{Chu} {et~al.}(2017){Chu}, {Sanders}, {Larson}, {Mazzarella},
  {Howell}, {D{\'\i}az-Santos}, {Xu}, {Paladini}, {Schulz}, {Shupe},
  {Appleton}, {Armus}, {Billot}, {Chan}, {Evans}, {Fadda}, {Frayer}, {Haan},
  {Ishida}, {Iwasawa}, {Kim}, {Lord}, {Murphy}, {Petric}, {Privon}, {Surace},
  \& {Treister}}]{chu17}
{Chu}, J.~K., {Sanders}, D.~B., {Larson}, K.~L., {et~al.} 2017, ApJS, 229, 25

\bibitem[{{Cid Fernandes} {et~al.}(2005){Cid Fernandes}, {Mateus}, {Sodr{\'e}},
  {Stasi{\'n}ska}, \& {Gomes}}]{fernandes05}
{Cid Fernandes}, R., {Mateus}, A., {Sodr{\'e}}, L., {Stasi{\'n}ska}, G., \&
  {Gomes}, J.~M. 2005, MNRAS, 358, 363

\bibitem[{{Ciesla} {et~al.}(2012){Ciesla}, {Boselli}, {Smith}, {Bendo},
  {Cortese}, {Eales}, {Bianchi}, {Boquien}, {Buat}, {Davies}, {Pohlen},
  {Zibetti}, {Baes}, {Cooray}, {De Looze}, {di Serego Alighieri}, {Galametz},
  {Gomez}, {Lebouteiller}, {Madden}, {Pappalardo}, {Remy}, {Spinoglio},
  {Vaccari}, {Auld}, \& {Clements}}]{ciesla12}
{Ciesla}, L., {Boselli}, A., {Smith}, M.~W.~L., {et~al.} 2012, A\&A, 543, A161

\bibitem[{{Ciesla} {et~al.}(2014){Ciesla}, {Boquien}, {Boselli}, {Buat},
  {Cortese}, {Bendo}, {Heinis}, {Galametz}, {Eales}, {Smith}, {Baes},
  {Bianchi}, {De Looze}, {di Serego Alighieri}, {Galliano}, {Hughes}, {Madden},
  {Pierini}, {R{\'e}my-Ruyer}, {Spinoglio}, {Vaccari}, {Viaene}, \&
  {Vlahakis}}]{ciesla14}
{Ciesla}, L., {Boquien}, M., {Boselli}, A., {et~al.} 2014, A\&A, 565, A128

\bibitem[{{Clark} {et~al.}(2016){Clark}, {Schofield}, {Gomez}, \&
  {Davies}}]{clark16}
{Clark}, C. J.~R., {Schofield}, S.~P., {Gomez}, H.~L., \& {Davies}, J.~I. 2016,
  \mnras, 459, 1646

\bibitem[{Coil {et~al.}(2015)Coil, Aird, Reddy, Shapley, Kriek, Siana,
  Mobasher, Freeman, Price, \& Shivaei}]{coil15}
Coil, A.~L., Aird, J., Reddy, N., {et~al.} 2015, ApJ, 801, 35

\bibitem[{Conroy {et~al.}(2009)Conroy, Gunn, \& White}]{conroy09}
Conroy, C., Gunn, J.~E., \& White, M. 2009, ApJ, 699, 486

\bibitem[{{Cortese} {et~al.}(2014){Cortese}, {Fritz}, {Bianchi}, {Boselli},
  {Ciesla}, {Bendo}, {Boquien}, {Roussel}, {Baes}, {Buat}, {Clemens}, {Cooray},
  {Cormier}, {Davies}, {De Looze}, {Eales}, {Fuller}, {Hunt}, {Madden},
  {Munoz-Mateos}, {Pappalardo}, {Pierini}, {R{\'e}my-Ruyer}, {Sauvage}, {di
  Serego Alighieri}, {Smith}, {Spinoglio}, {Vaccari}, \&
  {Vlahakis}}]{cortese14}
{Cortese}, L., {Fritz}, J., {Bianchi}, S., {et~al.} 2014, \mnras, 440, 942

\bibitem[{{Cullen} {et~al.}(2021){Cullen}, {Shapley}, {McLure}, {Dunlop},
  {Sanders}, {Topping}, {Reddy}, {Amor{\'\i}n}, {Begley}, {Bolzonella},
  {Calabr{\`o}}, {Carnall}, {Castellano}, {Cimatti}, {Cirasuolo}, {Cresci},
  {Fontana}, {Fontanot}, {Garilli}, {Guaita}, {Hamadouche}, {Hathi},
  {Mannucci}, {McLeod}, {Pentericci}, {Saxena}, {Talia}, \&
  {Zamorani}}]{cullen21}
{Cullen}, F., {Shapley}, A.~E., {McLure}, R.~J., {et~al.} 2021, MNRAS, 505, 903

\bibitem[{{Curti} {et~al.}(2017){Curti}, {Cresci}, {Mannucci}, {Marconi},
  {Maiolino}, \& {Esposito}}]{curti17}
{Curti}, M., {Cresci}, G., {Mannucci}, F., {et~al.} 2017, MNRAS, 465, 1384

\bibitem[{{Czekala} {et~al.}(2021){Czekala}, {Loomis}, {Teague}, {Booth},
  {Huang}, {Cataldi}, {Ilee}, {Law}, {Walsh}, {Bosman}, {Guzm{\'a}n}, {Gal},
  {{\"O}berg}, {Yamato}, {Aikawa}, {Andrews}, {Bae}, {Bergin}, {Bergner},
  {Cleeves}, {Kurtovic}, {M{\'e}nard}, {Nomura}, {P{\'e}rez}, {Qi}, {Schwarz},
  {Tsukagoshi}, {Waggoner}, {Wilner}, \& {Zhang}}]{czekala21}
{Czekala}, I., {Loomis}, R.~A., {Teague}, R., {et~al.} 2021, \apjs, 257, 2

\bibitem[{{da Cunha} {et~al.}(2008){da Cunha}, {Charlot}, \&
  {Elbaz}}]{dacunha08}
{da Cunha}, E., {Charlot}, S., \& {Elbaz}, D. 2008, MNRAS, 388, 1595

\bibitem[{{da Cunha} {et~al.}(2013){da Cunha}, {Groves}, {Walter}, {Decarli},
  {Weiss}, {Bertoldi}, {Carilli}, {Daddi}, {Elbaz}, {Ivison}, {Maiolino},
  {Riechers}, {Rix}, {Sargent}, \& {Smail}}]{dacunha13}
{da Cunha}, E., {Groves}, B., {Walter}, F., {et~al.} 2013, ApJ, 766, 13

\bibitem[{Dale \& Helou(2002)}]{dh02}
Dale, D.~A., \& Helou, G. 2002, ApJ, 576, 159

\bibitem[{{Dale} {et~al.}(2001){Dale}, {Helou}, {Contursi}, {Silbermann}, \&
  {Kolhatkar}}]{dale01}
{Dale}, D.~A., {Helou}, G., {Contursi}, A., {Silbermann}, N.~A., \&
  {Kolhatkar}, S. 2001, ApJ, 549, 215

\bibitem[{{Dale} {et~al.}(2007){Dale}, {Gil de Paz}, {Gordon}, {Hanson},
  {Armus}, {Bendo}, {Bianchi}, {Block}, {Boissier}, {Boselli}, {Buckalew},
  {Buat}, {Burgarella}, {Calzetti}, {Cannon}, {Engelbracht}, {Helou},
  {Hollenbach}, {Jarrett}, {Kennicutt}, {Leitherer}, {Li}, {Madore}, {Martin},
  {Meyer}, {Murphy}, {Regan}, {Roussel}, {Smith}, {Sosey}, {Thilker}, \&
  {Walter}}]{dale07}
{Dale}, D.~A., {Gil de Paz}, A., {Gordon}, K.~D., {et~al.} 2007, ApJ, 655, 863

\bibitem[{{Dale} {et~al.}(2012){Dale}, {Aniano}, {Engelbracht}, {Hinz},
  {Krause}, {Montiel}, {Roussel}, {Appleton}, {Armus}, {Beir{\~a}o}, {Bolatto},
  {Brandl}, {Calzetti}, {Crocker}, {Croxall}, {Draine}, {Galametz}, {Gordon},
  {Groves}, {Hao}, {Helou}, {Hunt}, {Johnson}, {Kennicutt}, {Koda}, {Leroy},
  {Li}, {Meidt}, {Miller}, {Murphy}, {Rahman}, {Rix}, {Sandstrom}, {Sauvage},
  {Schinnerer}, {Skibba}, {Smith}, {Tabatabaei}, {Walter}, {Wilson}, {Wolfire},
  \& {Zibetti}}]{dale12}
{Dale}, D.~A., {Aniano}, G., {Engelbracht}, C.~W., {et~al.} 2012, ApJ, 745, 95

\bibitem[{{De Rossi} {et~al.}(2018){De Rossi}, {Rieke}, {Shivaei}, {Bromm}, \&
  {Lyu}}]{derossi18}
{De Rossi}, M.~E., {Rieke}, G.~H., {Shivaei}, I., {Bromm}, V., \& {Lyu}, J.
  2018, The Astrophysical Journal, 869, 4

\bibitem[{{De Vis} {et~al.}(2019){De Vis}, {Jones}, {Viaene}, {Casasola},
  {Clark}, {Baes}, {Bianchi}, {Cassara}, {Davies}, {De Looze}, {Galametz},
  {Galliano}, {Lianou}, {Madden}, {Manilla-Robles}, {Mosenkov}, {Nersesian},
  {Roychowdhury}, {Xilouris}, \& {Ysard}}]{devis19}
{De Vis}, P., {Jones}, A., {Viaene}, S., {et~al.} 2019, A\&A, 623, A5

\bibitem[{{Decarli} {et~al.}(2020){Decarli}, {Aravena}, {Boogaard}, {Carilli},
  {Gonz{\'a}lez-L{\'o}pez}, {Walter}, {Cortes}, {Cox}, {da Cunha}, {Daddi},
  {D{\'\i}az-Santos}, {Hodge}, {Inami}, {Neeleman}, {Novak}, {Oesch},
  {Popping}, {Riechers}, {Smail}, {Uzgil}, {van der Werf}, {Wagg}, \&
  {Weiss}}]{decarli20}
{Decarli}, R., {Aravena}, M., {Boogaard}, L., {et~al.} 2020, ApJ, 902, 110

\bibitem[{{D{\'\i}az-Santos} {et~al.}(2010){D{\'\i}az-Santos}, {Charmandaris},
  {Armus}, {Petric}, {Howell}, {Murphy}, {Mazzarella}, {Veilleux}, {Bothun},
  {Inami}, {Appleton}, {Evans}, {Haan}, {Marshall}, {Sanders}, {Stierwalt}, \&
  {Surace}}]{diazsantos10}
{D{\'\i}az-Santos}, T., {Charmandaris}, V., {Armus}, L., {et~al.} 2010, ApJ,
  723, 993

\bibitem[{{D{\'\i}az-Santos} {et~al.}(2013){D{\'\i}az-Santos}, {Armus},
  {Charmandaris}, {Stierwalt}, {Murphy}, {Haan}, {Inami}, {Malhotra},
  {Meijerink}, {Stacey}, {Petric}, {Evans}, {Veilleux}, {van der Werf}, {Lord},
  {Lu}, {Howell}, {Appleton}, {Mazzarella}, {Surace}, {Xu}, {Schulz},
  {Sanders}, {Bridge}, {Chan}, {Frayer}, {Iwasawa}, {Melbourne}, \&
  {Sturm}}]{diazsantos13}
{D{\'\i}az-Santos}, T., {Armus}, L., {Charmandaris}, V., {et~al.} 2013, ApJ,
  774, 68

\bibitem[{{D{\'\i}az-Santos} {et~al.}(2017){D{\'\i}az-Santos}, {Armus},
  {Charmandaris}, {Lu}, {Stierwalt}, {Stacey}, {Malhotra}, {van der Werf},
  {Howell}, {Privon}, {Mazzarella}, {Goldsmith}, {Murphy}, {Barcos-Mu{\~n}oz},
  {Linden}, {Inami}, {Larson}, {Evans}, {Appleton}, {Iwasawa}, {Lord},
  {Sanders}, \& {Surace}}]{diazsantos17}
---. 2017, ApJ, 846, 32

\bibitem[{{Dole} {et~al.}(2006){Dole}, {Lagache}, {Puget}, {Caputi},
  {Fern{\'a}ndez-Conde}, {Le Floc'h}, {Papovich}, {P{\'e}rez-Gonz{\'a}lez},
  {Rieke}, \& {Blaylock}}]{dole06}
{Dole}, H., {Lagache}, G., {Puget}, J.~L., {et~al.} 2006, A\&A, 451, 417

\bibitem[{{Draine}(2003)}]{draine03}
{Draine}, B.~T. 2003, ARA\&A, 41, 241

\bibitem[{Draine \& Li(2007)}]{draine07a}
Draine, B.~T., \& Li, A. 2007, ApJ, 657, 810

\bibitem[{Draine {et~al.}(2007b)Draine, Dale, Bendo, Gordon, Smith, Armus,
  Engelbracht, Helou, Kennicutt, Li, Roussel, Walter, Calzetti, Moustakas,
  Murphy, Rieke, Bot, Hollenbach, Sheth, \& Teplitz}]{draine07b}
Draine, B.~T., Dale, D.~A., Bendo, G., {et~al.} 2007b, ApJ, 663, 866

\bibitem[{{Dunne} {et~al.}(2000){Dunne}, {Eales}, {Edmunds}, {Ivison},
  {Alexander}, \& {Clements}}]{dunne00}
{Dunne}, L., {Eales}, S., {Edmunds}, M., {et~al.} 2000, MNRAS, 315, 115

\bibitem[{{Dwek} {et~al.}(1997){Dwek}, {Arendt}, {Fixsen}, {Sodroski},
  {Odegard}, {Weiland}, {Reach}, {Hauser}, {Kelsall}, {Moseley}, {Silverberg},
  {Shafer}, {Ballester}, {Bazell}, \& {Isaacman}}]{dwek97}
{Dwek}, E., {Arendt}, R.~G., {Fixsen}, D.~J., {et~al.} 1997, ApJ, 475, 565

\bibitem[{{Elbaz} {et~al.}(2011){Elbaz}, {Dickinson}, {Hwang},
  {D{\'{\i}}az-Santos}, {Magdis}, {Magnelli}, {Le Borgne}, {Galliano},
  {Pannella}, {Chanial}, {Armus}, {Charmandaris}, {Daddi}, {Aussel}, {Popesso},
  {Kartaltepe}, {Altieri}, {Valtchanov}, {Coia}, {Dannerbauer}, {Dasyra},
  {Leiton}, {Mazzarella}, {Alexander}, {Buat}, {Burgarella}, {Chary}, {Gilli},
  {Ivison}, {Juneau}, {Le Floc'h}, {Lutz}, {Morrison}, {Mullaney}, {Murphy},
  {Pope}, {Scott}, {Brodwin}, {Calzetti}, {Cesarsky}, {Charlot}, {Dole},
  {Eisenhardt}, {Ferguson}, {F{\"o}rster Schreiber}, {Frayer}, {Giavalisco},
  {Huynh}, {Koekemoer}, {Papovich}, {Reddy}, {Surace}, {Teplitz}, {Yun}, \&
  {Wilson}}]{elbaz11}
{Elbaz}, D., {Dickinson}, M., {Hwang}, H.~S., {et~al.} 2011, A\&A, 533, A119

\bibitem[{{Engelbracht} {et~al.}(2005){Engelbracht}, {Gordon}, {Rieke},
  {Werner}, {Dale}, \& {Latter}}]{engelbracht05}
{Engelbracht}, C.~W., {Gordon}, K.~D., {Rieke}, G.~H., {et~al.} 2005, ApJL,
  628, L29

\bibitem[{{Faisst} {et~al.}(2020){Faisst}, {Fudamoto}, {Oesch}, {Scoville},
  {Riechers}, {Pavesi}, \& {Capak}}]{faisst20}
{Faisst}, A.~L., {Fudamoto}, Y., {Oesch}, P.~A., {et~al.} 2020, MNRAS, 498,
  4192

\bibitem[{{Faisst} {et~al.}(2017){Faisst}, {Capak}, {Yan}, {Pavesi},
  {Riechers}, {Bari{\v{s}}i{\'c}}, {Cooke}, {Kartaltepe}, \&
  {Masters}}]{faisst17}
{Faisst}, A.~L., {Capak}, P.~L., {Yan}, L., {et~al.} 2017, ApJ, 847, 21

\bibitem[{{Farren} {et~al.}(2021){Farren}, {Partridge}, {Kneissl}, {Aiola},
  {Datta}, {Gralla}, \& {Li}}]{farren21}
{Farren}, G.~S., {Partridge}, B., {Kneissl}, R., {et~al.} 2021, ApJS, 256, 19

\bibitem[{{Finke} {et~al.}(2010){Finke}, {Razzaque}, \& {Dermer}}]{finke10}
{Finke}, J.~D., {Razzaque}, S., \& {Dermer}, C.~D. 2010, ApJ, 712, 238

\bibitem[{{Freeman} {et~al.}(2019){Freeman}, {Siana}, {Kriek}, {Shapley},
  {Reddy}, {Coil}, {Mobasher}, {Muratov}, {Azadi}, {Leung}, {Sanders},
  {Shivaei}, {Price}, {DeGroot}, \& {Kere{\v{s}}}}]{freeman19}
{Freeman}, W.~R., {Siana}, B., {Kriek}, M., {et~al.} 2019, ApJ, 873, 102

\bibitem[{{Fudamoto} {et~al.}(2019){Fudamoto}, {Oesch}, {Magnelli},
  {Schinnerer}, {Liu}, {Lang}, {Jim{\'e}nez-Andrade}, {Groves}, {Leslie}, \&
  {Sargent}}]{fudamoto19}
{Fudamoto}, Y., {Oesch}, P.~A., {Magnelli}, B., {et~al.} 2019, \mnras, 2849

\bibitem[{{Fujimoto} {et~al.}(2016){Fujimoto}, {Ouchi}, {Ono}, {Shibuya},
  {Ishigaki}, {Nagai}, \& {Momose}}]{fujimoto16}
{Fujimoto}, S., {Ouchi}, M., {Ono}, Y., {et~al.} 2016, ApJS, 222, 1

\bibitem[{{Galametz} {et~al.}(2011){Galametz}, {Madden}, {Galliano}, {Hony},
  {Bendo}, \& {Sauvage}}]{galametz11}
{Galametz}, M., {Madden}, S.~C., {Galliano}, F., {et~al.} 2011, A\&A, 532, A56

\bibitem[{{Gallazzi} {et~al.}(2005){Gallazzi}, {Charlot}, {Brinchmann},
  {White}, \& {Tremonti}}]{gallazzi05}
{Gallazzi}, A., {Charlot}, S., {Brinchmann}, J., {White}, S. D.~M., \&
  {Tremonti}, C.~A. 2005, MNRAS, 362, 41

\bibitem[{{Galliano} {et~al.}(2018){Galliano}, {Galametz}, \&
  {Jones}}]{galliano18}
{Galliano}, F., {Galametz}, M., \& {Jones}, A.~P. 2018, ARAA, 56, 673

\bibitem[{{Galliano} {et~al.}(2005){Galliano}, {Madden}, {Jones}, {Wilson}, \&
  {Bernard}}]{galliano05}
{Galliano}, F., {Madden}, S.~C., {Jones}, A.~P., {Wilson}, C.~D., \& {Bernard},
  J.~P. 2005, A\&A, 434, 867

\bibitem[{{Galliano} {et~al.}(2011){Galliano}, {Hony}, {Bernard}, {Bot},
  {Madden}, {Roman-Duval}, {Galametz}, {Li}, {Meixner}, {Engelbracht},
  {Lebouteiller}, {Misselt}, {Montiel}, {Panuzzo}, {Reach}, \&
  {Skibba}}]{galliano11b}
{Galliano}, F., {Hony}, S., {Bernard}, J.~P., {et~al.} 2011, A\&A, 536, A88

\bibitem[{{Gonz{\'a}lez-L{\'o}pez} {et~al.}(2020){Gonz{\'a}lez-L{\'o}pez},
  {Novak}, {Decarli}, {Walter}, {Aravena}, {Carilli}, {Boogaard}, {Popping},
  {Weiss}, {Assef}, {Bauer}, {Bouwens}, {Cortes}, {Cox}, {Daddi}, {Cunha},
  {D{\'\i}az-Santos}, {Ivison}, {Magnelli}, {Riechers}, {Smail}, {van der
  Werf}, \& {Wagg}}]{gonzalez20}
{Gonz{\'a}lez-L{\'o}pez}, J., {Novak}, M., {Decarli}, R., {et~al.} 2020, ApJ,
  897, 91

\bibitem[{{Hao} {et~al.}(2011){Hao}, {Kennicutt}, {Johnson}, {Calzetti},
  {Dale}, \& {Moustakas}}]{hao11}
{Hao}, C.-N., {Kennicutt}, R.~C., {Johnson}, B.~D., {et~al.} 2011, ApJ, 741,
  124

\bibitem[{{Hatsukade} {et~al.}(2016){Hatsukade}, {Kohno}, {Umehata},
  {Aretxaga}, {Caputi}, {Dunlop}, {Ikarashi}, {Iono}, {Ivison}, {Lee},
  {Makiya}, {Matsuda}, {Motohara}, {Nakanishi}, {Ohta}, {Tadaki}, {Tamura},
  {Wang}, {Wilson}, {Yamaguchi}, \& {Yun}}]{hatsukade16}
{Hatsukade}, B., {Kohno}, K., {Umehata}, H., {et~al.} 2016, PASJ, 68, 36

\bibitem[{{Hirashita}(2015)}]{hirashita15}
{Hirashita}, H. 2015, MNRAS, 447, 2937

\bibitem[{{Hirashita} {et~al.}(2008){Hirashita}, {Nozawa}, {Takeuchi}, \&
  {Kozasa}}]{hirashita08}
{Hirashita}, H., {Nozawa}, T., {Takeuchi}, T.~T., \& {Kozasa}, T. 2008, MNRAS,
  384, 1725

\bibitem[{{Howell} {et~al.}(2010){Howell}, {Armus}, {Mazzarella}, {Evans},
  {Surace}, {Sanders}, {Petric}, {Appleton}, {Bothun}, {Bridge}, {Chan},
  {Charmandaris}, {Frayer}, {Haan}, {Inami}, {Kim}, {Lord}, {Madore},
  {Melbourne}, {Schulz}, {U}, {Vavilkin}, {Veilleux}, \& {Xu}}]{howell10}
{Howell}, J.~H., {Armus}, L., {Mazzarella}, J.~M., {et~al.} 2010, ApJ, 715, 572

\bibitem[{{Hughes} {et~al.}(2013){Hughes}, {Cortese}, {Boselli}, {Gavazzi}, \&
  {Davies}}]{hughes13}
{Hughes}, T.~M., {Cortese}, L., {Boselli}, A., {Gavazzi}, G., \& {Davies},
  J.~I. 2013, A\&A, 550, A115

\bibitem[{{Hunt} {et~al.}(2010){Hunt}, {Thuan}, {Izotov}, \&
  {Sauvage}}]{hunt10}
{Hunt}, L.~K., {Thuan}, T.~X., {Izotov}, Y.~I., \& {Sauvage}, M. 2010, ApJ,
  712, 164

\bibitem[{{Jones} {et~al.}(1996){Jones}, {Tielens}, \& {Hollenbach}}]{jones96}
{Jones}, A.~P., {Tielens}, A.~G.~G.~M., \& {Hollenbach}, D.~J. 1996, ApJ, 469,
  740

\bibitem[{{Jorsater} \& {van Moorsel}(1995)}]{JvM95}
{Jorsater}, S., \& {van Moorsel}, G.~A. 1995, AJ, 110, 2037

\bibitem[{{Kaasinen} {et~al.}(2020){Kaasinen}, {Walter}, {Novak}, {Neeleman},
  {Smail}, {Boogaard}, {Cunha}, {Weiss}, {Liu}, {Decarli}, {Popping},
  {Diaz-Santos}, {Cort{\'e}s}, {Aravena}, {Werf}, {Riechers}, {Inami}, {Hodge},
  {Rix}, \& {Cox}}]{kaasinen20}
{Kaasinen}, M., {Walter}, F., {Novak}, M., {et~al.} 2020, \apj, 899, 37

\bibitem[{{Kashino} {et~al.}(2019){Kashino}, {Silverman}, {Sanders},
  {Kartaltepe}, {Daddi}, {Renzini}, {Rodighiero}, {Puglisi}, {Valentino},
  {Juneau}, {Arimoto}, {Nagao}, {Ilbert}, {Le F{\`e}vre}, \&
  {Koekemoer}}]{kashino19}
{Kashino}, D., {Silverman}, J.~D., {Sanders}, D., {et~al.} 2019, The
  Astrophysical Journal Supplement Series, 241, 10

\bibitem[{{Kennicutt} \& {De Los Reyes}(2021)}]{kennicutt21}
{Kennicutt}, Robert~C., J., \& {De Los Reyes}, M. A.~C. 2021, ApJ, 908, 61

\bibitem[{Kennicutt(1998)}]{kennicutt98}
Kennicutt, R.~C. 1998, Annual Review of Astronomy and Astrophysics, 36, 189

\bibitem[{Kennicutt \& Evans(2012)}]{kennicutt12}
Kennicutt, R.~C., \& Evans, N.~J. 2012, Annual Review of Astronomy and
  Astrophysics, 50, 531

\bibitem[{{Kennicutt} {et~al.}(2011){Kennicutt}, {Calzetti}, {Aniano},
  {Appleton}, {Armus}, {Beir{\~a}o}, {Bolatto}, {Brandl}, {Crocker}, {Croxall},
  {Dale}, {Donovan Meyer}, {Draine}, {Engelbracht}, {Galametz}, {Gordon},
  {Groves}, {Hao}, {Helou}, {Hinz}, {Hunt}, {Johnson}, {Koda}, {Krause},
  {Leroy}, {Li}, {Meidt}, {Montiel}, {Murphy}, {Rahman}, {Rix}, {Roussel},
  {Sandstrom}, {Sauvage}, {Schinnerer}, {Skibba}, {Smith}, {Srinivasan},
  {Vigroux}, {Walter}, {Wilson}, {Wolfire}, \& {Zibetti}}]{kennicutt11}
{Kennicutt}, R.~C., {Calzetti}, D., {Aniano}, G., {et~al.} 2011, PASP, 123,
  1347

\bibitem[{{Kewley} {et~al.}(2013){Kewley}, {Dopita}, {Leitherer}, {Dav{\'e}},
  {Yuan}, {Allen}, {Groves}, \& {Sutherland}}]{kewley13}
{Kewley}, L.~J., {Dopita}, M.~A., {Leitherer}, C., {et~al.} 2013, ApJ, 774, 100

\bibitem[{{Kirkpatrick} {et~al.}(2015){Kirkpatrick}, {Pope}, {Sajina},
  {Roebuck}, {Yan}, {Armus}, {D{\'\i}az-Santos}, \&
  {Stierwalt}}]{kirkpatrick15}
{Kirkpatrick}, A., {Pope}, A., {Sajina}, A., {et~al.} 2015, ApJ, 814, 9

\bibitem[{{Kirkpatrick} {et~al.}(2012){Kirkpatrick}, {Pope}, {Alexander},
  {Charmandaris}, {Daddi}, {Dickinson}, {Elbaz}, {Gabor}, {Hwang}, {Ivison},
  {Mullaney}, {Pannella}, {Scott}, {Altieri}, {Aussel}, {Bournaud}, {Buat},
  {Coia}, {Dannerbauer}, {Dasyra}, {Kartaltepe}, {Leiton}, {Lin}, {Magdis},
  {Magnelli}, {Morrison}, {Popesso}, \& {Valtchanov}}]{kirkpatrick12}
{Kirkpatrick}, A., {Pope}, A., {Alexander}, D.~M., {et~al.} 2012, ApJ, 759, 139

\bibitem[{{Kirkpatrick} {et~al.}(2017){Kirkpatrick}, {Pope}, {Sajina}, {Dale},
  {D{\'\i}az-Santos}, {Hayward}, {Shi}, {Somerville}, {Stierwalt}, {Armus},
  {Kartaltepe}, {Kocevski}, {McIntosh}, {Sanders}, \& {Yan}}]{kirkpatrick17}
{Kirkpatrick}, A., {Pope}, A., {Sajina}, A., {et~al.} 2017, ApJ, 843, 71

\bibitem[{{Kriek} {et~al.}(2009){Kriek}, {van Dokkum}, {Labb{\'e}}, {Franx},
  {Illingworth}, {Marchesini}, \& {Quadri}}]{kriek09a}
{Kriek}, M., {van Dokkum}, P.~G., {Labb{\'e}}, I., {et~al.} 2009, \apj, 700,
  221

\bibitem[{Kriek {et~al.}(2015)Kriek, Shapley, Reddy, Siana, Coil, Mobasher,
  Freeman, de~Groot, Price, Sanders, Shivaei, Brammer, Momcheva, Skelton, van
  Dokkum, Whitaker, Aird, Azadi, Kassis, Bullock, Conroy, Dave, Keres, \&
  Krumholz}]{kriek15}
Kriek, M., Shapley, A.~E., Reddy, N.~A., {et~al.} 2015, ApJS, 218, 15

\bibitem[{{Lebouteiller} {et~al.}(2011){Lebouteiller}, {Barry}, {Spoon},
  {Bernard-Salas}, {Sloan}, {Houck}, \& {Weedman}}]{lebouteiller11}
{Lebouteiller}, V., {Barry}, D.~J., {Spoon}, H.~W.~W., {et~al.} 2011, ApJS,
  196, 8

\bibitem[{{Lehnert} \& {Heckman}(1996)}]{lehnert96}
{Lehnert}, M.~D., \& {Heckman}, T.~M. 1996, \apj, 472, 546

\bibitem[{{Leiton} {et~al.}(2015){Leiton}, {Elbaz}, {Okumura}, {Hwang},
  {Magdis}, {Magnelli}, {Valtchanov}, {Dickinson}, {B{\'e}thermin},
  {Schreiber}, {Charmandaris}, {Dole}, {Juneau}, {Le Borgne}, {Pannella},
  {Pope}, \& {Popesso}}]{leiton15}
{Leiton}, R., {Elbaz}, D., {Okumura}, K., {et~al.} 2015, A\& A, 579, A93

\bibitem[{{Leroy} {et~al.}(2011){Leroy}, {Bolatto}, {Gordon}, {Sand strom},
  {Gratier}, {Rosolowsky}, {Engelbracht}, {Mizuno}, {Corbelli}, {Fukui}, \&
  {Kawamura}}]{leroy11}
{Leroy}, A.~K., {Bolatto}, A., {Gordon}, K., {et~al.} 2011, ApJ, 737, 12

\bibitem[{{Leung} {et~al.}(2019){Leung}, {Coil}, {Aird}, {Azadi}, {Kriek},
  {Mobasher}, {Reddy}, {Shapley}, {Siana}, {Fetherolf}, {Fornasini}, {Freeman},
  {Price}, {Sanders}, {Shivaei}, \& {Zick}}]{leung19}
{Leung}, G. C.~K., {Coil}, A.~L., {Aird}, J., {et~al.} 2019, ApJ, 886, 11

\bibitem[{{Li}(2020)}]{li20}
{Li}, A. 2020, Nature Astronomy, 4, 339

\bibitem[{{Li} \& {Draine}(2001)}]{li01}
{Li}, A., \& {Draine}, B.~T. 2001, APJ, 554, 778

\bibitem[{{Liu} {et~al.}(2019){Liu}, {Lang}, {Magnelli}, {Schinnerer},
  {Leslie}, {Fudamoto}, {Bondi}, {Groves}, {Jim{\'e}nez-Andrade}, {Harrington},
  {Karim}, {Oesch}, {Sargent}, {Vardoulaki}, {B{\v{a}}descu}, {Moser},
  {Bertoldi}, {Battisti}, {da Cunha}, {Zavala}, {Vaccari}, {Davidzon},
  {Riechers}, \& {Aravena}}]{liu19}
{Liu}, D., {Lang}, P., {Magnelli}, B., {et~al.} 2019, ApJS, 244, 40

\bibitem[{{Lutz} {et~al.}(2011){Lutz}, {Poglitsch}, {Altieri}, {Andreani},
  {Aussel}, {Berta}, {Bongiovanni}, {Brisbin}, {Cava}, {Cepa}, {Cimatti},
  {Daddi}, {Dominguez-Sanchez}, {Elbaz}, {F{\"o}rster Schreiber}, {Genzel},
  {Grazian}, {Gruppioni}, {Harwit}, {Le Floc'h}, {Magdis}, {Magnelli},
  {Maiolino}, {Nordon}, {P{\'e}rez Garc{\'{\i}}a}, {Popesso}, {Pozzi},
  {Riguccini}, {Rodighiero}, {Saintonge}, {Sanchez Portal}, {Santini}, {Shao},
  {Sturm}, {Tacconi}, {Valtchanov}, {Wetzstein}, \& {Wieprecht}}]{lutz11}
{Lutz}, D., {Poglitsch}, A., {Altieri}, B., {et~al.} 2011, A\&A, 532, A90

\bibitem[{{Lyu} \& {Rieke}(2017)}]{lyu17}
{Lyu}, J., \& {Rieke}, G.~H. 2017, ApJ, 841, 76

\bibitem[{{Lyu} \& {Rieke}(2018)}]{lyu18}
---. 2018, ApJ, 866, 92

\bibitem[{{Lyu} {et~al.}(2016){Lyu}, {Rieke}, \& {Alberts}}]{lyu16}
{Lyu}, J., {Rieke}, G.~H., \& {Alberts}, S. 2016, ApJ, 816, 85

\bibitem[{Madau \& Dickinson(2014)}]{madau14}
Madau, P., \& Dickinson, M. 2014, Annual Review of Astronomy and Astrophysics,
  52, 415

\bibitem[{{Madden} {et~al.}(2013){Madden}, {R{\'e}my-Ruyer}, {Galametz},
  {Cormier}, {Lebouteiller}, {Galliano}, {Hony}, {Bendo}, {Smith}, {Pohlen},
  {Roussel}, {Sauvage}, {Wu}, {Sturm}, {Poglitsch}, {Contursi}, {Doublier},
  {Baes}, {Barlow}, {Boselli}, {Boquien}, {Carlson}, {Ciesla}, {Cooray},
  {Cortese}, {de Looze}, {Irwin}, {Isaak}, {Kamenetzky}, {Karczewski}, {Lu},
  {MacHattie}, {O'Halloran}, {Parkin}, {Rangwala}, {Schirm}, {Schulz},
  {Spinoglio}, {Vaccari}, {Wilson}, \& {Wozniak}}]{madden13}
{Madden}, S.~C., {R{\'e}my-Ruyer}, A., {Galametz}, M., {et~al.} 2013, PASP,
  125, 600

\bibitem[{{Magdis} {et~al.}(2012){Magdis}, {Daddi}, {B{\'e}thermin}, {Sargent},
  {Elbaz}, {Pannella}, {Dickinson}, {Dannerbauer}, {da Cunha}, {Walter},
  {Rigopoulou}, {Charmandaris}, {Hwang}, \& {Kartaltepe}}]{magdis12}
{Magdis}, G.~E., {Daddi}, E., {B{\'e}thermin}, M., {et~al.} 2012, ApJ, 760, 6

\bibitem[{{Magnelli} {et~al.}(2013){Magnelli}, {Popesso}, {Berta}, {Pozzi},
  {Elbaz}, {Lutz}, {Dickinson}, {Altieri}, {Andreani}, {Aussel},
  {B{\'e}thermin}, {Bongiovanni}, {Cepa}, {Charmandaris}, {Chary}, {Cimatti},
  {Daddi}, {F{\"o}rster Schreiber}, {Genzel}, {Gruppioni}, {Harwit}, {Hwang},
  {Ivison}, {Magdis}, {Maiolino}, {Murphy}, {Nordon}, {Pannella}, {P{\'e}rez
  Garc{\'{\i}}a}, {Poglitsch}, {Rosario}, {Sanchez-Portal}, {Santini}, {Scott},
  {Sturm}, {Tacconi}, \& {Valtchanov}}]{magnelli13}
{Magnelli}, B., {Popesso}, P., {Berta}, S., {et~al.} 2013, A\&A, 553, A132

\bibitem[{{Maiolino} {et~al.}(2008){Maiolino}, {Nagao}, {Grazian}, {Cocchia},
  {Marconi}, {Mannucci}, {Cimatti}, {Pipino}, {Ballero}, {Calura}, {Chiappini},
  {Fontana}, {Granato}, {Matteucci}, {Pastorini}, {Pentericci}, {Risaliti},
  {Salvati}, \& {Silva}}]{maiolino08}
{Maiolino}, R., {Nagao}, T., {Grazian}, A., {et~al.} 2008, A\&A, 488, 463

\bibitem[{{Marble} {et~al.}(2010){Marble}, {Engelbracht}, {van Zee}, {Dale},
  {Smith}, {Gordon}, {Wu}, {Lee}, {Kennicutt}, {Skillman}, {Johnson}, {Block},
  {Calzetti}, {Cohen}, {Lee}, \& {Schuster}}]{marble10}
{Marble}, A.~R., {Engelbracht}, C.~W., {van Zee}, L., {et~al.} 2010, The
  Astrophysical Journal, 715, 506

\bibitem[{{McMullin} {et~al.}(2007){McMullin}, {Waters}, {Schiebel}, {Young},
  \& {Golap}}]{mcmullin07}
{McMullin}, J.~P., {Waters}, B., {Schiebel}, D., {Young}, W., \& {Golap}, K.
  2007, in Astronomical Society of the Pacific Conference Series, Vol. 376,
  Astronomical Data Analysis Software and Systems XVI, ed. R.~A. {Shaw},
  F.~{Hill}, \& D.~J. {Bell}, 127

\bibitem[{{Mosleh} {et~al.}(2012){Mosleh}, {Williams}, {Franx}, {Gonzalez},
  {Bouwens}, {Oesch}, {Labbe}, {Illingworth}, \& {Trenti}}]{mosleh12}
{Mosleh}, M., {Williams}, R.~J., {Franx}, M., {et~al.} 2012, ApJL, 756, L12

\bibitem[{{M{\"u}ller} {et~al.}(2014){M{\"u}ller}, {Balog}, {Nielbock}, {Lim},
  {Teyssier}, {Olberg}, {Klaas}, {Linz}, {Altieri}, {Pearson}, {Bendo}, \&
  {Vilenius}}]{muller14}
{M{\"u}ller}, T., {Balog}, Z., {Nielbock}, M., {et~al.} 2014, Experimental
  Astronomy, 37, 253

\bibitem[{{Nanni} {et~al.}(2020){Nanni}, {Burgarella}, {Theul{\'e}},
  {C{\^o}t{\'e}}, \& {Hirashita}}]{nanni20}
{Nanni}, A., {Burgarella}, D., {Theul{\'e}}, P., {C{\^o}t{\'e}}, B., \&
  {Hirashita}, H. 2020, A\&A, 641, A168

\bibitem[{{Oliver} {et~al.}(2012){Oliver}, {Bock}, {Altieri}, {Amblard},
  {Arumugam}, {Aussel}, {Babbedge}, {Beelen}, {B{\'e}thermin}, {Blain},
  {Boselli}, {Bridge}, {Brisbin}, {Buat}, {Burgarella},
  {Castro-Rodr{\'\i}guez}, {Cava}, {Chanial}, {Cirasuolo}, {Clements},
  {Conley}, {Conversi}, {Cooray}, {Dowell}, {Dubois}, {Dwek}, {Dye}, {Eales},
  {Elbaz}, {Farrah}, {Feltre}, {Ferrero}, {Fiolet}, {Fox}, {Franceschini},
  {Gear}, {Giovannoli}, {Glenn}, {Gong}, {Gonz{\'a}lez Solares}, {Griffin},
  {Halpern}, {Harwit}, {Hatziminaoglou}, {Heinis}, {Hurley}, {Hwang}, {Hyde},
  {Ibar}, {Ilbert}, {Isaak}, {Ivison}, {Lagache}, {Le Floc'h}, {Levenson},
  {Faro}, {Lu}, {Madden}, {Maffei}, {Magdis}, {Mainetti}, {Marchetti},
  {Marsden}, {Marshall}, {Mortier}, {Nguyen}, {O'Halloran}, {Omont}, {Page},
  {Panuzzo}, {Papageorgiou}, {Patel}, {Pearson}, {P{\'e}rez-Fournon}, {Pohlen},
  {Rawlings}, {Raymond}, {Rigopoulou}, {Riguccini}, {Rizzo}, {Rodighiero},
  {Roseboom}, {Rowan-Robinson}, {S{\'a}nchez Portal}, {Schulz}, {Scott},
  {Seymour}, {Shupe}, {Smith}, {Stevens}, {Symeonidis}, {Trichas}, {Tugwell},
  {Vaccari}, {Valtchanov}, {Vieira}, {Viero}, {Vigroux}, {Wang}, {Ward},
  {Wardlow}, {Wright}, {Xu}, \& {Zemcov}}]{oliver12}
{Oliver}, S.~J., {Bock}, J., {Altieri}, B., {et~al.} 2012, MNRAS, 424, 1614

\bibitem[{{Pettini} \& {Pagel}(2004)}]{pp04}
{Pettini}, M., \& {Pagel}, B.~E.~J. 2004, MNRAS, 348, L59

\bibitem[{{Planck Collaboration} {et~al.}(2011){Planck Collaboration}, {Ade},
  {Aghanim}, {Arnaud}, {Ashdown}, {Aumont}, {Baccigalupi}, {Balbi}, {Banday},
  {Barreiro}, {Bartlett}, {Battaner}, {Benabed}, {Beno{\^\i}t}, {Bernard},
  {Bersanelli}, {Bhatia}, {Bock}, {Bonaldi}, {Bond}, {Borrill}, {Bot},
  {Bouchet}, {Boulanger}, {Bucher}, {Burigana}, {Cabella}, {Cardoso},
  {Catalano}, {Cay{\'o}n}, {Challinor}, {Chamballu}, {Chiang}, {Chiang},
  {Christensen}, {Clements}, {Colombi}, {Couchot}, {Coulais}, {Crill},
  {Cuttaia}, {Danese}, {Davies}, {Davis}, {de Bernardis}, {de Gasperis}, {de
  Rosa}, {de Zotti}, {Delabrouille}, {Delouis}, {D{\'e}sert}, {Dickinson},
  {Dobashi}, {Donzelli}, {Dor{\'e}}, {D{\"o}rl}, {Douspis}, {Dupac},
  {Efstathiou}, {En{\ss}lin}, {Finelli}, {Forni}, {Frailis}, {Franceschi},
  {Fukui}, {Galeotta}, {Ganga}, {Giard}, {Giardino}, {Giraud-H{\'e}raud},
  {Gonz{\'a}lez-Nuevo}, {G{\'o}rski}, {Gratton}, {Gregorio}, {Gruppuso},
  {Harrison}, {Helou}, {Henrot-Versill{\'e}}, {Herranz}, {Hildebrandt},
  {Hivon}, {Hobson}, {Holmes}, {Hovest}, {Hoyland}, {Huffenberger}, {Jaffe},
  {Jones}, {Juvela}, {Kawamura}, {Keih{\"a}nen}, {Keskitalo}, {Kisner},
  {Kneissl}, {Knox}, {Kurki-Suonio}, {Lagache}, {L{\"a}hteenm{\"a}ki},
  {Lamarre}, {Lasenby}, {Laureijs}, {Lawrence}, {Leach}, {Leonardi}, {Leroy},
  {Linden-V{\o}rnle}, {L{\'o}pez-Caniego}, {Lubin}, {Mac{\'\i}as-P{\'e}rez},
  {MacTavish}, {Madden}, {Maffei}, {Mandolesi}, {Mann}, {Maris},
  {Mart{\'\i}nez-Gonz{\'a}lez}, {Masi}, {Matarrese}, {Matthai}, {Mazzotta},
  {Meinhold}, {Melchiorri}, {Mendes}, {Mennella}, {Miville-Desch{\^e}nes},
  {Moneti}, {Montier}, {Morgante}, {Mortlock}, {Munshi}, {Murphy}, {Naselsky},
  {Nati}, {Natoli}, {Netterfield}, {N{\o}rgaard-Nielsen}, {Noviello},
  {Novikov}, {Novikov}, {Onishi}, {Osborne}, {Pajot}, {Paladini}, {Paradis},
  {Pasian}, {Patanchon}, {Perdereau}, {Perotto}, {Perrotta}, {Piacentini},
  {Piat}, {Plaszczynski}, {Pointecouteau}, {Polenta}, {Ponthieu}, {Poutanen},
  {Pr{\'e}zeau}, {Prunet}, {Puget}, {Reach}, {Rebolo}, {Reinecke}, {Renault},
  {Ricciardi}, {Riller}, {Ristorcelli}, {Rocha}, {Rosset}, {Rowan-Robinson},
  {Rubi{\~n}o-Mart{\'\i}n}, {Rusholme}, {Sandri}, {Savini}, {Scott},
  {Seiffert}, {Smoot}, {Starck}, {Stivoli}, {Stolyarov}, {Sudiwala}, {Sygnet},
  {Tauber}, {Terenzi}, {Toffolatti}, {Tomasi}, {Torre}, {Tristram}, {Tuovinen},
  {Umana}, {Valenziano}, {Varis}, {Vielva}, {Villa}, {Vittorio}, {Wade},
  {Wandelt}, {Wilkinson}, {Ysard}, {Yvon}, {Zacchei}, \& {Zonca}}]{planck11}
{Planck Collaboration}, {Ade}, P.~A.~R., {Aghanim}, N., {et~al.} 2011, A\&A,
  536, A17

\bibitem[{{Planck Collaboration} {et~al.}(2014){Planck Collaboration}, {Ade},
  {Aghanim}, {Armitage-Caplan}, {Arnaud}, {Ashdown}, {Atrio-Barandela},
  {Aumont}, {Baccigalupi}, {Banday}, {Barreiro}, {Bartlett}, {Battaner},
  {Benabed}, {Beno{\^\i}t}, {Benoit-L{\'e}vy}, {Bernard}, {Bersanelli},
  {Bethermin}, {Bielewicz}, {Blagrave}, {Bobin}, {Bock}, {Bonaldi}, {Bond},
  {Borrill}, {Bouchet}, {Boulanger}, {Bridges}, {Bucher}, {Burigana}, {Butler},
  {Cardoso}, {Catalano}, {Challinor}, {Chamballu}, {Chen}, {Chiang}, {Chiang},
  {Christensen}, {Church}, {Clements}, {Colombi}, {Colombo}, {Couchot},
  {Coulais}, {Crill}, {Curto}, {Cuttaia}, {Danese}, {Davies}, {Davis}, {de
  Bernardis}, {de Rosa}, {de Zotti}, {Delabrouille}, {Delouis}, {D{\'e}sert},
  {Dickinson}, {Diego}, {Dole}, {Donzelli}, {Dor{\'e}}, {Douspis}, {Dupac},
  {Efstathiou}, {En{\ss}lin}, {Eriksen}, {Finelli}, {Forni}, {Frailis},
  {Franceschi}, {Galeotta}, {Ganga}, {Ghosh}, {Giard}, {Giraud-H{\'e}raud},
  {Gonz{\'a}lez-Nuevo}, {G{\'o}rski}, {Gratton}, {Gregorio}, {Gruppuso},
  {Hansen}, {Hanson}, {Harrison}, {Helou}, {Henrot-Versill{\'e}},
  {Hern{\'a}ndez-Monteagudo}, {Herranz}, {Hildebrandt}, {Hivon}, {Hobson},
  {Holmes}, {Hornstrup}, {Hovest}, {Huffenberger}, {Jaffe}, {Jaffe}, {Jones},
  {Juvela}, {Kalberla}, {Keih{\"a}nen}, {Kerp}, {Keskitalo}, {Kisner},
  {Kneissl}, {Knoche}, {Knox}, {Kunz}, {Kurki-Suonio}, {Lacasa}, {Lagache},
  {L{\"a}hteenm{\"a}ki}, {Lamarre}, {Langer}, {Lasenby}, {Laureijs},
  {Lawrence}, {Leonardi}, {Le{\'o}n-Tavares}, {Lesgourgues}, {Liguori},
  {Lilje}, {Linden-V{\o}rnle}, {L{\'o}pez-Caniego}, {Lubin},
  {Mac{\'\i}as-P{\'e}rez}, {Maffei}, {Maino}, {Mandolesi}, {Maris}, {Marshall},
  {Martin}, {Mart{\'\i}nez-Gonz{\'a}lez}, {Masi}, {Massardi}, {Matarrese},
  {Matthai}, {Mazzotta}, {Melchiorri}, {Mendes}, {Mennella}, {Migliaccio},
  {Mitra}, {Miville-Desch{\^e}nes}, {Moneti}, {Montier}, {Morgante},
  {Mortlock}, {Munshi}, {Murphy}, {Naselsky}, {Nati}, {Natoli}, {Netterfield},
  {N{\o}rgaard-Nielsen}, {Noviello}, {Novikov}, {Novikov}, {Osborne},
  {Oxborrow}, {Paci}, {Pagano}, {Pajot}, {Paladini}, {Paoletti}, {Partridge},
  {Pasian}, {Patanchon}, {Perdereau}, {Perotto}, {Perrotta}, {Piacentini},
  {Piat}, {Pierpaoli}, {Pietrobon}, {Plaszczynski}, {Pointecouteau}, {Polenta},
  {Ponthieu}, {Popa}, {Poutanen}, {Pratt}, {Pr{\'e}zeau}, {Prunet}, {Puget},
  {Rachen}, {Reach}, {Rebolo}, {Reinecke}, {Remazeilles}, {Renault},
  {Ricciardi}, {Riller}, {Ristorcelli}, {Rocha}, {Rosset}, {Roudier},
  {Rowan-Robinson}, {Rubi{\~n}o-Mart{\'\i}n}, {Rusholme}, {Sandri}, {Santos},
  {Savini}, {Scott}, {Seiffert}, {Serra}, {Shellard}, {Spencer}, {Starck},
  {Stolyarov}, {Stompor}, {Sudiwala}, {Sunyaev}, {Sureau}, {Sutton},
  {Suur-Uski}, {Sygnet}, {Tauber}, {Tavagnacco}, {Terenzi}, {Toffolatti},
  {Tomasi}, {Tristram}, {Tucci}, {Tuovinen}, {T{\"u}rler}, {Valenziano},
  {Valiviita}, {Van Tent}, {Vielva}, {Villa}, {Vittorio}, {Wade}, {Wandelt},
  {Welikala}, {White}, {White}, {Winkel}, {Yvon}, {Zacchei}, \&
  {Zonca}}]{planck14}
---. 2014, A\&A, 571, A30

\bibitem[{{Popping} {et~al.}(2017){Popping}, {Somerville}, \&
  {Galametz}}]{popping17}
{Popping}, G., {Somerville}, R.~S., \& {Galametz}, M. 2017, \mnras, 471, 3152

\bibitem[{{Popping} {et~al.}(2019){Popping}, {Pillepich}, {Somerville},
  {Decarli}, {Walter}, {Aravena}, {Carilli}, {Cox}, {Nelson}, {Riechers},
  {Weiss}, {Boogaard}, {Bouwens}, {Contini}, {Cortes}, {da Cunha}, {Daddi},
  {D{\'\i}az-Santos}, {Diemer}, {Gonz{\'a}lez-L{\'o}pez}, {Hernquist},
  {Ivison}, {Le F{\`e}vre}, {Marinacci}, {Rix}, {Swinbank}, {Vogelsberger},
  {van der Werf}, {Wagg}, \& {Yung}}]{popping19}
{Popping}, G., {Pillepich}, A., {Somerville}, R.~S., {et~al.} 2019, ApJ, 882,
  137

\bibitem[{{Popping} {et~al.}(2021){Popping}, {Pillepich}, {Calistro Rivera},
  {Schulz}, {Hernquist}, {Kaasinen}, {Marinacci}, {Nelson}, \&
  {Vogelsberger}}]{popping21}
{Popping}, G., {Pillepich}, A., {Calistro Rivera}, G., {et~al.} 2021, arXiv
  e-prints, arXiv:2101.12218

\bibitem[{Reddy {et~al.}(2012a)Reddy, Dickinson, Elbaz, Morrison, Giavalisco,
  Ivison, Papovich, Scott, Buat, Burgarella, Charmandaris, Daddi, Magdis,
  Murphy, Altieri, Aussel, Dannerbauer, Dasyra, Hwang, Kartaltepe, Leiton,
  Magnelli, \& Popesso}]{reddy12a}
Reddy, N., Dickinson, M., Elbaz, D., {et~al.} 2012a, ApJ, 744, 154

\bibitem[{Reddy {et~al.}(2010)Reddy, Erb, Pettini, Steidel, \&
  Shapley}]{reddy10}
Reddy, N.~A., Erb, D.~K., Pettini, M., Steidel, C.~C., \& Shapley, A.~E. 2010,
  ApJ, 712, 1070

\bibitem[{Reddy {et~al.}(2012b)Reddy, Pettini, Steidel, Shapley, Erb, \&
  Law}]{reddy12b}
Reddy, N.~A., Pettini, M., Steidel, C.~C., {et~al.} 2012b, ApJ, 754, 25

\bibitem[{Reddy {et~al.}(2015)Reddy, Kriek, Shapley, Freeman, Siana, Coil,
  Mobasher, Price, Sanders, \& Shivaei}]{reddy15}
Reddy, N.~A., Kriek, M., Shapley, A.~E., {et~al.} 2015, ApJ, 806, 259

\bibitem[{{Reddy} {et~al.}(2018{\natexlab{a}}){Reddy}, {Oesch}, {Bouwens},
  {Montes}, {Illingworth}, {Steidel}, {van Dokkum}, {Atek}, {Carollo},
  {Cibinel}, {Holden}, {Labb{\'e}}, {Magee}, {Morselli}, {Nelson}, \&
  {Wilkins}}]{reddy18a}
{Reddy}, N.~A., {Oesch}, P.~A., {Bouwens}, R.~J., {et~al.} 2018{\natexlab{a}},
  ApJ, 853, 56

\bibitem[{{Reddy} {et~al.}(2018{\natexlab{b}}){Reddy}, {Shapley}, {Sanders},
  {Kriek}, {Coil}, {Shivaei}, {Freeman}, {Mobasher}, {Siana}, {Azadi},
  {Fetherolf}, {Fornasini}, {Leung}, {Price}, {Zick}, \& {Barro}}]{reddy18b}
{Reddy}, N.~A., {Shapley}, A.~E., {Sanders}, R.~L., {et~al.}
  2018{\natexlab{b}}, ApJ, 869, 92

\bibitem[{{Reddy} {et~al.}(2020){Reddy}, {Shapley}, {Kriek}, {Steidel},
  {Shivaei}, {Sanders}, {Mobasher}, {Coil}, {Siana}, {Freeman}, {Azadi},
  {Fetherolf}, {Leung}, {Price}, \& {Zick}}]{reddy20}
{Reddy}, N.~A., {Shapley}, A.~E., {Kriek}, M., {et~al.} 2020, ApJ, 902, 123

\bibitem[{{Reddy} {et~al.}(2021){Reddy}, {Topping}, {Shapley}, {Steidel},
  {Sanders}, {Du}, {Coil}, {Mobasher}, \& {Price}}]{reddy21}
{Reddy}, N.~A., {Topping}, M.~W., {Shapley}, A.~E., {et~al.} 2021, arXiv
  e-prints, arXiv:2108.05363

\bibitem[{{Remjian} {et~al.}(2019){Remjian}, {Biggs}, {Cortes}, {Dent}, {Di
  Franceso}, {Fomalont}, {Hales}, {Kameno}, {Mason}, {Philips}, {Saini}, {Vila
  Vilaro}, \& {Villard}}]{remjian19}
{Remjian}, A., {Biggs}, A., {Cortes}, P.~A., {et~al.} 2019, {ALMA Technical
  Handbook,ALMA Doc. 7.3, ver. 1.1}, 2019, doi:10.5281/zenodo.4511522

\bibitem[{{R{\'e}my-Ruyer} {et~al.}(2013){R{\'e}my-Ruyer}, {Madden},
  {Galliano}, {Hony}, {Sauvage}, {Bendo}, {Roussel}, {Pohlen}, {Smith},
  {Galametz}, {Cormier}, {Lebouteiller}, {Wu}, {Baes}, {Barlow}, {Boquien},
  {Boselli}, {Ciesla}, {De Looze}, {Karczewski}, {Panuzzo}, {Spinoglio},
  {Vaccari}, \& {Wilson}}]{remyruyer13}
{R{\'e}my-Ruyer}, A., {Madden}, S.~C., {Galliano}, F., {et~al.} 2013, A\&A,
  557, A95

\bibitem[{{R{\'e}my-Ruyer} {et~al.}(2014){R{\'e}my-Ruyer}, {Madden},
  {Galliano}, {Galametz}, {Takeuchi}, {Asano}, {Zhukovska}, {Lebouteiller},
  {Cormier}, {Jones}, {Bocchio}, {Baes}, {Bendo}, {Boquien}, {Boselli},
  {DeLooze}, {Doublier-Pritchard}, {Hughes}, {Karczewski}, \&
  {Spinoglio}}]{remyruyer14}
---. 2014, A\&A, 563, A31

\bibitem[{{R{\'e}my-Ruyer} {et~al.}(2015){R{\'e}my-Ruyer}, {Madden},
  {Galliano}, {Lebouteiller}, {Baes}, {Bendo}, {Boselli}, {Ciesla}, {Cormier},
  {Cooray}, {Cortese}, {De Looze}, {Doublier-Pritchard}, {Galametz}, {Jones},
  {Karczewski}, {Lu}, \& {Spinoglio}}]{remyruyer15}
---. 2015, A\&A, 582, A121

\bibitem[{{Rezaee} {et~al.}(2021){Rezaee}, {Reddy}, {Shivaei}, {Fetherolf},
  {Emami}, \& {Khostovan}}]{rezaee21}
{Rezaee}, S., {Reddy}, N., {Shivaei}, I., {et~al.} 2021, MNRAS, 506, 3588

\bibitem[{{Rieke} {et~al.}(2009){Rieke}, {Alonso-Herrero}, {Weiner},
  {P{\'e}rez-Gonz{\'a}lez}, {Blaylock}, {Donley}, \& {Marcillac}}]{rieke09}
{Rieke}, G.~H., {Alonso-Herrero}, A., {Weiner}, B.~J., {et~al.} 2009, ApJ, 692,
  556

\bibitem[{{Rieke} {et~al.}(2008){Rieke}, {Blaylock}, {Decin}, {Engelbracht},
  {Ogle}, {Avrett}, {Carpenter}, {Cutri}, {Armus}, {Gordon}, {Gray}, {Hinz},
  {Su}, \& {Willmer}}]{rieke08}
{Rieke}, G.~H., {Blaylock}, M., {Decin}, L., {et~al.} 2008, AJ, 135, 2245

\bibitem[{{Rodighiero} {et~al.}(2006){Rodighiero}, {Lari}, {Pozzi},
  {Gruppioni}, {Fadda}, {Franceschini}, {Lonsdale}, {Surace}, {Shupe}, \&
  {Fang}}]{rodighiero06}
{Rodighiero}, G., {Lari}, C., {Pozzi}, F., {et~al.} 2006, \mnras, 371, 1891

\bibitem[{{Rujopakarn} {et~al.}(2011){Rujopakarn}, {Rieke}, {Eisenstein}, \&
  {Juneau}}]{rujopakarn11}
{Rujopakarn}, W., {Rieke}, G.~H., {Eisenstein}, D.~J., \& {Juneau}, S. 2011,
  ApJ, 726, 93

\bibitem[{{Rujopakarn} {et~al.}(2013){Rujopakarn}, {Rieke}, {Weiner},
  {P{\'e}rez-Gonz{\'a}lez}, {Rex}, {Walth}, \& {Kartaltepe}}]{rujopakarn13}
{Rujopakarn}, W., {Rieke}, G.~H., {Weiner}, B.~J., {et~al.} 2013, ApJ, 767, 73

\bibitem[{{Rujopakarn} {et~al.}(2019){Rujopakarn}, {Daddi}, {Rieke}, {Puglisi},
  {Schramm}, {P{\'e}rez-Gonz{\'a}lez}, {Magdis}, {Alberts}, {Bournaud},
  {Elbaz}, {Franco}, {Kawinwanichakij}, {Kohno}, {Narayanan}, {Silverman},
  {Wang}, \& {Williams}}]{rujopakarn19}
{Rujopakarn}, W., {Daddi}, E., {Rieke}, G.~H., {et~al.} 2019, ApJ, 882, 107

\bibitem[{{Safarzadeh} {et~al.}(2015){Safarzadeh}, {Ferguson}, {Lu}, {Inami},
  \& {Somerville}}]{safarzadeh15}
{Safarzadeh}, M., {Ferguson}, H.~C., {Lu}, Y., {Inami}, H., \& {Somerville},
  R.~S. 2015, \apj, 798, 91

\bibitem[{{Salim} {et~al.}(2007){Salim}, {Rich}, {Charlot}, {Brinchmann},
  {Johnson}, {Schiminovich}, {Seibert}, {Mallery}, {Heckman}, {Forster},
  {Friedman}, {Martin}, {Morrissey}, {Neff}, {Small}, {Wyder}, {Bianchi},
  {Donas}, {Lee}, {Madore}, {Milliard}, {Szalay}, {Welsh}, \& {Yi}}]{salim07}
{Salim}, S., {Rich}, R.~M., {Charlot}, S., {et~al.} 2007, ApJS, 173, 267

\bibitem[{{Sanders} {et~al.}(2007){Sanders}, {Salvato}, {Aussel}, {Ilbert},
  {Scoville}, {Surace}, {Frayer}, {Sheth}, {Helou}, {Brooke}, {Bhattacharya},
  {Yan}, {Kartaltepe}, {Barnes}, {Blain}, {Calzetti}, {Capak}, {Carilli},
  {Carollo}, {Comastri}, {Daddi}, {Ellis}, {Elvis}, {Fall}, {Franceschini},
  {Giavalisco}, {Hasinger}, {Impey}, {Koekemoer}, {Le F{\`e}vre}, {Lilly},
  {Liu}, {McCracken}, {Mobasher}, {Renzini}, {Rich}, {Schinnerer}, {Shopbell},
  {Taniguchi}, {Thompson}, {Urry}, \& {Williams}}]{sanders07}
{Sanders}, D.~B., {Salvato}, M., {Aussel}, H., {et~al.} 2007, ApJS, 172, 86

\bibitem[{{Sanders} {et~al.}(2015){Sanders}, {Shapley}, {Kriek}, {Reddy},
  {Freeman}, {Coil}, {Siana}, {Mobasher}, {Shivaei}, {Price}, \& {de
  Groot}}]{sanders15}
{Sanders}, R.~L., {Shapley}, A.~E., {Kriek}, M., {et~al.} 2015, ApJ, 799, 138

\bibitem[{{Sanders} {et~al.}(2016a){Sanders}, {Shapley}, {Kriek}, {Reddy},
  {Freeman}, {Coil}, {Siana}, {Mobasher}, {Shivaei}, {Price}, \& {de
  Groot}}]{sanders16a}
---. 2016a, ApJ, 816, 23

\bibitem[{{Sanders} {et~al.}(2018){Sanders}, {Shapley}, {Kriek}, {Freeman},
  {Reddy}, {Siana}, {Coil}, {Mobasher}, {Dav{\'e}}, {Shivaei}, {Azadi},
  {Price}, {Leung}, {Fetherolf}, {de Groot}, {Zick}, {Fornasini}, \&
  {Barro}}]{sanders18}
---. 2018, ApJ, 858, 99

\bibitem[{{Sanders} {et~al.}(2020){Sanders}, {Shapley}, {Reddy}, {Kriek},
  {Siana}, {Coil}, {Mobasher}, {Shivaei}, {Freeman}, {Azadi}, {Price}, {Leung},
  {Fetherolf}, {de Groot}, {Zick}, {Fornasini}, \& {Barro}}]{sanders20a}
{Sanders}, R.~L., {Shapley}, A.~E., {Reddy}, N.~A., {et~al.} 2020, MNRAS, 491,
  1427

\bibitem[{{Sanders} {et~al.}(2021){Sanders}, {Shapley}, {Jones}, {Reddy},
  {Kriek}, {Siana}, {Coil}, {Mobasher}, {Shivaei}, {Dav{\'e}}, {Azadi},
  {Price}, {Leung}, {Freeman}, {Fetherolf}, {de Groot}, {Zick}, \&
  {Barro}}]{sanders21}
{Sanders}, R.~L., {Shapley}, A.~E., {Jones}, T., {et~al.} 2021, ApJ, 914, 19

\bibitem[{{Santini} {et~al.}(2014){Santini}, {Maiolino}, {Magnelli}, {Lutz},
  {Lamastra}, {Li Causi}, {Eales}, {Andreani}, {Berta}, {Buat}, {Cooray},
  {Cresci}, {Daddi}, {Farrah}, {Fontana}, {Franceschini}, {Genzel}, {Granato},
  {Grazian}, {Le Floc'h}, {Magdis}, {Magliocchetti}, {Mannucci}, {Menci},
  {Nordon}, {Oliver}, {Popesso}, {Pozzi}, {Riguccini}, {Rodighiero}, {Rosario},
  {Salvato}, {Scott}, {Silva}, {Tacconi}, {Viero}, {Wang}, {Wuyts}, \&
  {Xu}}]{santini14}
{Santini}, P., {Maiolino}, R., {Magnelli}, B., {et~al.} 2014, A\&A, 562, A30

\bibitem[{{Schinnerer} {et~al.}(2016){Schinnerer}, {Groves}, {Sargent},
  {Karim}, {Oesch}, {Magnelli}, {LeFevre}, {Tasca}, {Civano}, {Cassata}, \&
  {Smol{\v{c}}i{\'c}}}]{schinnerer16}
{Schinnerer}, E., {Groves}, B., {Sargent}, M.~T., {et~al.} 2016, \apj, 833, 112

\bibitem[{{Schreiber} {et~al.}(2018){Schreiber}, {Elbaz}, {Pannella}, {Ciesla},
  {Wang}, \& {Franco}}]{schreiber18}
{Schreiber}, C., {Elbaz}, D., {Pannella}, M., {et~al.} 2018, Astronomy and
  Astrophysics, 609, A30

\bibitem[{{Scoville} {et~al.}(2016){Scoville}, {Sheth}, {Aussel}, {Vanden
  Bout}, {Capak}, {Bongiorno}, {Casey}, {Murchikova}, {Koda},
  {{\'A}lvarez-M{\'a}rquez}, {Lee}, {Laigle}, {McCracken}, {Ilbert}, {Pope},
  {Sanders}, {Chu}, {Toft}, {Ivison}, \& {Manohar}}]{scoville16}
{Scoville}, N., {Sheth}, K., {Aussel}, H., {et~al.} 2016, ApJ, 820, 83

\bibitem[{{Shapley} {et~al.}(2020){Shapley}, {Cullen}, {Dunlop}, {McLure},
  {Kriek}, {Reddy}, \& {Sanders}}]{shapley20}
{Shapley}, A.~E., {Cullen}, F., {Dunlop}, J.~S., {et~al.} 2020, ApJL, 903, L16

\bibitem[{{Shapley} {et~al.}(2015){Shapley}, {Reddy}, {Kriek}, {Freeman},
  {Sanders}, {Siana}, {Coil}, {Mobasher}, {Shivaei}, {Price}, \& {de
  Groot}}]{shapley15}
{Shapley}, A.~E., {Reddy}, N.~A., {Kriek}, M., {et~al.} 2015, ApJ, 801, 88

\bibitem[{{Shapley} {et~al.}(2021){Shapley}, {Sanders}, {Salim}, {Reddy},
  {Kriek}, {Mobasher}, {Coil}, {Siana}, {Price}, {Shivaei}, {Dunlop}, {McLure},
  \& {Cullen}}]{shapley21}
{Shapley}, A.~E., {Sanders}, R.~L., {Salim}, S., {et~al.} 2021, arXiv e-prints,
  arXiv:2109.14630

\bibitem[{{Shipley} {et~al.}(2016){Shipley}, {Papovich}, {Rieke}, {Brown}, \&
  {Moustakas}}]{shipley16}
{Shipley}, H.~V., {Papovich}, C., {Rieke}, G.~H., {Brown}, M.~J.~I., \&
  {Moustakas}, J. 2016, ApJ, 818, 60

\bibitem[{{Shivaei} {et~al.}(2020{\natexlab{a}}){Shivaei}, {Darvish},
  {Sattari}, {Chartab}, {Mobasher}, {Scoville}, \& {Rieke}}]{shivaei20b}
{Shivaei}, I., {Darvish}, B., {Sattari}, Z., {et~al.} 2020{\natexlab{a}},
  \apjl, 903, L28

\bibitem[{{Shivaei} {et~al.}(2015b){Shivaei}, {Reddy}, {Shapley}, {Kriek},
  {Siana}, {Mobasher}, {Coil}, {Freeman}, {Sanders}, {Price}, {de Groot}, \&
  {Azadi}}]{shivaei15b}
{Shivaei}, I., {Reddy}, N.~A., {Shapley}, A.~E., {et~al.} 2015b, ApJ, 815, 98

\bibitem[{{Shivaei} {et~al.}(2016){Shivaei}, {Kriek}, {Reddy}, {Shapley},
  {Barro}, {Conroy}, {Coil}, {Freeman}, {Mobasher}, {Siana}, {Sanders},
  {Price}, {Azadi}, {Pasha}, \& {Inami}}]{shivaei16}
{Shivaei}, I., {Kriek}, M., {Reddy}, N.~A., {et~al.} 2016, ApJL, 820, L23

\bibitem[{{Shivaei} {et~al.}(2017){Shivaei}, {Reddy}, {Shapley}, {Siana},
  {Kriek}, {Mobasher}, {Coil}, {Freeman}, {Sanders}, {Price}, {Azadi}, \&
  {Zick}}]{shivaei17}
{Shivaei}, I., {Reddy}, N.~A., {Shapley}, A.~E., {et~al.} 2017, ApJ, 837, 157

\bibitem[{{Shivaei} {et~al.}(2018){Shivaei}, {Reddy}, {Siana}, {Shapley},
  {Kriek}, {Mobasher}, {Freeman}, {Sanders}, {Coil}, {Price}, {Fetherolf},
  {Azadi}, {Leung}, \& {Zick}}]{shivaei18}
{Shivaei}, I., {Reddy}, N.~A., {Siana}, B., {et~al.} 2018, ApJ, 855, 42

\bibitem[{{Shivaei} {et~al.}(2020{\natexlab{b}}){Shivaei}, {Reddy}, {Rieke},
  {Shapley}, {Kriek}, {Battisti}, {Mobasher}, {Sanders}, {Fetherolf}, {Azadi},
  {Coil}, {Freeman}, {de Groot}, {Leung}, {Price}, {Siana}, \&
  {Zick}}]{shivaei20a}
{Shivaei}, I., {Reddy}, N., {Rieke}, G., {et~al.} 2020{\natexlab{b}}, ApJ, 899,
  117

\bibitem[{Skelton {et~al.}(2014)Skelton, Whitaker, Momcheva, Brammer, van
  Dokkum, Labb\'{e}, Franx, van~der Wel, Bezanson, Da~Cunha, Fumagalli,
  F\"{o}rster~Schreiber, Kriek, Leja, Lundgren, Magee, Marchesini, Maseda,
  Nelson, Oesch, Pacifici, Patel, Price, Rix, Tal, Wake, \& Wuyts}]{skelton14}
Skelton, R.~E., Whitaker, K.~E., Momcheva, I.~G., {et~al.} 2014, ApJS, 214, 24

\bibitem[{{Smith} {et~al.}(2012){Smith}, {Dunne}, {da Cunha}, {Rowlands},
  {Maddox}, {Gomez}, {Bonfield}, {Charlot}, {Driver}, {Popescu}, {Tuffs},
  {Dunlop}, {Jarvis}, {Seymour}, {Symeonidis}, {Baes}, {Bourne}, {Clements},
  {Cooray}, {De Zotti}, {Dye}, {Eales}, {Scott}, {Verma}, {van der Werf},
  {Andrae}, {Auld}, {Buttiglione}, {Cava}, {Dariush}, {Fritz}, {Hopwood},
  {Ibar}, {Ivison}, {Kelvin}, {Madore}, {Pohlen}, {Rigby}, {Robotham},
  {Seibert}, \& {Temi}}]{smith12}
{Smith}, D.~J.~B., {Dunne}, L., {da Cunha}, E., {et~al.} 2012, MNRAS, 427, 703

\bibitem[{{Smith} {et~al.}(2017){Smith}, {Ibar}, {Maddox}, {Valiante}, {Dunne},
  {Eales}, {Dye}, {Furlanetto}, {Bourne}, {Cigan}, {Ivison}, {Gomez}, {Smith},
  \& {Viaene}}]{smith17}
{Smith}, M. W.~L., {Ibar}, E., {Maddox}, S.~J., {et~al.} 2017, ApJS, 233, 26

\bibitem[{{Somerville} \& {Dav{\'e}}(2015)}]{somerville15}
{Somerville}, R.~S., \& {Dav{\'e}}, R. 2015, ARAA, 53, 51

\bibitem[{Speagle {et~al.}(2014)Speagle, Steinhardt, Capak, \&
  Silverman}]{speagle14}
Speagle, J.~S., Steinhardt, C.~L., Capak, P.~L., \& Silverman, J.~D. 2014,
  ApJS, 214, 15

\bibitem[{{Steidel} {et~al.}(2016){Steidel}, {Strom}, {Pettini}, {Rudie},
  {Reddy}, \& {Trainor}}]{steidel16}
{Steidel}, C.~C., {Strom}, A.~L., {Pettini}, M., {et~al.} 2016, ApJ, 826, 159

\bibitem[{Steidel {et~al.}(2014)Steidel, Rudie, Strom, Pettini, Reddy, Shapley,
  Trainor, Erb, Turner, Konidaris, Kulas, Mace, Matthews, \&
  McLean}]{steidel14}
Steidel, C.~C., Rudie, G.~C., Strom, A.~L., {et~al.} 2014, ApJ, 795, 165

\bibitem[{{Strom} {et~al.}(2018){Strom}, {Steidel}, {Rudie}, {Trainor}, \&
  {Pettini}}]{strom18}
{Strom}, A.~L., {Steidel}, C.~C., {Rudie}, G.~C., {Trainor}, R.~F., \&
  {Pettini}, M. 2018, ApJ, 868, 117

\bibitem[{{Tabatabaei} {et~al.}(2014){Tabatabaei}, {Braine}, {Xilouris},
  {Kramer}, {Boquien}, {Combes}, {Henkel}, {Relano}, {Verley}, {Gratier},
  {Israel}, {Wiedner}, {R{\"o}llig}, {Schuster}, \& {van der
  Werf}}]{tabatabaei14}
{Tabatabaei}, F.~S., {Braine}, J., {Xilouris}, E.~M., {et~al.} 2014, A\&A, 561,
  A95

\bibitem[{{Tacconi} {et~al.}(2013){Tacconi}, {Neri}, {Genzel}, {Combes},
  {Bolatto}, {Cooper}, {Wuyts}, {Bournaud}, {Burkert}, {Comerford}, {Cox},
  {Davis}, {F{\"o}rster Schreiber}, {Garc{\'\i}a-Burillo}, {Gracia-Carpio},
  {Lutz}, {Naab}, {Newman}, {Omont}, {Saintonge}, {Shapiro Griffin}, {Shapley},
  {Sternberg}, \& {Weiner}}]{tacconi13}
{Tacconi}, L.~J., {Neri}, R., {Genzel}, R., {et~al.} 2013, ApJ, 768, 74

\bibitem[{{Tacconi} {et~al.}(2018){Tacconi}, {Genzel}, {Saintonge}, {Combes},
  {Garc{\'\i}a-Burillo}, {Neri}, {Bolatto}, {Contini}, {F{\"o}rster Schreiber},
  {Lilly}, {Lutz}, {Wuyts}, {Accurso}, {Boissier}, {Boone}, {Bouch{\'e}},
  {Bournaud}, {Burkert}, {Carollo}, {Cooper}, {Cox}, {Feruglio}, {Freundlich},
  {Herrera-Camus}, {Juneau}, {Lippa}, {Naab}, {Renzini}, {Salome}, {Sternberg},
  {Tadaki}, {{\"U}bler}, {Walter}, {Weiner}, \& {Weiss}}]{tacconi18}
{Tacconi}, L.~J., {Genzel}, R., {Saintonge}, A., {et~al.} 2018, ApJ, 853, 179

\bibitem[{{Tasca} {et~al.}(2015){Tasca}, {Le F{\`e}vre}, {Hathi}, {Schaerer},
  {Ilbert}, {Zamorani}, {Lemaux}, {Cassata}, {Garilli}, {Le Brun}, {Maccagni},
  {Pentericci}, {Thomas}, {Vanzella}, {Zucca}, {Amorin}, {Bardelli},
  {Cassar{\`a}}, {Castellano}, {Cimatti}, {Cucciati}, {Durkalec}, {Fontana},
  {Giavalisco}, {Grazian}, {Paltani}, {Ribeiro}, {Scodeggio}, {Sommariva},
  {Talia}, {Tresse}, {Vergani}, {Capak}, {Charlot}, {Contini}, {de la Torre},
  {Dunlop}, {Fotopoulou}, {Koekemoer}, {L{\'o}pez-Sanjuan}, {Mellier}, {Pforr},
  {Salvato}, {Scoville}, {Taniguchi}, \& {Wang}}]{tasca15}
{Tasca}, L.~A.~M., {Le F{\`e}vre}, O., {Hathi}, N.~P., {et~al.} 2015, A\&A,
  581, A54

\bibitem[{{Theios} {et~al.}(2019){Theios}, {Steidel}, {Strom}, {Rudie},
  {Trainor}, \& {Reddy}}]{theios19}
{Theios}, R.~L., {Steidel}, C.~C., {Strom}, A.~L., {et~al.} 2019, The
  Astrophysical Journal, 871, 128

\bibitem[{{Topping} {et~al.}(2020{\natexlab{a}}){Topping}, {Shapley}, {Reddy},
  {Sanders}, {Coil}, {Kriek}, {Mobasher}, \& {Siana}}]{topping20b}
{Topping}, M.~W., {Shapley}, A.~E., {Reddy}, N.~A., {et~al.}
  2020{\natexlab{a}}, MNRAS, 499, 1652

\bibitem[{{Topping} {et~al.}(2020{\natexlab{b}}){Topping}, {Shapley}, {Reddy},
  {Sanders}, {Coil}, {Kriek}, {Mobasher}, \& {Siana}}]{topping20a}
---. 2020{\natexlab{b}}, MNRAS, 495, 4430

\bibitem[{{Toribio San Cipriano} {et~al.}(2017){Toribio San Cipriano},
  {Dom{\'\i}nguez-Guzm{\'a}n}, {Esteban}, {Garc{\'\i}a-Rojas}, {Mesa-Delgado},
  {Bresolin}, {Rodr{\'\i}guez}, \& {Sim{\'o}n-D{\'\i}az}}]{cipriano17}
{Toribio San Cipriano}, L., {Dom{\'\i}nguez-Guzm{\'a}n}, G., {Esteban}, C.,
  {et~al.} 2017, MNRAS, 467, 3759

\bibitem[{{Troncoso} {et~al.}(2014){Troncoso}, {Maiolino}, {Sommariva},
  {Cresci}, {Mannucci}, {Marconi}, {Meneghetti}, {Grazian}, {Cimatti},
  {Fontana}, {Nagao}, \& {Pentericci}}]{troncoso14}
{Troncoso}, P., {Maiolino}, R., {Sommariva}, V., {et~al.} 2014, A\&A, 563, A58

\bibitem[{{van der Wel} {et~al.}(2014){van der Wel}, {Franx}, {van Dokkum},
  {Skelton}, {Momcheva}, {Whitaker}, {Brammer}, {Bell}, {Rix}, {Wuyts},
  {Ferguson}, {Holden}, {Barro}, {Koekemoer}, {Chang}, {McGrath},
  {H{\"a}ussler}, {Dekel}, {Behroozi}, {Fumagalli}, {Leja}, {Lundgren},
  {Maseda}, {Nelson}, {Wake}, {Patel}, {Labb{\'e}}, {Faber}, {Grogin}, \&
  {Kocevski}}]{vanderwel14}
{van der Wel}, A., {Franx}, M., {van Dokkum}, P.~G., {et~al.} 2014, ApJ, 788,
  28

\bibitem[{{Ventura} {et~al.}(2012{\natexlab{a}}){Ventura}, {di Criscienzo},
  {Schneider}, {Carini}, {Valiante}, {D'Antona}, {Gallerani}, {Maiolino}, \&
  {Tornamb{\'e}}}]{ventura12b}
{Ventura}, P., {di Criscienzo}, M., {Schneider}, R., {et~al.}
  2012{\natexlab{a}}, MNRAS, 424, 2345

\bibitem[{{Ventura} {et~al.}(2012{\natexlab{b}}){Ventura}, {di Criscienzo},
  {Schneider}, {Carini}, {Valiante}, {D'Antona}, {Gallerani}, {Maiolino}, \&
  {Tornamb{\'e}}}]{ventura12a}
---. 2012{\natexlab{b}}, MNRAS, 420, 1442

\bibitem[{{Walter} {et~al.}(2008){Walter}, {Brinks}, {de Blok}, {Bigiel},
  {Kennicutt}, {Thornley}, \& {Leroy}}]{walter08}
{Walter}, F., {Brinks}, E., {de Blok}, W.~J.~G., {et~al.} 2008, AJ, 136, 2563

\bibitem[{{Walter} {et~al.}(2016){Walter}, {Decarli}, {Aravena}, {Carilli},
  {Bouwens}, {da Cunha}, {Daddi}, {Ivison}, {Riechers}, {Smail}, {Swinbank},
  {Weiss}, {Anguita}, {Assef}, {Bacon}, {Bauer}, {Bell}, {Bertoldi}, {Chapman},
  {Colina}, {Cortes}, {Cox}, {Dickinson}, {Elbaz}, {G{\'o}nzalez-L{\'o}pez},
  {Ibar}, {Inami}, {Infante}, {Hodge}, {Karim}, {Le Fevre}, {Magnelli}, {Neri},
  {Oesch}, {Ota}, {Popping}, {Rix}, {Sargent}, {Sheth}, {van der Wel}, {van der
  Werf}, \& {Wagg}}]{walter16}
{Walter}, F., {Decarli}, R., {Aravena}, M., {et~al.} 2016, ApJ, 833, 67

\bibitem[{{Whitaker} {et~al.}(2017){Whitaker}, {Pope}, {Cybulski}, {Casey},
  {Popping}, \& {Yun}}]{whitaker17}
{Whitaker}, K.~E., {Pope}, A., {Cybulski}, R., {et~al.} 2017, \apj, 850, 208

\bibitem[{{Ysard} {et~al.}(2019){Ysard}, {Koehler}, {Jimenez-Serra}, {Jones},
  \& {Verstraete}}]{ysard19}
{Ysard}, N., {Koehler}, M., {Jimenez-Serra}, I., {Jones}, A.~P., \&
  {Verstraete}, L. 2019, A\&A, 631, A88

\bibitem[{{Zheng} {et~al.}(2006){Zheng}, {Bell}, {Rix}, {Papovich}, {Le
  Floc'h}, {Rieke}, \& {P{\'e}rez-Gonz{\'a}lez}}]{zheng06}
{Zheng}, X.~Z., {Bell}, E.~F., {Rix}, H.-W., {et~al.} 2006, \apj, 640, 784

\bibitem[{{Zick} {et~al.}(2018){Zick}, {Kriek}, {Shapley}, {Reddy}, {Freeman},
  {Siana}, {Coil}, {Azadi}, {Barro}, {Fetherolf}, {Fornasini}, {de Groot},
  {Leung}, {Mobasher}, {Price}, {Sand ers}, \& {Shivaei}}]{zick18}
{Zick}, T.~O., {Kriek}, M., {Shapley}, A.~E., {et~al.} 2018, ApJL, 867, L16

\bibitem[{{Zubko} {et~al.}(2004){Zubko}, {Dwek}, \& {Arendt}}]{zubko04}
{Zubko}, V., {Dwek}, E., \& {Arendt}, R.~G. 2004, ApJS, 152, 211

\end{thebibliography}

\end{document}